\documentclass[preprint,rmp,aps]{revtex4}
\usepackage{graphics}

\begin{document}


\title{The world according to the Hubble Space Telescope}

\author{Mario Livio}
\email[]{mlivio@stsci.edu}
\homepage[]{http://www-int.stsci.edu/~mlivio/}

\affiliation{Space Telescope Science Institute, Baltimore, MD 21218}

\date{\today}

\begin{abstract}
The Hubble Space Telescope (HST), in its thirteen years of operation, has allowed us to observe properties of the universe humans have been able, until very recently, to probe only with their thoughts. This review presents a brief summary of a few of the highlights of HST discoveries, discusses their physical implications, and identifies unsolved problems. A broad range of topics is covered, from our own solar system to cosmology.  The topics fall into the general categories of: planets (including both in the solar system and extrasolar), stellar evolution, black holes (including both of stellar-mass and supermassive), galaxy formation and evolution, the determination of cosmological parameters, and the nature of the recently discovered ``dark energy.''
\end{abstract}


\maketitle

\tableofcontents
\clearpage

\section{INTRODUCTION}

When Galileo Galilei peered through his small telescope 400 years ago, it resulted in an unprecedented period of astronomical discovery.  There is no doubt that the Hubble Space Telescope (HST; Fig.~1) in its first 13~years of operation has had a similarly profound impact on astronomical research.  But the Hubble Space Telescope did much more that that.  It literally brought a glimpse of the wonders of the universe into millions of homes worldwide, thereby inspiring an unprecedented public curiosity and interest in science.

The Hubble Space Telescope has seen farther and sharper than any optical/UV/IR telescope before it.  Unlike astronomical experiments that were dedicated to a single, very specific goal (like the Cosmic Background Explorer or the Wilkinson Microwave Anisotropy Probe), the Hubble Space Telescope's achievements are generally not of the type of singular discoveries.  More often, the Hubble Space Telescope has taken what were existing hints and suspicions from ground-based observations and has turned them into certainty.  In other cases, the level of detail that HST has provided forced theorists to re-think previous broad-brush models and to construct new ones that would be consistent with the superior emerging data.  In a few instances, the availability of HST's razor-sharp resoution at critical events provided unique insights into inividual phenomena.  In total, by observing tens of thousands of astronomical targets, the Hubble Space Telescope has contributed significantly to essentially all the topics of current astronomical resarch, covering objects from our own solar system to the most distant galaxies.

The 12.5-ton orbiting observatory was launched into orbit on April 24, 1990.  It circles the Earth every 90~minutes in a 330~nmi (607~km) orbit and operates around the clock above all but the thinnest remnants of the Earth's atmosphere.  The telescope provides information to many individual astronomers and teams of scientists worldwide, and it is engaged in the study of virtually all the astronomical constituents of the universe.

Crucial to fulfilling the objective of a 20-year mission is a series of on-orbit manned servicing missions. The first servicing mission (SM1) took place in 1993, the second (SM2) was flown in 1997, and the third was separated into two parts, one (SM3A) in 1999 and the second (SM3B) in 2002.  A fourth mission is currently planned for the end of 2004 (pending the investigation into the shuttle Columbia disaster).  During these missions, shuttle astronauts upgrade the the observatory's capabilities (by installing new instruments) and perform planned maintenance activities and necessary repairs.  To facilitate this process, the telescope was designed so that its science instruments and several vital engineering subsystems have been configured as modular packages with standardized fittings accessible to astronauts in pressurized suits.

Hubble was designed to provide three basic capabilities, all of which are essential for innovative astronomical research:
\begin{enumerate}
\item High angular resolution---the ability to image fine details.  The spatial resolution of HST in the optical regime is about 0.05 arcseconds.
\item Ultraviolet performance---using the fact that the observatory is above the Earth's atmosphere to produce ultraviolet images and spectra.  The Hubble Space Telescope is sensitive down to a wavelength of 1150~\AA\ (below the Lyman Alpha line of hydrogen).
\item High sensitivity---the ability to detect and take spectra of even very faint objects.  The Hubble Space Telescope is about a hundred to a thousand times more sensitive in the ultraviolet than the previous space observatory (the International Ultraviolet Explorer). 
\end{enumerate}
Astronomers and astrophysicists using HST data have published over 3000 scientific papers to date.  Hubble's findings are thus far too numerous to be described even briefly in one article.  The following sections therefore simply highlight a few of the astronomical discoveries which, in my clearly biased view, have significantly advanced our understanding of the cosmos. In the spirit of progressing by ``powers of ten'' in cosmic distance, I shall start with the solar system and finish with cosmological distances. I apologize in advance to the many astronomers whose important work will not be presented in this article.  I also regretfully acknowledge the incompleteness of the list of references.  My list should be regarded as representative rather than comprehensive. Finally, the different sections are not intended to be exhaustive reviews---they are concise summaries of HST contributions to major astrophysical poblems.

\section{OUR OWN BACKYARD---THE SOLAR SYSTEM}

The Hubble Space Telescope has proved well suited for the study of objects in our own solar system.  In fact, it has produced images of the outer planets that approach the clarity of those obtained from Voyager flybys.  The Hubble Space Telescope's images of Mars have also been surpassed only by close-up shots taken by visiting space probes. The Hubble Space Telescope's images of the auroras of Jupiter and Saturn, of the disintegration of comet LINEAR, of the binary Kuiper belt object KBO~1998~WW31, and of a global dust storm on Mars have been truly spectacular and informative. However, no observation could compete with the drama provided by Comet Shoemaker-Levy~9 (SL-9) slamming into Jupiter.

\subsection{The collision of Comet SL-9 with Jupiter}

From a cosmic perspective, the impact of the ninth periodic comet discovered by the team of Carolyn and Gene Shoemaker and David Levy on Jupiter was unremarkable.  The cratered surfaces of many planetary bodies and satellites and the prehistoric mass extinctions on Earth are a testament to the fact that such impacts are ubiquitous in the history of the solar system. Furthermore, even the ``seeds'' of life on Earth may have been planted by such a bombardment.

From a human perspective, however, this collision marked a ``once in a lifetime'' event, which caused the mobilization of all the astronomical resources for an unprecedented observational campaign in July 1994.

The Hubble Space Telescope played a crucial role in the SL-9 campaign, both in the initial characterization of the comet fragments and in producing unique images and spectra of the impact events themselves and later of the impact sites.

The multiple comet SL-9 was discovered in late-March 1993 
\cite{Shoemaker:1993}.  Subsequent images taken with HST in July of 1993 
\cite{Weaver:1994} revealed that the comet had spectacularly split into 20-odd fragments, probably because of tidal disruption (during a previous close approach to Jupiter), producing a train of fragments fittingly dubbed a ``string of pearls'' (Fig.~2).

The conditions for tidal break-up of a self-gravitating incompressible cometary sphere can generally be written in the form 
\cite[e.g.,][]{Dobrovolskis:1990,Sekanina:1994}
\begin{equation}
\frac{\rho_c}{\rho_p}\left( \frac{d}{R_p}\right)^3 \,{\raise-.5ex\hbox{$\buildrel<\over\sim$}}\,\ 
k\frac{P_c}{U}~,
\end{equation}
where $\rho_c$, $\rho_p$ are the densities of the comet and planet, respectively, $R_p$ is the planet's radius, $d$ is the distance of the comet from the planet, $P_c$ is the comet's central pressure, $U$ is a measure of the comet's tensile strength, and $k$ is a constant of order unity. The facts that: (i)~secondary fragmentation was observed by HST, and (ii)~the gradual disappearance of condensations was also well documented by HST 
\cite{Weaver:1995}, argued strongly for tidal splitting of a discrete nucleus as a result of gradual fissure propogation.  Computations of the orbit suggested that the comet  would collide with Jupiter in July 1994 
\cite{Nakano:1993,Yeomans:1993}.

The impact of the first fragment (fragment~``A'') occurred on July~16, 1994, followed by all the other fragments smashing into Jupiter's atmosphere over the following week. The Hubble Space Telescope's images taken during the impact events 
\cite{Hammel:1995} showed spectacular plumes resembling nuclear ``mushrooms'' above the limb of Jupiter (Fig.~3).  Typically, the plumes attained altitudes of about 3000~km within 6--8 minutes of impact, and were seen falling and spreading within 10~minutes of impact.  The manner in which the height of the plume scales with SL-9 parameters and the ambient pressure can be obtained from the following simple considerations 
\cite[e.g.,][]{Chyba:1993,Field:1995,Zahnle:1994}. Once aerodynamic forces overcome material strength, the incoming fragment deforms to form a flattened ``pancake'' due to the difference in ram pressure acting on the front and on the sides.  Due to the increase in the cross-section (and concomitantly in the drag), the fragment is brought to an abrupt halt, accompanied by an explosive energy release.  If we denote the energy being released per steradian by $E_i/4\pi$, then the obtained ejection velocity is given by
\begin{equation}
V_{ej}\simeq \sqrt{\frac{E_i}{4\pi\rho_a H^3}} = \sqrt{\frac{E_ig}{4\pi P_a H^2}}~,
\end{equation}
where $\rho_a$, $P_a$ are the ambient density and pressure, respectively, $H$ is the pressure scale height in the atmosphere and $g$ is the gravitational acceleration.  The plume height, therefore, scales like
\begin{equation}
h_{\rm plume}\simeq\frac{V^2_{ej}}{2g}\sim\frac{E_i}{P_a H^2}.
\end{equation}
Detailed numerical simulations have confirmed the basic physical picture described above 
\cite[e.g.,][]{MacLow:1996}.

A combination of the observations from the Keck Telescope, HST, Palomar, and the Galileo spacecraft, eventually produced a comprehensive explanation for the observed $2.3\mu$ lightcurve 
\cite[shown in Fig.~4 for fragment~R;][]{Graham:1995}.
The sequence of events can be described phenomenologically by the cartoon in Fig.~5 
\cite{Zahnle:1996}.

The first precursor marked the meteor trail (with a brightening timescale of order $H/v\sim1s)$. As explained above, most of the fragment's kinetic energy was released at the last second of its existence, resulting in an explosive fireball.  The second precursor marks the rise of the fireball above Jupiter's limb, having already cooled to $\sim500$--700~K 
\cite{Carlson:1995}.  As the fireball expanded, silicates condensed, followed by carbonaceous matter. The sunlit parts of these condensates were the tracers that marked the plume seen by HST. The main peak in the light curve was caused by the ejecta plume falling back onto the atmosphere 
\cite{Nicholson:1995}.  The rise and fallback occurred on a timescale of $\sqrt{8h_{\rm plume}/g}\sim 10^3$~s (for a plume height of $\sim3000$~km). The final, smaller peak, was probably caused by material bouncing off the atmosphere and falling back.

Another interesting phenomenon revealed by the HST images of the impact \textit{sites} was the propagation of waves (Fig.~6).  The radius of the prominent dark ring in Fig.~6 is 3700~km, and a smaller ring, of radius 1750~km is also visible 
\cite{Hammel:1995}.

Measurements of the radii of the circular rings as a function of time showed that the ``fast'' wave was seen to spread at a constant speed of 450~m~s$^{-1}$, while the speed of the ``slow'' wave (corresponding to the fainter ring) was somewhat less constrained.  
\textcite{Ingersoll:1996} examined many possibilities for the identification of the fast wave, including $f$-modes, $p$-modes, $r$-modes and $g$-modes.  These waves are classified by the different restoring forces that cause them to propagate.  In acoustic waves ($p$-modes) the restoring force is compressibility; in surface waves (like those in the ocean; $f$-modes) it is gravity; in inertial oscillations ($r$-modes) it is the Coriolis force, and in internal gravity waves ($g$-modes) it is gravity. Ingersoll and Kanamori showed that the only waves that match the observed speed of 450~m~s$^{-1}$ are $g$-modes propagating in a water cloud which has an O/H ratio that is ten times higher than that ratio in the Sun, $e_\mathrm{H_2O}=10$ (the waves were too slow to be acoustic waves and moved at too constant a speed to be rings of debris).  The propagation speed of these modes varies as the square root of the water abundance. The requirement of a high O/H ratio was found to be in conflict with data from the Galileo probe, which suggested that from above the cloud tops down to the 20~bar level the water was \textit{less} abundant than in the Sun.  The cause for this discrepancy is still unknown, although it has been suggested that the probe may have entered a clearing in the clouds (a 
``5-micron hot spot'') associated with a dry downdraft.  This uncertainty demonstrates that even the highest quality data do not provide all the answers. Nevertheless, there is no question that the impact of comet SL-9 on Jupiter provided a unique opportunity to study such events.

\section{EXTRASOLAR PLANETS}

The solar system was not the only place to provide for planetary drama.  During the past eight years we have witnessed the number of known extrasolar planets orbiting Sun-like stars going from \textit{zero} to about one hundred at the time of this writing! The human fascination with the possibility of discovering extraterrestrial life has helped turn planet detection into one of the major frontiers of today's astronomy.

The Hubble Space Telescope's contributions to this exciting field have been related mostly to the discovery of circumstellar disks from which planets form, and these will be described in Section~IV. In two cases, however, HST has produced three truly unique sets of observations, one in a globular star cluster, and two following the light curve and spectrum of a transiting planet.

\subsection{Where have all the planets gone?}

The discovery of a planet orbiting the star 51~Peg 
\cite{Mayor:1995} in a remarkably tight orbit of 4.231 days challenged prevailing theoretical views and marked the start of a ``Golden Age'' in planet discoveries by radial velocity surveys 
\cite[see, for example,][]{Marcy:2000}.  The existence of gas giant planets in very close orbits (at 
0.04--0.05~AU) enables even \textit{photometric} searches for these planets (``hot Jupiters'').  At such small separations, the probability of the planet transiting (``eclipsing'') the parent star, given random orbital inclinations, is about 10\%. More precisely, from Kepler's laws, the \textit{duration} of a transit in hours (reduced by $\pi/4$ for average chord length) is given by
\begin{equation}
\tau_\mathrm{tran}\simeq1.4 M_*^{-1/3} R_* P_\mathrm{orb}^{1/3}~\mathrm{hr}~,
\end{equation}
where $M_*$, $R_*$ are the stellar mass and radius in solar units and $P_\mathrm{orb}$ is the orbital period in days.  The \textit{probability} of a transit per system is given by 
\begin{equation}
Pr_\mathrm{tran}\simeq23.8\% M_*^{-1/3} R_* P_\mathrm{orb}^{-2/3}~.
\end{equation}

Transits are very important because of the following reasons: (a)~With transits one can immediately obtain the \textit{size} of the planet [the transit depth is equal to $(R_p/R_*)^2$].  (b)~In the case of transits, measurements of radial velocities immediately provide the planet's \textit{mass} (and not just $m_p\sin i$, where $i$ is the otherwise unknown orbital inclination). (c)~A knowledge of the mass and radius can be used to test theoretical 
mass-radius relations for close-orbiting planets.

In 1998, a team of astronomers led by Ron Gilliland of the Space Telescope Science Institute embarked on the ambitious program of searching for transits in some 40,000 stars in the core of the globular cluster 47~Tucanae (Fig.~7).  This cluster is a particularly interesting target for such a search for several reasons. (i)~For a typical star in the cluster, $M_*=0.81$, $R_*=0.92$, and a typical ``hot Jupiter'' orbital period (e.g., $P_\mathrm{orb}=3.8$ days), the probability of transit is 9.6\% and the duration is 2.2~hours.  For a planet of radius $R_p=1.3R_J$ 
\cite[where $R_J$ is Jupiter's radius, as observed for the transiting system HD~209458;][]{Charbonneau:2000,Henry:2000},
the expected transit depth is about 2\%, easily detectable by HST. (ii)~Given that the frequency of ``hot Jupiters'' with orbital periods shorter than 5~days (ten are known at the time of writing) in the solar neighborhood is about 0.8\%--1\%, with a 10\% chance of transit one expects roughly one in every 1000 surveyed stars in 47~Tuc to show a transit (and therefore a few tens of detections when monitoring $\sim40,000$ stars). (iii)~The globular cluster 47~Tuc is about 11~billion years old and it has a total metallicity (abundance of elements heavier than helium) of about one-third that of the Sun 
\cite{Salaris:1998}.  The frequency of planets in such an environment can therefore have important implications for models of planet formation.

The Hubble Space Telescope observed 34,091 stars in 47~Tuc, obtaining time series photometry over a period of 8.3~days (continuously, except for Earth occultations and passages through the South Atlantic Anomaly).  A more detailed calculation of the expected number of transits, taking into account the actual distribution of $V$~magnitudes of the stars in the sample 
\cite{Gilliland:2000} and time series noise, showed that \textit{17~transits should have been seen}, assuming the same frequency of hot Jupiters as in the solar neighborhood.  None was detected!  Undoubtedly, this was not a problem with the detection method. As many as 75~variable stars have actually been detected.  Furthermore, an eclipsing system with a period of 1.34~days and an eclipse depth of $\sim3$\% was actually discovered.  However, the presence of a secondary eclipse in the system showed that this was a grazing eclipse by a low-mass (K~dwarf) star and not a planet.

The direct implication is therefore that with a very high degree of confidence the frequency of hot Jupiters in 47~Tuc is \textit{at most 1/10} that in the solar neighborhood.

The main question is then:  why are there much fewer orbiting planets in 47~Tuc?  There are, in principle, two possible explanations for the paucity of planets in orbits around globular cluster stars: (i)~Planets do not form at all (or form much more rarely) in globular clusters. (ii)~Planets are torn from their parent stars.

Planet \textit{formation} may be suppressed in old stellar clusters because of two main reasons:  low metallicity, or photoevaporation of protoplanetary disks.

Conventional wisdom of giant planet formation suggests that this is a two-step process.  In the first, the collisional accumulation of rocky and big  planetesimals leads to a runaway growth of a solid core 
\cite[e.g.,][]{Lissauer:1987} of about ten Earth masses (10~M$_{\oplus}$). In the second, the core acquires a gaseous atmosphere, which eventually collapses, leading to the rapid accretion of hydrogen and helium from the protoplanetary disk.  One might suspect, therefore, that in low-metallicity environments, the nucleation of dust grains would be greatly suppressed, and concomitantly planets would form more rarely. Furthermore, it has generally been noted that stars which bear planets are, on average, enriched in metals in comparison to the Sun 
\cite[e.g.,][]{Gonzalez:1997,Butler:2000,Laughlin:2000,Reid:2002}, and this trend is even more pronounced for the short-period planets.  Thus, the paucity of hot Jupiters in 47~Tuc may simply reflect the cluster's low metallicity. 

There is another reason why giant planet \textit{formation} may be suppressed in dense stellar clusters.  The disk around any star in a cluster is exposed to ultraviolet radiation from massive cluster stars.  This radiation heats the disk surface, raises the sound speed, $c_s$, and causes gas beyond a radius of
\begin{equation}
R_g\simeq0.5\frac{GM_*}{c_s^2}
\end{equation}
to become unbound and flow away as a thermal wind. The mass-loss rate in such a photoevaporation flow scales as 
\cite[e.g.,][]{Bertoldi:1990,Johnstone:1998}
\begin{equation}
\dot{M}_\mathrm{outflow}\sim\Phi^{1/2} d^{-1}~,
\end{equation}
where $\Phi$ is the flux of ionizing photons and $d$ is the disk's distance from the emitting source.  In a star forming region like Orion, the ionizing flux is of order $\Phi\sim10^{49}$~s$^{-1}$, of the same order as the flux expected in a cluster with $n_\mathrm{cluster}\sim10^3$ members. The mass loss rate from one of the disks in Orion, HST 182$-$413, has been estimated from HST observations to be $\dot{M}_\mathrm{outflow}\sim4\times10^{-7}$~M$_{\odot}$~yr$^{-1}$. Since the mass of that disk is estimated to be 
\cite{Johnstone:1998} $M_\mathrm{disk}\sim0.04$~M$_{\odot}$, the disk lifetime is expected to be $\tau_\mathrm{disk}\sim M_\mathrm{disk}/\dot{M}_\mathrm{outflow}\sim10^5$~yr.  For somewhat smaller disk sizes (e.g., like that in the solar system $\sim30$~AU) the disk lifetime can be somewhat extended (the EUV-induced flow scales as $r_\mathrm{disk}^{3/2}$).  Nevertheless, the disk lifetime may be significantly shorter than the planet formation timescale, estimated to be of order $10^6$--$10^7$~yrs 
\cite[e.g.,][]{Shu:1993,Pollack:1996}. 
\textcite{Armitage:2000} calculated the expected disk lifetime as a function of the cluster richness. He found that giant planet formation is strongly suppressed in custers with $\sim10^5$ stars out to around 1~pc or more.

There are two ways in which, in principle at least, the conclusion of a low probability for planet formation in clusters, due to disk destruction, may be avoided: (i)~If giant planet formation occurs not via a core accretion mechanism, but rather through direct and rapid (within $\sim10^3$~yr) fragmentation (due to a gravational instability) into self-gravitating clumps of gas 
\cite[e.g.,][]{Boss:2000,Mayer:2002}.
(ii)~If the formation of high-mass stars (which produce the photoevaporating radiation) is significantly  delayed (by $\sim10^6$~yr) compared to the formation of low mass stars. At present both of these processes are sufficiently uncertain that no definitive conclusion about their viability can be drawn.

The second possible explanation for the absence of ``hot Jupiters'' in 47~Tuc is that even if giant planets do form, they are torn away from their parent stars because of encounters in the crowded cluster environment, and thus cannot exhibit transits. This possibility has been investigated in some detail by 
\textcite{Davies:2001}. The cross-section for two stars with a relative velocity at infinity $V_{\infty}$ to pass within $R_\mathrm{min}$ from each other is given by
\begin{equation}
\sigma=\pi R_\mathrm{min}^2 
\left(1+ \frac{V^2}{V_{\infty}^2} \right)~,
\end{equation}
where $V$ is the relative velocity ($V^2\propto (M_1+M_2)/R_\mathrm{min}$) at closest approach.  The second term expresses the effect of gravitational focusing.  When $V\gg V_{\infty}$, as is expected in globular clusters, $\sigma\propto R_\mathrm{min}$. The timescale for an encounter between a star-planet system and a star is of order ($\tau_\mathrm{enc}=1/n\sigma V)$
\begin{equation}
\tau_\mathrm{enc}\simeq3\times10^8\ \mathrm{yr}
\left( \frac{n}{10^5\ \mathrm{pc}^{-3}} \right)^{-1}
\left( \frac{V_{\infty}}{10\ \mathrm{km~s}^{-1}} \right)
\left( \frac{d}{1~\mathrm{AU}} \right)^{-1}
\left( \frac{M}{\mathrm{M}_{\odot}} \right)^{-1}~,
\end{equation}
where $n$ is the number density of stars of mass $M$ and $d$ is the semimajor axis of the planetary orbit. Encounters therefore occur on a timescale that is much shorter than a Hubble time in a cluster like 47~Tuc, which has a density of about $1.5\times10^5$~pc$^{-3}$ at its center and about an order of magnitude lower at its half-mass radius 
\cite[e.g.,][]{Howell:2000}. Generally, binaries (with masses $M_1$ and $M_2$ for the primary and secondary, respectively) in which the binding energy is lower than the kinetic energy of the colliding star (of mass $M_3$) are disrupted. The condition for breakup is
\begin{equation}
V_{\infty}^2\,{\raise-.5ex\hbox{$\buildrel>\over\sim$}}\,
\frac{G M_1 M_2(M_1+M_2+M_3)}{M_3(M_1+M_2)d}~.
\end{equation}
Since $M_2\ll M_1,M_3$, in the case of a star-planet system, breakup of the system is expected to occur with a high probability when the third star passes within a distance $d$. 
\textcite{Davies:2001} performed a large number of simulations for various values of $V_{\infty}/V_\mathrm{orb}$ and considered both encounters (of star-planet systems) with single stars and with binaries.  They found that wide ($d\,{\raise-.5ex\hbox{$\buildrel>\over\sim$}}\,$0.3~AU) planetary systems are likely to be broken up within the half-mass radius of 47~Tuc, but tighter systems 
($d\,{\raise-.5ex\hbox{$\buildrel<\over\sim$}}\,$0.1~AU) or systems in less dense regions may survive. The results of Davies and Sigurdsson still do not provide a complete explanation for the absence of transits in 47~Tuc. Future observations, perhaps of less dense, somewhat higher metallicity clusters (like NGC~6352) will be required to explain which factor is responsible for the dearth of planetary transits in 47~Tuc.  An attempt to detect transits associated with Galactic bulge and disk stars (that span 1.5~dex in metallicity, and are in an environment that is orders of magnitude less dense than the center of a globular cluster)  would also be very valuable. An understanding of the environments that are either conducive to or prohibitive for the existence of planetary transits may constitute an important step in the study of planet formation.

\subsection{The transiting planet HD 209458}

The low-mass companion to the star HD~209458 was the first extrasolar planet found to transit the optical disk of its parent star 
\cite{Charbonneau:2000,Henry:2000,Mazeh:2000}. 
As I explained in the previous section, transits offer a unique opportunity to determine the properties of the orbiting planet. The Hubble Space Telescope followed four transits \cite{Brown:2001}
and obtained an exquisitely detailed light curve (Fig.~8; note that the depth of the eclipse is only 1.7\%!).  The orbital period was determined to be 3.5247 days, the lower limit on the mass of the planet $M_p\sin i =0.69\pm0.05M_J$ (with the inclination angle $i= 86^\mathrm{o}\llap{.}6\pm9^\mathrm{o}\llap{.}14$), and the radius of the planet $R_p=1.347\pm0.060~R_J$. Most impressively, the precision of the HST light curve allowed even for searches for \textit{rings} around the planet, for constraints on planetary \textit{satellites}, and for probing the planet's \textit{atmosphere}. 

A ring system with significant opacity around the planet would cause distortions of the light curve relative to that of a spherical body. In particular, one would expect small dips in the light curve before the first and after the fourth contact. The observations were consistent with \textit{no rings} (maximum ring radius consistent with observations was 1.8~$R_p$). 

A satellite orbiting the planet might be detectable either from its photometric signature (the satellite would block light in addition to that obstructed by the planet), or from its influence on the orbital motion of the planet.  \textcite{Brown:2001} 
showed that satellites larger than 1.2~R$_{\oplus}$ (where R$_{\oplus}$ is the Earth radius), or with masses larger than 3~M$_{\oplus}$ are excluded by the data.

Finally, precision spectrophotometric observations in the region of the sodium resonance doublet at 589.3~nm revealed that the photometric dimming during transit was deeper by ($2.32\pm0.57)\times10^{-4}$ relative to simultaneous observations in adjacent bands \cite{Charbonneau:2002}.
This additional dimming has been interpreted as absorption by sodium in the planet's atmosphere.  In fact, the existence of a detectable sodium feature in the spectrum has been predicted by theoretical models \cite[e.g.,][]{Seager:2000}. 
Furthermore, observations of brown dwarfs (stellar objects below the hydrogen burning limit) with similar surface temperatures show strong absorption in alkali metal lines 
\cite[e.g.,][]{Burrows:2000}.
HD~209458 therefore represents the first \emph{detection of an atmosphere} of an extrasolar planet. While not a shocking result in itself, since giant planets are clearly expected to have atmospheres, the actual detection marks the beginning of a new era in extrasolar planets research.  

Even more interestingly perhaps, HST observations of HD~209458 during three transits revealed atomic hydrogen absorption of $\sim$15\% over the stellar Lyman~$\alpha$ line 
\cite{Vidal:2003}. A comparison of the observations with models showed that the observations could be explained in terms of hydrogen atoms evaporating and escaping the planet in an asymmetric cometary-like tail. To account for the observed absorption depth, the simulations implied a minimum escape flow rate of $\sim10^{10}$~g~s$^{-1}$.

In the future, other spectral features (e.g., water vapor; with implications for the searches for extraterrestrial life) can be searched for. Furthermore, in principle, HST can detect even the secondary eclipse (when the planet is eclipsed by the star), expected to be characterized by a $\sim10^{-5}$ decrease in the luminosity. This will allow for a determination of the planet's albedo, and for a comparison with models of planetary atmospheres. All of these results, whether already obtained, or expected, are somewhat of a surprise, since HST has been heralded primarily as a ``cosmology machine,'' not as an exquisite tool in exo-planetary research.

Moving now from planets to stars, the Hubble Space Telescope has been particularly instrumental in revealing new details of stellar births and deaths.

\section{STARS AT BIRTH AND DEATH}

Using HST's high resolution on one hand and its infrared capabilities on the other, astronomers have been able to probe the dusty environments of young stellar objects (YSOs) and of stars in their late evolutionary stages with unprecedented detail.  While these observations have not revolutionized the field, they have, on one hand, led to a much deeper understanding of the processes involved, and on the other, opened an entirely new set of questions, while producing some of the most spectacular images.

\subsection{Outflows and jets from young stellar objects}

Molecular clouds are the reservoirs of mass and angular momentum from which stars are born. In the initial phase, dynamical infall occurs. The finite angular momentum of infalling cloud material leads to the formation of an accretion disk. Angular momentum is transported outward in this disk resulting in the accretion of mass (and some angular momentum) onto the central object 
\cite[e.g.,][]{Najita:2000,Shu:1994,Camenzind:1990}.

For a long time it has been known that outflows and collimated jets are signposts of stellar birth \cite[see, for example,][]{Konigl:1989,Reipurth:1993}.
In particular, many of the radiative shocks known as Herbig-Haro (HH) objects were found to be associated with highly collimated jets emanating from the vicinity of YSOs 
\cite[e.g.,][]{Reipurth:1999}. 
In some cases, these jets were found to be several parsecs in length \cite{Reipurth:1998}.
Figure~9 shows a few remarkable HST images of jets from YSOs.

In spite of the ubiquity of astrophysical jets, we still lack a comprehensive theory of their acceleration and collimation 
\cite[see e.g., review by][]{Livio:2000}. The most promising mechanism relies on an accretion disk (around a central compact object like a YSO or a black hole) threaded by a large-scale poloidal magnetic field.  Some magnetic flux is assumed to be in open field lines, inclined by an angle of more than $30^{\circ}$ to the vertical to the disk's surface. Ionized gas is forced to flow along field lines. Since the foot points of these lines are anchored in the disk and rotate with it, material is accelerated centrifugally like beads on rotating wires \cite{Blandford:1982,Ogilvie:2001}. In this picture, the acceleration basically stops at the Alfv\'en surface, where the kinetic energy density in the jet becomes comparable to the magnetic energy density.  Collimation, however, occurs primarily outside the Alfv\'en surface. There, the field gets wound up by the rotation, generating toroidal loops. The curvature force exerted by the toroidal field acts in the direction of the rotation axis, thus achieving collimation by ``hoop stresses.'' Alternatively, in a vertical field of the form $B_z\sim(r^2/R_\mathrm{in}+1)^{-1/2}$ (where $R_\mathrm{in}$ is the radius at the accretion disk's inner edge), in which the flux is largest at the outer disk, the field lines have a naturally collimating shape \cite[e.g.,][]{Spruit:1996}. Physically, poloidal collimation is achieved in this case as the material encounters the high flux in the outer disk (especially if $R_\mathrm{disk}\sim R_\mathrm{Alfv\acute{e}n}$).

While this theoretical picture of jets being accelerated and collimated by accretion disks has been largely developed prior to any HST observations, HST has provided the first \textit{direct} evidence for the fact that jets indeed originate at the centers of disks (at least in the case of YSOs). Observations of the objects HH~30 \cite[][Fig.~10]{Burrows:1996}, 
DG~Tau (Fig.~9), and a few other YSOs clearly show the jet emanating from the disk center, and they reveal even the illuminated upper and lower disk surfaces. 

Furthermore, the high resolution of HST has enabled the determination of proper motions in YSO jets, from images obtained over a time interval of a few years \cite[e.g.,][]{Bally:2003}. 
The highest velocities have been observed along the jet axis. Typically, the velocities are of order 200 to 400 km~s$^{-1}$. Since these are precisely of the order of the Keplerian velocity in the inner disk around a YSO, the proper motion observations provide additional support for the jet formation scenario described above. Incidentally, similar observations of the optical jet in the active galaxy M87 also showed proper motion. In this case at the apparent ``superluminal'' speed was of 4c--6c 
\cite{Biretta:1999}, again corresponding to the fact that one expects relativistic speeds from the vicinity of a central black hole.

Another related phenomenon that only HST could discover is the existence of externally irradiated small jets \cite[``microjets'';][]{Bally:2000}.
Many of these microjets (Fig.~11) are only about $0.1''$ wide, and are therefore usually undetectable against the nebular background in ground-based observations. For irradiated jets, observations of the H$\alpha$ surface brightness, the emission measure ($EM=\int n_e^2 dx$; where $n_e$ is the electron density and $x$ is the linear size of the emitting region), and the jet width, allow for a determination of the electron density. Together with the jet speed one can therefore obtain the rate of mass loss in the jet, $\dot{M}_j$. Typically, the irradiated small jets in the Orion nebula are characterized by $\dot{M}_j\sim
10^{-9}$~M$_{\odot}$~yr$^{-1}$, at least an order of magnitude lower than the rates observed in the long jets associated with Herbig-Haro objects.  Generally, the rate of mass loss in jets from YSOs is found to be about 1--10\% of the rate at which mass is accreted through the disk onto the young stellar object.

In addition to the important disk-jet connection, HST observations of YSOs have provided another interesting new element, this time related to planet formation.

\subsection{Protoplanetary disks}

Contrary perhaps to the expectation that protoplanetary disks would be deeply embedded within the clouds from which they form, and they would therefore be inaccessible to optical observations, HST revealed many dozens of protoplanetary disks 
\cite[``proplyds''; e.g.,][following the initial correct identification by Churchwell \emph{et~al.}\ 1987 and Meaburn 1988]{ODell:2001,ODell:1993,ODell:1996,Bally:2000}. Many of these disks are seen silhouetted against the background nebular light (when they are shielded from photoionization), with some possessing ionized skins and tails 
\cite[e.g.,][Fig.~12]{Henney:1999,Bally:2000}.

The ubiquity of the protoplanetary dust disks \cite[they are seen in 55\%--97\% of stars;][]{Hillebrand:1998,Lada:2000} demonstrates that at least the raw materials for planet formation are in place around many young stars. Indeed, in a few cases, like the dust ring and disk in HR~4796A and the nearly edge-on disk surrounding Beta Pictoris, the detailed HST images reveal gaps and warping (respectively) that could represent the effects of orbiting planets 
\cite{Schneider:1999,Kalas:2000}.

Another aspect of the protoplanetary disks that is significant for planet formation is the discovery of evaporating disks in the Orion Nebula. As was noted in Section IIIA, some of the Orion proplyds were shown to be evaporating (due to photo-ablation by UV radiation from young, nearby stars) at rates of 
$\dot{M}\sim10^{-7}$ to $10^{-6}$~M$_\odot$ yr$^{-1}$ 
\cite[e.g.,][]{Henney:1999}. Given that the masses of these disks are typically of order 
$10^{-2}$~M$_\odot$ (if normal interstellar grains are assumed, so that the observed dust emission can be scaled to the total mass), this implies lifetimes for these disks of $10^5$~years or less. There exists, however, some evidence that the grain sizes in Orion's disks may, in fact, be relatively large---perhaps of the order of millimeters 
\cite{Throop:2000}. The latter conclusion is based on the fact that the outer portions of the disks appear to be gray (they do not redden background light), and on the failure to detect the disks at radio wavelengths in spite of the implied large extinction in the infrared (hiding the central star in some cases). The observations are thus consistent with grain sizes in excess of the radio wavelength used, of 1.3~mm.  When we think about the potential implications of these two findings (about disk lifetimes and grain sizes), we realize that they may have interesting consequences for the \textit{demographics} of planets in Orion. The relatively short disk lifetimes but relative large grain sizes may mean that while rocky (terrestrial) planets can form in these strongly irradiated environments, giant planets (that require the accretion of hydrogen and helium from the protoplanetary disk) cannot \cite[unless their formation process is extremely fast;][]{Boss:2000,Mayer:2002}. It is nevertheless clear from the many observations of ``hot Jupiters'' (giant planets with orbital radii 
${\raise-.5ex\hbox{$\buildrel<\over\sim$}}\,0.05$~AU) that less extreme environments do exist, in which giant planets not only form, but also have sufficient time to gravitationally interact with their parent disk and migrate inward, to produce the distribution in orbital separations we observe today 
\cite[see, for example,][]{Lin:1996,Armitage:2002}.

While disks around young stars produce jets and form planets, similar structures around old stars help perhaps to shape incredible ``sculptures'' around dying stars.

\subsection{Morphology of stellar deaths}

The evolution of stars is determined primarily by their mass, because evolutionary timescales are set by the rate of consumption of the nuclear fuel. Figure~13 shows the evolutionary tracks of 1~M$_{\odot}$, 
5~M$_{\odot}$, and 25~M$_{\odot}$ stellar models in the luminosity-effective temperature plane. It is clear that certain phases of stellar evolution are characterized by considerable mass ejection, either via relatively gentle processes such as radiation pressure on atomic lines or on dust, or through stellar explosions. A process of the former type produces the objects known as \textit{planetary nebulae}---shells of ejected matter from giant stars that are ionized by the radiation emitted by the hot 
({\raise-.5ex\hbox{$\buildrel>\over\sim$}}\,30,000~K) central core. The explosions known as Type~II supernovae are the result of the dynamical collapse of the iron cores of massive stars, and the ejected nebulae form the objects known as \textit{supernova remnants}.

One of the most striking phenomena revealed by HST observations of stellar deaths and of stars in their late stages of evolution is the fact that \textit{axisymmetric} nebulae are extremely common.  This observation remains true even after recognizing that the morphologies are blurred by projection effects, and by the lines used to obtain the image (i.e., the [O~III] image may look different than the [N~II] image).  Furthermore, axisymmetry is found not only in planetary nebulae (PNe), that are formed by intermediate-mass stars (1--8~M$_\odot$), but also around Luminous Blue Variables (LBVs), that are extremely massive stars ($\sim100$~M$_\odot$), and supernovae (SNe), that represent the death of massive stars (8--30~M$_\odot$).

Eta Carinae ($\eta$~Car), for example, belongs to the small class of very massive stars known as Luminous Blue Variables.  These stars are believed to represent a rapid evolutionary phase, in which the stars experience severe mass loss (losing many solar masses in $\sim10^4$ years), sometimes via giant outbursts 
\cite[e.g.,][]{Davidson:1997}.  In the 1840s $\eta$~Car suffered such an outburst, increasing in brightness by several magnitudes and ejecting a large amount of material.  The star stabilized around 1870 (except for a minor eruption between 1887 and 1895). The precise cause for this giant outburst is still unknown, but it has produced a spectacular, bipolar nebula commonly referred to as the Homunculus (Fig.~14). In addition to the Homunculus, the HST images revealed for the first time the presence of a ragged ejecta disk around the ``waist'' of the hourglass structure, composed, at least partly, of radial streaks.  The total mass in the equatorial disk has been estimated (based on conventional gas to dust ratios) to be at least 
0.1--0.2~M$_\odot$, but this could underestimate the mass significantly.

Recent spectroscopic observations of $\eta$~Car by the Space Telescope Imaging Spectrograph (STIS) on board HST showed that the spectrum could be fitted well using a model with a mass-loss rate of 
$\sim10^{-3}$~M$_\odot$ yr$^{-1}$.  The minimum mass of the system is currently estimated to be 120~M$_\odot$ 
\cite{Hillier:2001}.

On the basis of photometric and radial velocity variations (in particular the disappearance of 
high-excitation lines like He~I, Fe~III) and x-ray observations, it appears that $\eta$~Car is a binary system, with an orbital period of 5.52~years 
\cite{Damineli:1997,Damineli:2000,Ishibashi:1999}. The mass of the secondary is not known, but it may be less the 30~M$_\odot$ 
\cite{Hillier:2001}.

Eta Carinae is an extremely enigmatic object on many fronts. However, the morphology of its ejected nebula becomes particularly intriguing when we realize that HST observations reveal almost identical nebular morphologies in objects of very different mass and evolutionary history. One of the best known of these is the supernova 1987A (SN~1987A) in the Large Magellanic Clouds 
\cite[see, for example,][for a review]{McCray:2003}. The supernova was first observed in February 1987 (and hence 3~years prior to the launch of HST) and was immediately classified as a Type~II Supernova (SN~II; representing the collapse of the iron core of a massive star) by virtue of its strong hydrogen lines (coming from the hydrogen-rich envelope).  For the first time, the detection of neutrino events (formed copiously as matter rapidly neutronizes) directly confirmed the association between SNe~II and core collapses of massive stars.  The exploding star itself, SK~$-69^\circ202$, had actually been observed prior to the explosion to be a B3 blue supergiant, with a luminosity of $L\simeq1.1\times10^5$~L$_\odot$.  Since the HST launch, SN~1987A has become a prime target for the telescope, being the nearest supernova in modern times. The HST observations have revealed a remarkable system of circumstellar rings surrounding the bright center 
\cite[Fig.~15;][]{Burrows:1995,Pun:1997}. These rings reflect the morphology of material ejected by the supernova progenitor a couple of tens of thousands of years before the explosion 
\cite{Burrows:1995}. This can be inferred from the fact that the central ring is expanding at about 10~km~s$^{-1}$ 
\cite{Crotts:1991} and it currently has a radius of about $6.3\times10^{17}$~cm. While a full explanation for the formation of the rings is still lacking, there is very little doubt that what we are observing is a bipolar structure, in which the inner ring marks the narrow ``waist,'' while the larger rings are somehow ``painted'' on the bipolar lobes, or possibly mark their edges. The entire structure is thus very similar to the one observed in $\eta$~Car.

Bipolar structures have not been restricted only to massive stars, however.  Some planetary nebulae and symbiotic nebulae exhibit morphologies that are almost identical to those of $\eta$~Car and SN~1987A. Some of the best examples are probably My~Cn18 (Fig.~16; the ``hourglass'' nebula, usually classified as a planetary nebula), and the ``Southern Crab'' (Fig.~17), now recognized as a symbiotic nebula 
\cite{Corradi:2001}.

Planetary nebulae represent the late stages in the lives of stars of about 1--8~M$_\odot$. At that phase, the stars eject their outer envelopes exposing the hot cores 
($T\,{\raise-.5ex\hbox{$\buildrel>\over\sim$}}\,30,000$~K), which, in turn, ionize the nebulae causing them to fluoresce.  Symbiotic nebulae have at their centers symbiotic binary systems, consisting typically of a red supergiant and a white dwarf that provides the ionizing radiation.

The main question that arises, therefore, is: What is (are) the mechanism(s) that is (are) capable of producing such bipolar morphologies in stars of different masses, and different ages and evolutionary histories? The two main mechanisms that have been proposed are: (1)~\textit{Interacting winds} in the presence of an equatorial to polar \textit{density contrast}, and (2)~\textit{Magnetic tension} of a toroidal field 
\cite[see][for a more extensive discussion]{Balick:2002}.

Let us first examine the interacting winds model. The original ``interacting winds'' model for planetary nebulae 
\cite{Kwok:1982,Kahn:1982} suggested that old, intermediate-mass stars, in the phase of evolution known as the asymptotic giant branch (AGB), first emit a slow ($\sim20$~km~s$^{-1}$; of the order of the escape velocity from the AGB star's surface) wind, followed by a fast wind ($\sim1000$~km~s$^{-1}$), once the hot and compact nucleus (the AGB star's core) is exposed. The fast wind catches up with the slowly moving material and shocks it. 
\textcite{Balick:1987} proposed that when the interacting winds are allowed to operate in the presence of a \textit{density contrast} between the equator and the pole, a variety of axially symmetric morphologies can be obtained. The idea is that, for reasons that will be explored below, the slow wind contains a non-spherical density distribution, with material being denser around the equator than in the polar direction.  Consequently, the (spherically symmetric) fast wind can penetrate more easily at the poles, forming an axisymmetric nebula. Numerical simulations have shown that when a range of density contrasts (between the equatorial and polar directions) is used, and, in addition, the nebular inclination with respect to the line of sight is taken into consideration, most of the observed morphologies can be reproduced 
\cite{Soker:1989,Icke:1992,Frank:1993,Mellema:1995,Dwarkadas:1996}.

The second class of models involves the action of a magnetic field.

The toroidal field component in an outflow from a star is given by 
\cite{Parks:1991}
\begin{equation}
B_\phi = B_S \frac{V_\mathrm{rot}}{V_W(\theta)} \left(\frac{R_S}{r}\right)^2 \left(\frac{r}{R_S}-1\right) \sin \theta~~,
\end{equation}
where $R_S$, $B_S$ are the stellar radius and the surface magnetic field, $V_\mathrm{rot}$ is the equatorial rotational velocity and $V_W(\theta)$ is the wind terminal velocity.  Consequently, the ratio of the toroidal to radial component increases like ($r/R_S$) at large distances, leading potentially to an axisymmetric configuration, by the fact that magnetic stresses can slow down the flow in the equatorial direction (while not interfering with the polar direction). The key physical parameters determining the obtained morphology are the stellar rotation rate, the ionizing radiation, and the stellar magnetic field.  These can be expressed by: $\Omega_\mathrm{rot}/\Omega_\mathrm{crit}$, $F_*$, and $\sigma$ 
\cite{Chevalier:1994,Garcia:1999}. Here $\Omega_\mathrm{crit}$ is the critical (Keplerian) angular velocity, $\Omega_\mathrm{crit}=(GM_S/R_S^3)^{1/2}$, $F_*$ is the average flux of Ly$\alpha$ photons (of order $10^{45}$--$10^{47}$~s$^{-1}$ for typical PNe nuclei), and $\sigma$ is the ratio of the magnetic energy density to the kinetic energy density in the wind 
\cite[e.g.,][]{Begelman:1992}
\begin{equation}
\sigma =
\frac{B^2}{4\pi\rho V_W^2} =
\frac{B_S^2R_S^2}{\dot{M}V_W} 
\left(\frac{V_\mathrm{rot}}{V_W}\right)^2~~.
\end{equation}

Let us now examine the effects of each one of these parameters.

For the rotation to have a significant effect and produce a bipolar morphology, the star needs to rotate at a significant fraction of its breakup speed ($\Omega_\mathrm{rot}/\Omega_\mathrm{crit}\,{\raise-.5ex\hbox{$\buildrel>\over\sim$}}\,0.5$). Under these conditions, conservation of angular momentum results in a significant focusing of the wind towards the equatorial plane, leading to an equatorially compressed outflow 
\cite[e.g.,][]{Bjorkman:1993,Owocki:1994,Livio:1994}. The concomitant equator to pole density contrast leads to bipolar morphologies.

Ionization does not, in itself, produce a bipolar morphology. Rather, ionization fronts tend to excite instabilities 
\cite[similar to the Rayleigh-Taylor instability or the instability discussed by][]{Vishniac:1983}, which, in turn, produce finger-shaped structures and dense ``knots.'' Observations of a number of relatively nearby planetary nebulae, and in particular of the ``Helix'' nebula, reveal that such dense knots are probably very common \cite{ODell:2003,Speck:2002}.

The toroidal magnetic field is carried by the fast wind and it can, in principle, produce a bipolar morphology even if the slow wind is spherically  symmetric. Basically, as the magnetic energy density of the shocked wind becomes larger than the thermal energy density, the flow becomes (due to the increasing importance of the toroidal component) bipolar, with an increased collimation as the value of $\sigma$ is increased. Numerical simulations 
\cite{Garcia:1999} show that qualitatively, the morphologies obtained for different combinations of $\Omega_\mathrm{rot}/\Omega_\mathrm{crit}$ and $\sigma$ are as shown in Fig.~18. The minimum field required to produce magnetic shaping was found to be of order 
\cite{Chevalier:1994}
\begin{equation}
B_S^\mathrm{min}\simeq11G
\left(\frac{\sigma}{10^{-4}}\right)^{1/2}
\left(\frac{\dot{M}}{10^{-8}~\mathrm{M}_\odot~\mathrm{yr}^{-1}}\right)^{1/2}
\left(\frac{V_\mathrm{FW}}{2000~\mathrm{km~s}^{-1}}\right)^{1/2}
\left(\frac{R_S}{10^{11}~\mathrm{cm}}\right)^{-1}
\left(\frac{V_\mathrm{rot}/V_\mathrm{FW}}{0.1}\right)^{-1}~~,
\end{equation}
where $V_\mathrm{FW}$ is the velocity of the fast wind.

From eq.~13 we see that if the star rotates too slowly, the minimum required field may be unattainable. Similarly, if the star does not rotate at a significant fraction of its breakup speed, an equatorially compressed outflow is not formed. The question of: What is the mechanism that produces highly bipolar outflows? can therefore be reduced to:  What causes the star to rotate close to breakup? or: What can produce a strong density contrast between the equatorial and polar directions in the slow wind?

One obvious possibility is: binary companions!

Companions to the central star can act in several ways to aid in the formation of bipolar morphologies:
(1)~For binaries that were initially relatively close (separation less than \mbox{$\sim1000$~R$_\odot$}), so that the primary could fill its Roche lobe (the critical potential surface beyond which mass transfer onto the companion occurs) during the asymptotic giant branch (AGB) phase, an unstable mass transfer ensues. As a result, the companion and the AGB star's core start spiralling-in inside a common envelope
\cite[see, e.g., a review by][]{Iben:1993}. This has two effects. First, the envelope of the primary can be spun-up to angular velocities of the order of
\begin{equation}
\frac{\Omega_\mathrm{rot}}{\Omega_\mathrm{crit}}\simeq0.1
\left(\frac{M_C}{0.01~\mathrm{M}_\odot}\right)
\left(\frac{M_\mathrm{env}}{\mathrm{M}_\odot}\right)^{-1}
\left(\frac{K_g^2}{0.1}\right)^{-1}
\left(\frac{a}{R_S}\right)^{1/2}~~,
\end{equation}
where $M_C$ is the companion's mass, $M_\mathrm{env}$ is the mass of the giant's envelope, $M_SK_g^2R_S^2$ is the star's moment of inertia and $a$ is the initial separation between the giant and the companion. Equation~14 shows that even brown dwarf (sub-stellar) companions can bring the envelope close to critical rotation. Second, since the envelope mass is ejected (due to orbital energy deposition) primarily close to the orbital plane (because angular momentum is also deposited into the envelope), the common envelope phase can generate an equator-to-pole density contrast. Hydrodynamic simulations of common envelope evolution reveal that during the late stages about 80\% of the mass is ejected within 30$^\circ$ of the binary orbital plane 
\cite[e.g.,][]{Terman:1994,Terman:1995,Rasio:1996}.  One can expect that due to cooling, the mass will sink even more toward the orbital plane at later times.  

Another possibility for the presence of a higher equatorial density, in principle at least, that does not involve binary companions, is the inner rim of the protostellar disk. If the outer part of the protostellar disk survives till late stages in the stellar life (which may not be difficult in the case of massive, 
short-lived stars), then the fast wind could interact with the inner rim of this disk 
\cite{Pringle:1989}. In a few planetary nebulae 
\cite[for example, the ``Red Rectangle'' and the ``Egg Nebula,''][respectively]{Bond:1996,Thompson:1997}, HST observations reveal the presence of relatively large disks, similar to the ones observed in young stellar objects.

In at least some symbiotic nebulae (e.g., M2$-$9), the bipolar morphology may reflect the action of the white dwarf companion.  The white dwarf accretes from the wind of the AGB star, an accretion disk is formed, and the disk powers a mildly collimated fast wind, which in turn produces the bipolar morphology 
\cite{Soker:2001,LivioSoker:2001}.

The conclusion from this discussion is that several mechanisms are capable, in principle at least, to produce the observed bipolar morphologies. Different mechanisms may be operating in different systems. In some cases, we can look forward to the future and expect more definitive answers to emerge. For example, in SN~1987A, the supernova blast wave will eventually hit the entire inner ring, and the luminosity that will be generated by this interaction will illuminate the entire SN vicinity 
\cite{McCray:2003}. A few brightening spots, where the blast wave has already hit protrusions on the ring, have been observed by HST 
\cite[Fig.~19; see also][]{Panagia:2002}. Generally, a transmitted shock at normal incidence is expected to propagate into the ring with a speed of $V_\mathrm{ring}\simeq(n_o/n_\mathrm{ring})^{1/2} V_\mathrm{blast}$ (where $n_o$, $n_\mathrm{ring}$ are the number densities of the circumstellar matter and the ring, respectively, and  $V_\mathrm{blast}$ is the blast wave velocity). For SN~1987A, $V_\mathrm{blast}\sim4000$~km~s$^{-1}$, $n_o\sim150$~cm$^{-3}$, $n_\mathrm{ring}\sim10^4$~cm$^{-3}$ giving $V_\mathrm{ring}\sim500$~km~s$^{-1}$. Eventually, the H$\alpha$ flux from the ring is expected to be more than 30~times higher than today ($F_\mathrm{H \alpha}\,{\raise-.5ex\hbox{$\buildrel>\over\sim$}}\,3\times10^{-12}$~erg cm$^{-2}$ s$^{-1}$), and even brighter in UV lines 
\cite{Luo:1994}. This ionizing flux will turn the circumstellar matter into an emission nebula, thus revealing its distribution and velocity field, and hopefully allowing for a reconstruction of the mass-loss history of the system. In all of this, HST will provide a front seat view.

The collapses of very massive stars are believed to produce stellar-mass black holes. In some cases, these collapses produce the most spectacular explosions in the universe since the ``big bang''---Gamma Ray Bursts. The Hubble Space Telescope has played an important role in the attempts to understand the nature of these dramatic explosions. However, much more massive black holes are also produced in the universe quite commonly, and the questions of how they form and evolve have intrigued astronomers for decades. The initial answers had to come from large collections of stars like galaxies and stellar clusters.

Given the unprecedented resolution that HST provides, very crowded fields, like the centers of galaxies, were obvious targets. In particular, the early suggestive evidence that galactic centers harbor supermassive black holes 
\cite[see][for a review]{Kormendy:1995}, virtually ensured that centers of galaxies will be observed extensively with HST.  Indeed, these observations were carried out and proved to be extremely fruitful.

\section{BLACK HOLES---FROM SUPERMASSIVE TO STELLAR}

The idea that active galactic nuclei (AGNs) are powered by accretion at high rates (up to a few solar masses per year) onto supermassive ($10^6$--$10^8$~M$_\odot$) central black holes has been around for a long time 
\cite[e.g.,][]{Zeldovich:1964}. Furthermore, it has long been realized that collapse into a black hole may be inevitable in the center of many dense stellar environments, once the central potential gets sufficiently deep 
\cite[for example,][]{Lynden:1969,Rees:1978}. Consequently, one might expect supermassive black holes to reside at the center of most galaxies.

In addition, searches for the host galaxies of quasars (thought to represent accreting supermassive black holes) have been going on for a long time, following the pioneering works of 
\textcite{Kristian:1973}, 
\textcite{Wyckoff:1980},  
\textcite{Hutchings:1983}, and 
\textcite{Hutchings:1989}. The ground-based studies not only detected host galaxies, but also suggested that some of the hosts have experienced tidal interactions 
\cite{Boroson:1982,Kukula:1996}. 

These tentative detections and suggestions have turned into reality with the spectacular, high-resolution images of quasars obtained with HST 
\cite[e.g.,][]{Bahcall:1994,Bahcalletal:1997,Kirhakos:1999}. The HST images revealed unambiguously that nine radio-loud quasars reside either in bright elliptical hosts or in strongly interacting systems. These findings probably indicate that interactions with neighboring galaxies play an important role in the fueling of the central black hole in active galaxies \cite[see][]{Krolik:1999a}.

\subsection{Search techniques for supermassive black holes}

The actual \textit{searches} for supermassive black holes rely primarily on stellar- and gas-dynamical evidence 
\cite[see, e.g.,][for a review]{Kormendy:1995}. In particular, the idea is to unambiguously show that the mass-to-light ratio, $M/L$, increases toward the galactic center to values that are difficult to accomodate with other types of stellar populations.  Ideally, one would want to follow this up with the detection of relativistic speeds, but at present even HST cannot resolve orbits at a few gravitational (Schwarzschild) radii ($R_S=2GM_\mathrm{BH}/c^2$; where $M_\mathrm{BH}$ is the black hole mass, $G$ is the gravitational constant and $c$ is the speed of light).

Generally, the basic principle behind the early stellar-dyamical search techniques can be explained by the following, simplified picture.  Taking the first velocity moment of the collisionless Boltzmann equation gives
\begin{equation}
M(r)=
\frac{V^2_\mathrm{rot}r}{G}+
\frac{\sigma^2_r r}{G}
\left[ - \frac{d ln \rho_t}{d ln r} -
\frac{d ln \sigma^2_r}{d ln r} -
\left( 1-\frac{\sigma_\theta^2}{\sigma^2_r}\right)-
\left( 1-\frac{\sigma^2_\phi}{\sigma^2_r}\right) \right]~~,
\end{equation}
where $M(r)$ is the mass enclosed within radius $r$, $V_\mathrm{rot}$ is the rotational velocity, $\sigma_r$,  $\sigma_\theta$, $\sigma_\phi$ are the components of the velocity dispersion, and $\rho_t$ is the density of the tracer stars being observed (usually assumed to be proportional to the volume brightness). A direct measurement of $\vec{V}_\mathrm{rot}$ and $\vec{\sigma}$, therefore can determine the central mass. To actually use eq.~15, however, the ranges of \textit{unprojected} quantities (e.g., $\vec{V}_\mathrm{rot}$, $\vec{\sigma}$) need to be derived, and various techniques to achieve that have been developed 
\cite[see for example,][]{Kormendy:1988,Dressler:1988,Marel:1994,Gerhard:1993}. 

The more recent search techniques fit axisymmetric, three-integral dynamical models of the galaxy (using the line-of-sight velocity distribution), to the observed light distribution. Basically, the orbits in the ($R$, $z$) plane in realistic galactic potentials are often found to have in addition to the two integrals of motion $E$ (the energy) and $L_z$ (the $z$-component of the angular momentum, where $z$ is the symmetry axis), a third integral, $I_3$, that can be associated with the approximately conserved total angular momentum, $L$.

The dynamical model assumes axisymmetry and an inclination for the galaxy and first determines the optimal density distributrion that is consistent with the surface brightness distribution (making certain assumptions about $M/L_V$). Then, a central point mass is added, and the potential calculated. Orbits that sample the phase space of the three integrals of motion are then calculated, and the data of the full line-of-sight velocity distribution are fitted to the models.

Since galactic gas is affected also by forces which are non-gravitational (e.g., radiation pressure),  stellar kinematics are considered more secure than gas dynamics in black hole searches. Nevertheless, in a few cases, the presence of the black hole may be revealed by gas-dynamical searches. The prototype of this technique is provided by the radio galaxy M87. Early stellar-dynamical observations revealed that the velocity dispersion continues to rise inward, to $r\simeq1''\llap{.}5$ 
\cite[e.g.,][]{Sargent:1978,Lauer:1992}. However, due to the expected anisotropy in the velocity dispersion, models without a black hole could also fit the observations 
\cite[e.g.,][]{Binney:1982,Dressler:1990,Marel:1994}. In particular, the last author found $\sigma\simeq400$~km~s$^{-1}$ at $r \,{\raise-.5ex\hbox{$\buildrel<\over\sim$}}\, 0''\llap{.}5$, which could nevertheless be fitted with anisotropic models that do not include a black hole.

A more definitive answer, however, came in this case from the gas. High-resolution HST observations of the nucleus revealed the presence of a small gas disk (about 20~pc in radius), with a major axis perpendicular to the optical jet (Fig.~20). Spectra taken at two opposite points along the major axis (at a 
luminosity-weighted mean radius of 16~pc) found emission lines separated by $2V=916$~km~s$^{-1}$ 
\cite{Ford:1994,Harms:1994}. For an inclination angle of the disk of $\sim42^{\circ}$ (implied by the observed axis ratio), the observed velocity amplitude (if interpreted as a circular Keplerian motion) corresponds to a dark mass of $M_\mathrm{BH}\simeq3\times10^9$~M$_\odot$ at the center of M87. These results have been further confirmed by 
\textcite{Macchetto:1997}, making the black hole in M87 the most massive observed so far (with a relatively secure mass determination).  I should caution that an airtight case for circular Keplerian motion is still to be made for M87. Incidentally, the best case for a black hole based on gas dynamics is for the modest active galactic nucleus NGC~4258. In that case, Very Long Baseline Array (VLBA) observations of high-velocity masers show a rotation curve that can be fitted remarkably well with 
$V_\mathrm{rot}(r)\simeq(832\pm2)(r/0.25~\mathrm{pc})^{-1/2}$~km~s$^{-1}$, implying a black hole mass (mass interior to 0.18~pc) of $M_\mathrm{BH}\simeq4.1\times10^7$~M$_\odot$.

\subsection{The \emph{M}\lower.4ex\hbox{\scriptsize{BH}} -- $\sigma$ relationship}

Using three-integral models and HST spectroscopy, 
\textcite{Gebhardt:2000} were able to determine black hole masses in 16~galaxies. Adding to these two galaxies with maser mass determinations (NGC~4258 and NGC~1068), six galaxies with black hole masses determined by gas dynamics, and our own galaxy 
\cite{Ghez:1998,Genzel:2000} and M31 
\cite{Dressler:1988}, they were able to discover a tight correlation between the black hole mass and the velocity dispersion in the galactic bulge. Specifically, they found that
\begin{equation}
M_\mathrm{BH}=1.2\times10^8~\mathrm{M}_\odot
\left(\frac{\sigma_e}{200~\mathrm{km~s}^{-1}}\right)^{3.75}~~,
\end{equation}
where $\sigma_e$ is the line of sight aperture dispersion within the half-light radius $R_e$. The obtained relation is much tighter (Fig.~21, right) than a previously determined relation 
\cite[e.g.,][]{Magorrian:1998} between black hole mass and the bulge luminosity (Fig.~21, left). Based on a smaller sample (12~galaxies), 
\textcite{Ferrarese:2000} found independently a 
$M_\mathrm{BH}\propto\sigma^\alpha$ relation, with a somewhat steeper slope of $\alpha\sim4.8$.

Clearly, the observed tight $M_\mathrm{BH}-\sigma$ correlation strongly suggests that the formation and evolution of the bulge and of the black hole are causally connected. The precise nature of this connection, however, is still a matter of considerable uncertainty. A relatively simple theoretical model for the relation has been suggested by 
\textcite{Adams:2001}. 

In this model, a slowly rotating isothermal sphere (with a seed central black hole) collapses to form the bulge (the dark matter is assumed to move in tandem with the baryons). The density distribution is assumed to be of the form
\begin{equation}
\rho(r)=\frac{c_s^2}{2\pi G r^2}~~,
\end{equation}
where $c_s$ is the speed of sound. Note that for dissipationless collapse the velocity dispersion is roughly given by $\sigma^2\simeq2c_s^2$. The region is assumed to rotate rigidly (due to effective tidal torques) at an angular speed $\Omega$, and the specific angular momentum is $j=r^2_\infty\Omega\sin^2\theta$, where 
$r_\infty$ is the initial radius and $\theta$ is the polar angle. For zero-energy orbits, the pericenter distance in the equatorial plane is therefore
\begin{equation}
p=\frac{j^2}{2GM}=
\frac{r^4_\infty \Omega^2}{2GM}=
\frac{(GM)^3 \Omega^2}{32 c_s^8}~~,
\end{equation}
where in the last equality we used the fact that for the assumed density distribution $M(r)=2c^2_s r/G$. For material to be captured by the black hole we need $p\leq4R_S$, where $R_S$ is the gravitational (Schwarzschild) radius $R_S=2GM/c^2$. Using eq.~18, the condition $p=4R_S$ therefore reads (assuming that the capture condition defines the black hole mass)
\begin{equation}
M_\textrm{BH}=
\frac{16c_s^4}{Gc\Omega}=
\frac{4}{Gc\Omega}\sigma^4~~,
\end{equation}
in good agreement with the observed relation (eq.~16).

Other considerations result in similar expressions. For example, in a protogalaxy modeled as an isothermal sphere of cold dark matter, with $\rho(r)=\sigma^2/2\pi Gr^2$, with a fraction $f_\mathrm{gas}$ in the form of gas, a central, accreting black hole will generate an intense wind outflow. The black hole itself may be  assumed to form by coherent collapse before most of the bulge gas turns into stars. If the black hole radiates at the Eddington luminosity, $L_\mathrm{EDD}=4\pi cGM_\mathrm{BH}/\kappa$ (at which gravity is balanced by radiation pressure; where $\kappa$ is the electron scattering opacity),  and a fraction $f_\mathrm{out}$ is deposited into kinetic energy of the outflow, then a shell of swept-up material will be moving outward at a speed
\begin{equation}
V_\mathrm{out}=
\left( \frac{8\pi^2G f_\mathrm{out} L_\mathrm{EDD}}{f_\mathrm{gas}\sigma^2}\right)^{1/3}~~.
\end{equation}
The condition that the shell would escape, and therefore, that the black hole would unbind the bulge gas, requires $V_\mathrm{out}>\sigma$. This implies that the black hole mass is limited by 
\cite{Silk:1998}
\begin{equation}
M_\mathrm{BH}\simeq
\frac{1}{32\pi^3}
\frac{f_\mathrm{gas}}{f_\mathrm{out}}
\frac{\kappa}{G^2c}\sigma^5~~,
\end{equation}
similar to the relation found by 
\textcite{Ferrarese:2000}.

Semi-analytical, hierarchical galaxy formation models (see Section~VI), in which galaxies form by merging halos (and merging  central black holes), with simple prescriptions for gas cooling, star formation, and feedback from supernovae, also tend to produce scalings of the form $M_\mathrm{BH}\sim\sigma^4$ 
\cite[for example,][]{Haehnelt:2000}. Broadly speaking, this relation can be traced to the facts that: (i)~In mergers, the black hole mass scales with the halo mass. 
(ii)~$\sigma\sim\rho^{1/6}M_\mathrm{halo}^{1/3}$, and (iii)~$M_\mathrm{halo}$ scales like $\rho^{-2}$ in typical cold-dark-matter cosmologies (really like $\rho^{-2/(3+n)}$, where $n\sim -2$ is the slope of the dark matter fluctuations spectrum). Combining (i)--(iii) gives $M_\mathrm{BH}\sim\sigma^4$.

Somewhat more exotic scenarios, which involve the accretion of collisional dark matter 
\cite[invoked to make galactic halos less dense; for example,][]{Spergel:2000}, also produce black hole masses which scale roughly with $\sigma^{4.5}$ 
\cite{Ostriker:2000}.

\subsection{Intermediate-mass black holes?}

An interesting question is related to the \textit{range} of black hole masses over which the observed relation (16) applies. Most recently, high-spatial-resolution spectroscopy with HST, 
\cite{Marel:2002,Gerssen:2002,Gerssen:2003}, of the central part of the globular cluster M15, revealed evidence for the existence of a dense, central concentration of dark mass 
\cite[see also][]{Dull:2002,Baumgardt:2002}. The interpretation of the nature of this dark mass has been somewhat controversial.  \textcite{Dull:2002} and \textcite{Baumgardt:2002} have shown that the sharp rise in $M/L$ that was observed toward the center of M15 could be explained by a central concentration of neutron stars and massive white dwarfs. On the other hand, \textcite{Gerssen:2003} argued that if one were to allow for the fact that a large fraction of the neutron stars born within the cluster cannot be retained (due to ``kick'' velocities of a few hundred km~s$^{-1}$ at birth), the observed $M/L$ might require the existence of a central black hole of mass $M_\mathrm{BH}=1.7^{+2.7}_{-1.7}\times10^3$~M$_\odot$. Interestingly, the derived mass fits quite well onto the $M_\mathrm{BH}-\sigma$ relationship 
\cite{Tremaine:2002,Gerssen:2002}. Similarly, evidence for a central black hole with a mass of $\sim2\times10^4$~M$_{\odot}$ has been found in the stellar cluster G1 in the Andromeda galaxy M31
\cite{Gebhardt:2002}. The latter black hole also fits the $M_\mathrm{BH}-\sigma$ relation. In an entirely independent work 
\cite{DAmico:2002}, pulse timing observations determined positions for five millisecond pulsars in the cluster NGC~6752 with a 20~mas accuracy.  Three were found to have line-of-sight accelerations larger than the maximum value predicted on the basis of the central mass density derived from optical observations. The measured accelerations thus provided dynamical evidence for a central mass-to-light ratio of 
$M/L\,\raise-.5ex\hbox{$\buildrel>\over\sim$}\,10$. All of these findings, \textit{if confirmed}, are extremely exciting, since the implied masses make these black holes \textit{intermediate} between the stellar black holes (with $M_\mathrm{BH}\sim10$~M$_\odot$) and the suppermassive ones 
($M_\mathrm{BH}\sim10^6$--$10^9$~M$_\odot$).

The potential existence of a class of intermediate-mass black holes has also been suggested following the discovery of ultraluminous x-ray sources 
\cite[e.g.,][]{Colbert:1999}. The latter are x-ray sources outside the nuclei of external galaxies, with luminosities in excess of $10^{39}$~erg~s$^{-1}$. These sources were originally discovered by the Einstein satellite 
\cite{Fabbiano:1989}, but have been found in large numbers by the ROSAT and Chandra observatories 
\cite[e.g.,][]{Colbert:1999,Lira:2002,Jeltema:2002}. More recently, however, it has been pointed out that although the ultraluminous x-ray sources may form a heterogeneous class of systems, the most likely explanation for the majority of them is that they constitute the high-luminosity tail of the stellar-mass black-hole binary distribution 
\cite[e.g.,][]{King:2001,Roberts:2001,Podsiadlowski:2002}. In particular, \textcite{Podsiadlowski:2002} have shown that in binary systems in which the donor star becomes a giant and the evolution is driven by the nuclear evolution of the hydrogen-burning shell, luminosities that are potentially as high as $\sim10^{41}$~erg~s$^{-1}$ can be obtained. The existence of ultraluminous x-ray sources by itself, therefore, may not imply the existence of intermediate-mass black holes. 

The potential existence of intermediate-mass black holes also led to a new scenario for the formation of the supermassive ones 
\cite{Ebisuzaki:2001}. The idea is that first, intermediate-mass black holes form in young clusters, due to runaway mergers of massive stars. The massive stars are assumed to sink to the cluster center due to dynamical friction, on a timescale of 
\cite[e.g.,][]{Binney:1987}
\begin{equation}
t_\mathrm{df}=
\frac{1.17}{\log\Lambda}
\frac{r^2\sigma}{Gm}\simeq
2.7\times10^7
\left(\frac{r}{1~\mathrm{pc}}\right)^2
\left(\frac{r_h}{10~\mathrm{pc}}\right)^{-1/2}
\left(\frac{M}{10^6~\mathrm{M}_\odot}\right)^{1/2}
\left(\frac{m}{20~\mathrm{M}_\odot}\right)^{-1}~\mathrm{yr}~~,
\end{equation}
where $\log\Lambda$ is the Coulomb logarithm, $\sigma$ is the velocity dispersion, $r$ is the distance to the cluster center, $r_h$ is the half-mass radius and $M$ and $m$ are the masses of the cluster and the star, respectively. During the same time that the intermediate-mass black holes are forming, the host clusters themselves sink to the galactic center (again due to dynamical friction), evaporate, and deposit their black holes. The latter can then form black hole binaries which merge due to gravitational radiation, eventually leading to the formation of supermassive black holes. I should note that this is only one path out of the many that have been suggested over the years for the formation of supermassive black holes 
\cite{Rees:1984}.

The installation of HST's Advanced Camera for Surveys (ACS) in March 2002, with its superior sensitivity (by a factor of about 3--5 over the previous Wide Field Planetary Camera~2) and increased resolution (in the high-resolution channel), combined with X-ray observations by the Chandra and XMM-Newton Observatories, and infrared observations of dust-obscured active galactic nuclei by the Space Infrared Telescope Facility (SIRTF) promise that the study of black holes in clusters (if they exist) and in galactic centers is only beginning. In particular, future observations will help clarify the relation between galaxy evolution in general, and the formation and evolution of the cental black holes.

Observations with HST have provided important insights not only in the study of supermassive black holes, but also in researches related to the probable formation of stellar-mass black holes---the study of 
gamma-ray bursters.

\subsection{Gamma-ray bursts}

Gamma-ray bursts were first detected by the military Vela satellites in 1967.  These are short flashes of gamma-rays (typically with a peak energy around 100~keV), lasting between a few milliseconds and tens of minutes. Already in the early 1990s, the BATSE experiment on board the Compton Gamma-Ray Observatory demonstrated that the rate of gamma-ray bursts is about 1--2 per day, and that they are distributed isotropically in the sky.  The data from BATSE also showed that gamma-ray bursts (GRBs) come in two distinct classes:  (i)~short-duration ({\raise-.5ex\hbox{$\buildrel<\over\sim$}}\,2~sec) with hard spectra, and (ii)~long duration with softer spectra 
\cite{Kouveliotou:1993,Fishman:2001}.

Due to the relatively poor localization capabilities of gamma-ray detectors (which did not allow identification of the sources in other wavebands), two thousand bursts have been detected before it became possible to determine the distance to the bursts. The isotropic sky distribution argued for either a cosmological origin or an extremely local one (e.g., the Galactic halo). This situation changed dramatically since the operation of the Italian-Dutch BeppoSAX satellite in 1997. BeppoSAX was able to determine the positions of bursts to within arcminutes. The discovery of rapidly declining ``afterglows'' in the X-ray, optical and radio bands 
\cite{Costa:1997,vanParadijs:1997,Metzger:1997,Frail:1999} allowed for redshift determinations that immediately placed GRBs at cosmoligical distances. I should note that to date, only afterglows of the 
long-duration bursts have been observed. Thus, it is not even clear if the short bursts produce afterglows 
\cite[the best limit to date, for the burst GRB~020531, failed to detect an afterglow candidate \mbox{[down to $V\sim25$]} about 20~hours after the burst;][]{Salamanca:2002}. Typically, the afterglows decay with time as $t^{-\alpha}$, with $\alpha$ in the ranges 1--2. This behavior was \textit{predicted} by a model in which a fireball is expanding into a homogeneous external medium 
\cite{Meszaros:1997}.

At the time of this writing redshifts have been determined to more than two dozen GRBs, and they usually lie in the $z=0.5$--1.5 range (although redshifts as high as 4.5 have been recorded). The observed fluxes imply energies of up to $10^{54}$~ergs for isotropic emission. However, there is increasing evidence, in the form of kinks in the afterglow light curve, and in polarization detected in a few bursts, that 
gamma-ray bursts are in fact collimated into narrow jets. If the observer's line of sight is within the jet solid angle, $\Omega_j$, then as long as the Lorentz factor $\gamma$ satisfies 
$\gamma\,\raise-.5ex\hbox{$\buildrel>\over\sim$}\,\Omega_j^{-1/2}$, the light-cone is within the jet boundary. However, as the jet decelerates, $\gamma$ eventually drops below $\Omega_j^{-1/2}$.  Consequently, the (transverse) emitting area starts to grow more slowly (as $r_{||}\Omega_j^{-1/2}$ instead of $(r_{||}/\gamma)^2$), resulting in a break (faster decay) in the light curve 
\cite{Rhoads:1997,Meszaros:1999,LivioWaxman:2000}, consistent with observations in GRB~990123 
\cite{Kulkarni:1999,Fruchter:1999}. When collimation is taken into account, the average total energy of GRBs is estimated to be of the order of $2\times10^{51}$~ergs 
\cite{Frail:2001,Panaitescu:2001}.

Based on the energetics, and the fact that the production of astrophysical jets typically relies on the collimation and acceleration provided by an accretion disk around a compact object 
\cite{Livio:2000}, the most popular models for GRBs involve the formation of stellar-mass black holes, surrounded by a debris torus from which mass accretion onto the central object occurs 
\cite{Meszaros:2002}. The two most likely progenitors to produce such a configuration are the core collapse of massive stars 
\cite{Woosley:1993,Paczynski:1998}, or the ongoing merger of neutron star-neutron star or black 
hole-neutron star binaries 
\cite[e.g.,][]{Paczynski:1986,Goodman:1986,Eichler:1989}. In both cases the main source of energy is gravitational, even though tapping into the (large) spin energy of the black hole is also possible in principle 
\cite{Blandford:1977,Krolik:1999b}. A schematic describing how a GRB and its afterglow may be generated by internal and external shocks (respectively) in a collimated jet arising from a stellar collapse, is shown in Figure~22.

There are several pieces of evidence suggesting that the long duration GRBs may indeed be associated with stellar collapses. First, in merging neutron stars (or a black hole and a neutron star) the associated timescales (dictated at that point by the emission of gravitational wave radiation) are probably too short to produce multi-second-long bursts. On the other hand, core collapses that lead to a black hole and an accretion disk are naturally associated with longer timescales. Second, there exists quite strong evidence that at least some GRBs are associated with supernova explosions.

In particular, the supernova SN~1998bw \cite[of the relatively rare type~Ic, caused by the explosion of a massive star that had rid itself of its hydrogen envelope prior to exploding;][]{Filippenko:1986} was found to be approximately coincident in both position and time with the relatively weak GRB~980425, at redshift $z=0.0085$ 
\cite{Galama:1998}. Furthermore, the supernova light curve was found to be consistent with the formation of a black hole 
\cite{Iwamoto:1998}. In a few other bursts (e.g., GRB~011121) the optical and near infrared afterglow light curve exhibits a ``bump'' with a time delay and amplitude that are consistent with resulting from a supernova explosion occuring simultaneously with the GRB 
\cite[e.g.,][]{Greiner:2001}.

The contributions of HST to this exciting field have been in a few areas.

First, by resolving the host galaxies and being able to pinpoint the GRB on the host, HST has shown unambiguously that at least some GRBs are not associated with galactic nuclei. In fact, they sometimes occur far from the nucleus or in spiral arms 
\cite{Sahu:1997,Andersen:2002}. Second, the Hubble Space Telescope has shown \cite{Fruchter:1999,Fruch:2002} that the colors of the hosts of GRBs corresond to the bluest colors observed for galaxies in the Hubble Deep Fields (see Section~VI). This means that GRBs with afterglows occur preferentially in galaxies with high star formation rates---a finding that is consistent with GRBs being associated with collapses of massive stars. Furthermore, the fact that in almost all cases the GRB's position was found to be within the extent of the rest-frame ultraviolet image (where star formation is intense) of the hosts, also argues in favor of the ``collapsar'' model for the long-duration bursts. Neutron star binaries, on the other hand, are born with ``kicks'' of as much as a few hundred km/sec, which could drive them out of the star-forming regions.

The Hubble Space Telescope also helped to firm the association of at least some GRBs with supernova explosions.  In GRB~011121, for example, HST detected an intermediate-time flux excess (``red bump'') that was redder in color relative to the GRB afterglow.  This ``bump'' could be well described by a redshifted Type~Ic supernova that exploded approximately at the same time as the GRB  
\cite{Bloom:2002,Garnavich:2003}.  Near-infrared and radio observations of the afterglow further provided evidence for extensive mass loss ($\dot{M}\sim2\times10^{-7}$~M$_\odot$ yr$^{-1}$) from the massive stellar progenitor of GRB~011121 
\cite{Price:2002}.

To conclude this topic, it is very likely that \emph{every time a GRB goes off, a black hole is born}. Gamma Ray Bursts therefore offer, in principle at least, a direct measure of the massive-star formation rate in the early universe. Future observations with the HETE-2 spacecraft and with the Swift multi-wavelength GRB afterglow mission (equipped with \mbox{$\gamma$-ray}, X-ray and optical detectors), as well as follow-ups with HST and other observatories from the ground and space, will hopefully reveal the true nature of these most dramatic explosions (including the short-duration bursts).

\section{DEEP FIELDS AND A BRIEF COSMIC HISTORY}

One of the major goals of observational cosmology is to understand the processes involved in galaxy formation and evolution. A full theoretical treatment of this problem requires an understanding of the primordial density fluctuations, of the formation of dark matter halos, of the dissipation that occurs in the cooling gas, of the processes involved in star formation, of the feedback between stellar explosions and the interstellar medium, of galaxy mergers, and of the interactions of galaxies with intergalactic gas. 

Most current models for the formation of structure in the universe assume that dark matter halos build up hierarchically, with the assembly being controled by the cosmological parameters, the power spectrum of the density fluctuations, and the nature of the dark matter itself. The build-up of the stellar mass \emph{within} galaxies is a secondary process, in which gaseous dissipation, the more intricate physics of star formation, and feedback, all play a role.

Since galaxy formation involves non-linear processes on many scales, the problem has been addressed theoretically, mainly using large-scale $N$-body hydrodynamic simulations 
\cite[e.g.,][]{Pearce:2001}. On the observational side, HST, the Chandra Observatory and the Keck and VLT telescopes have allowed for an unprecedented view of the high-redshift universe, probing galaxies from infancy to old age.

Short of the awsome pillars of dust and molecular gas revealed by the HST images of the ``Eagle Nebula,'' the best known images that Hubble has produced are those of the ``Hubble Deep Fields'' (HDFs). In fact, the first HDF and the follow-up observations of the same field by other observatories probably represent the most concentrated research effort ever in astronomy into what was previously a blank piece of the sky!

Observations taken shortly after the first servicing mission, which restored the HST optics capabilities, demonstrated that HST could resolve distant galaxies spectacularly well. In particular, observations of the relatively distant clusters CL~0939+4713 
\cite{Dressler:1994} and the cluster around the radio galaxy 3C324 
\cite{Dickinson:1995} revealed faint galaxies of small angular sizes, the morphologies of which were entirely  inaccessible from the ground.

The idea of the Hubble Deep Field North---ten days of one continuous observation of a field 2.6 arcminutes on the side in December 1995---was conceived by the then Director of the Space Telescope Science Institute, Bob Williams. Williams decided to use his Director's Discretionary Time on HST to produce the deepest image of the universe in optical/UV wavelengths.

The northern field itself (Fig.~23) was carefully selected so as to be in HST's northern continuous viewing zone (CVZ; without interference by Earth occultations), to be free of bright stars, nearby galaxies and radio sources, and to have relatively low Galactic extinction. The southern field (Fig.~24) was selected so as to include a quasi-stellar object (QSO) that could be used to study absorption systems along the line of sight. The HDF-N observations were taken in December 1995 and the HDF-S in October 1998.

The filter selection for the observations was driven partly by the desire for depth and color information (to identify high-redshift galaxies by their Lyman-break), and partly by practical considerations involving scattered light within the telescope.  Accordingly, images were taken in four broad-band passes, spanning a wavelength range from about 2500~\AA\ to 9000~\AA.

In order to achieve a higher resolution than the detectors' pixel sizes, the pointing of the telescope was dithered (shifted slightly) during the observation, thus ensuring that images were recorded on different pixels. A ``drizzling'' (subpixel linear reconstruction) technique was developed by 
\textcite{Fruchter:2002}.  Details on the observational techniques and a review of some of the results can be found in the excellent reviews by 
\textcite{Ferguson:1998} and 
\textcite{Ferguson:2000}.  

As soon as the HDF-N image was obtained, it was obvious that a new era in astronomy has begun.  The image revealed some 3000 galaxies of different colors, shapes, and sizes.  A broad summary of key results can be found in 
\textcite{Livio:1998}.  Here I will concentrate only on two main topics, to which HST's contribution has been crucial:  (i)~galaxy sizes and morphologies (and their implications for galaxy formation and evolution), and (ii)~the global, cosmic star-formation history.  The dramatic discovery of a particular supernova in the HDF-N will be discussed in Section~VIII. 

\subsection{The morphology and sizes of high-redshift galaxies}

One of the key questions in galaxy formation and evolution is the importance of interactions and mergers. In particular, cold-dark-matter and dark-energy dominated models have as one of their natural consequences a higher merger rate in the past, since in these models, today's Hubble Sequence galaxies have been assembled via a process of hierarchical mergers 
\cite[e.g.,][]{Cole:2000,Kauffmann:1999,Somerville:2001}.

In the local universe, the luminosity density is dominated by ellipticals, lenticulars and spirals, even though dwarf ellipticals and dwarf irregulars (all of low luminosity) dominate by number.  The HDF-N allowed for the first time for a morphological classification of galaxies down to a magnitude of $I=25$
\cite[e.g.,][]{Abraham:1996}. These studies showed that the fraction of irregular, multiple-component, and peculiar-looking galaxies was indeed considerably higher ($\sim40$\%) than expected from a direct extrapolation of the numbers at $z\simeq0$. The trend of a rising fraction of faint systems with irregular morphology was already noted in HST's Medium Deep Survey 
\cite{Griffiths:1994,Windhorst:1995}, although there, the survey reached only down to $I\sim23$.

One worry one might have when examining the morphologies of galaxies at high redshift is that optical images reflect, in fact, the UV rest frame images of the galaxies. Since the UV light traces, in particular, pockets of intense star formation, one might expect a more irregular appearance for 
high-redshift galaxies 
\cite{Giavalisco:1996}. A study of the redshift distribution of galaxies with $17 < I < 21.5$ 
\cite{Im:1999}, showed indeed the irregular/peculiar class to consist of both low-redshift dwarfs and intermediate redshift 
($0.4 \,{\raise-.5ex\hbox{$\buildrel<\over\sim$}}\, z \,{\raise-.5ex\hbox{$\buildrel<\over\sim$}}\, 1$) galaxies that are by themselves a mix of starbursts (of relatively low mass) and more massive, interacting galaxies.

Nevertheless, 
\textcite{Brinchmann:1998} and \textcite{Abraham:1999a} have shown that the redshifting of UV wavelengths into the optical is not sufficient to explain the preponderance of irregular morphologies. Further confirmation that the rapid rise with redshift in the fraction of galaxies with irregular/peculiar morphologies is real, came from the Near Infrared Camera and Multi Object Spectrograph (NICMOS) observations. The NICMOS observations 
\cite{Thompson:1999,Dickinson:2000a,Dickinson:2000b} have demonstrated that with very few (interesting by themselves) exceptions, the irregularities seen in the optical (WFPC2) image persist in the infrared image (Fig.~25), even at redshifts as high as $z\sim3$. Since the NICMOS observations reflect the rest-frame optical light (up to $z\sim3$), the absence of any significant changes in the morphology (in the majority of galaxies) proves that the peculiarities are not artifacts of wavelength shifting.

A morphological trend that may be indicated by the HDFs is a significant decline in the fraction of barred spirals with redshift 
\cite[for $z \,{\raise-.5ex\hbox{$\buildrel>\over\sim$}}\, 0.5$; e.g.,][]{Abraham:1999b}. However, since bars are more difficult to detect at bluer rest wavelengths, optical surveys of galaxies at high-$z$ may be biased against finding bars 
\cite{Eskridge:2000,Bergh:2002}. The precise reason for this trend (if confirmed) is not known, but it may be related either to the state of the disk (the development of a bar instability requires a cold disk), or to the timescale needed to form long-lived bars \cite[perhaps through episodic growth, aided by spiral patterns;][]{Selwood:2000}.

Another extremely interesting result to have come out of the HDFs is related to the \textit{sizes} of galaxies. When examining the HDFs, one is immediately struck by the small angular diameters of the faint galaxies. In particular, No-Evolution models 
\cite[e.g.,][]{Bouwens:1997}, Pure Luminosity Evolution models
\cite[in which galaxies form at some redshift and are characterized by their star-formation history, 
but no merging occurs; e.g.,][]{Metcalfe:1996}, and models dominated by low-surface-brightness galaxies
\cite[e.g.,][]{Ferguson:1995}, all produce half-light radii that are considerably larger than those observed for $24 \,{\raise-.5ex\hbox{$\buildrel<\over\sim$}}\, I \,{\raise-.5ex\hbox{$\buildrel<\over\sim$}}\, 27.5$.
\textcite{Roche:1998}, who used profile fitting to determine half-light radii, also find that at $z > 0.35$ galaxies are significantly more compact than predicted by Pure Luminosity Evolution models (although no evolution in the size was found for $z \,{\raise-.5ex\hbox{$\buildrel<\over\sim$}}\, 0.35$), and more generally, that galaxies at $z \,{\raise-.5ex\hbox{$\buildrel>\over\sim$}}\, 2$ are more compact than today's $L^*$ galaxies.

There is no question that the most attractive explanation for the higher fraction of morphologically peculiar, and smaller galaxies in the past (beyond $z\simeq1$ smaller angular sizes correspond to smaller physical sizes, essentially irrespective of the cosmological model), is in terms of hierarchical galaxy formation and interactions. In the context of this model, the highly irregular morphologies are a consequence of interactions (that were more frequent in the denser past), direct collisions, and mergers. In this picture, the smaller size objects are essentially the ``building blocks'' of today's galaxies. 

Some attempts have been made to transform this qualitative statement into a quantitative tool. While the Hubble Sequence classification 
\cite{Hubble:1926} has proved extremely useful for analyzing the gross morphological properties of nearby galaxies, it becomes rather ineffective at high redshifts, when mergers and intense star formation become the rule, rather than the exception.

A method that is quite successful in distinguishing between irregular galaxies forming stars stochastically and irregularities induced by mergers is based on color-asymmetry diagrams 
\cite[e.g.,][]{Conselice:1997, Bershady:2000}. In these diagrams, galaxies are placed in the $[(B-V),A]$ plane, where $A$ is an asymmetry parameter, based on comparing the galaxy image with its counterpart obtained through a rotation by 180$^{\circ}$. Formally, the asymmetry parameter is defined by
\begin{equation}
A=min\left(\frac{\sum |I_o-I_{\phi}|}{\sum |I_o|} \right)
-min\left(\frac{\sum |B_o-B_{\phi}|}{\sum |I_o|} \right)~~,
\end{equation}
where $I_o$ is the intensity distribution in the image pixels, $I_{\phi}$ is the intensity distribution in the rotated image (by angle $\phi$), and $B$ is the intensity in background pixels (with the second term correcting for the noise). A calibration using nearby galaxies 
\cite{Conselice:2000} shows that all the galaxies with $A>0.35$ represent merging systems. Galaxies with enhanced star formation rates are bluer and may become asymmetric due to pockets of star formation. Nevertheless, their asymmetry parameter does not exceeed 0.35.

An examination of the color-asymmetry diagram of the HDF-N shows that the highest fraction of mergers is found in the highest redshift range, $1.5<z<2.5$. Using $A>0.35$ and $M_B<-18$ as criteria for mergers, gives for the merger fraction as a function of redshift $f\propto(1+z)^{2.1\pm0.5}$ 
\cite{Conselice:2001}. While considerable uncertainties still exist, this suggests that the evolution of galaxies may indeed be primarily dominated by mergers. Furthermore, the merger fraction appears to be starting to flatten (or possibly even to decline) for $z \,{\raise-.5ex\hbox{$\buildrel>\over\sim$}}\, 2$.  As we shall soon see, this may be related to the behavior of the star formation rate with redshift.

Additional information about \textit{hierarchical} vs.\ \textit{monolithic} formation models comes from studies of elliptical galaxies. In the former scenario, giant ellipticals form via mergers of galaxies of comparable mass that had already (prior to merger) used up at least some of their gas to form stars 
\cite[e.g.,][]{Kauffmann:1993}. In the monolithic scenario, on the other hand 
\cite[e.g.,][]{Eggen:1962,Tinsley:1976}, ellipticals form at high redshifts via a single collapse (and a concomitant starburst), and evolve passively thereafter. The two \textit{pure} models have rather distinct observational predictions. Clearly, in the hierarchical scenario the number density of ellipticals is expected to \textit{decrease} with increasing redshift. This can be contrasted with the predictions of the monolithic scenario: the number density of ellipticals should stay fairly constant with redshift, but the bolometric luminosity is expected in increase (up to the starburst phase). The observations, however, proved to be more ambiguous than one might have hoped. In particular, 
\textcite{Treu:1999} found that while Pure Luminosity Evolution models in which all ellipticals are assumed to form at $z\simeq5$ over-predict the observed counts, these models tend to under-predict the counts when the formation redshift is assumed to be $z\simeq2$ 
\cite[see also][]{Zepf:1997,Franceschini:1998,Benitez:1999}. Thus, the observations with both ground-based surveys and HST suggest a picture which is somewhat intermediate between the \textit{pure} hierarchical and monolithic scenarios. Namely, objects that resemble elliptical galaxies (red, obeying a $R^{1/4}$ law in their luminosity profile) existed already at moderately high ($z \,{\raise-.5ex\hbox{$\buildrel>\over\sim$}}\, 1.5$) redshifts 
\cite[e.g.,][]{McCarthy:2001,Moustakas:2002}. These objects continued, however, to experience mergers until $z\sim1$, from which time on the resulting giant ellipticals evolved mostly passively 
\cite[see also][]{Giavalisco:2002}.

The Advanced Camera for Surveys, installed on board HST in March 2002, is expected to produce a wealth of  
high-quality data on galaxy sizes and morphologies. We can therefore expect even more significant constraints on hierarchical formation models to emerge in the coming months.

As I noted above, the evolution of galaxies is also intimately related to the way they assemble their mass and, concomitantly, to their star formation rate. As it turned out, the HDF-N proved to be seminal in the discussion of the cosmic star formation history.

\subsection{The global cosmic star-formation history}

The rest-frame UV luminosity of galaxies traces nicely their metal production rate, because both are produced primarily by massive stars. On the other hand, the conversion from a UV luminosity density to an actual cosmic star formation rate is somewhat less straightforward, since it involves a knowledge of the Initial Mass Function (IMF).

Early estimates of the metal production rate as a function of redshift, $\dot{\rho}_Z(z)$, relied on 
ground-based redshift surveys 
\cite{Lilly:1996,Gallego:1995}
and data from QSO absorption line systems 
\cite[e.g.,][]{Lanzetta:1995,Pei:1995}.  These estimates indicated a monotonic increase in the metal production rate from $z=0$ to $z\simeq1$ (with the rate at $z\simeq1$ being about an order of magnitude higher than at the present).

One of the seminal results to have come out of the HDF-N was the attempt to estimate the metal production rate at high redshifts and thereby to generate a continuous plot of the star formation rate as a function of redshift 
\cite{Madau:1996}. In the original diagram, no corrections were made, for either dust extinction or for surface brightness effects. Consequently, the results were presented as lower limits. Both in the original work and in some of the subsequent work that followed (some of which included the effects of dust extinction), it was found that the star formation rate peaks at a redshift of $z\sim1$--2 and decreases or stays nearly constant at higher redshifts 
\cite[e.g.,][Fig.~26]{Steidel:1999,Hopkins:2000,Calzetti:1999,Pei:1999}. The question of whether there truly is a decrease in the star formation rate for $z \,{\raise-.5ex\hbox{$\buildrel>\over\sim$}}\, 2$ has become a focal point of the discussion of the cosmic star formation history. Some doubts were raised on the basis of dust extinction and selection effects on one hand, and cosmological surface brightness dimming effects on the other. Broadly speaking, the UV luminosity density \textit{could} be underestimated if a significant fraction of the star formation occurred in environments obscured by dust or in very 
low-surface-brightness galaxies. Several attemps were made to correct for dust attenuation 
\cite[using the empirical attenuation law of][]{Calzetti:1994}, by calibrating the relation between the UV spectral slope and the far IR emission 
\cite[e.g.,][]{Meurer:1999,Steidel:1999}. Others used constraints obtained from extragalactic background radiation and neutral gas 
\cite[e.g.,][]{Pei:1999,Calzetti:1999}. Most of these investigations concluded that the star formation rate rises from the present to $z\sim1$--2 and then stays approximately flat to $z\sim5$.

Another way to address the question of the history of mass assembly in galaxies is to try to measure the actual stellar \textit{masses} of galaxies (as opposed to the \textit{rates} of star formation; ideally one would want to do \textit{both}). To this goal, observations in the near-infrared are typically used, since the near-infrared luminosity traces the stellar mass reasonably well. 
\textcite{Dickinson:2003} used an infrared-selected sample of galaxies from the HDF-N to determine the global stellar mass density, $\Omega_*(z)$, for $0<z<3$. They found that $\Omega_*(z)$ increases with time from $z=3$ to the present (Fig.~27). Dickinson \textit{et~al.} concluded that by $z\sim1$, about 50--75\% of the present-day stellar mass density had already formed, but that the stellar mass density at $z\sim2.7$ was about 17 times lower than today.  These observations appear to be in clear contradiction with scenarios in which most stars in today's spheroids formed at $z\gg2$, but the observations are in general agreement with a global star formation rate that rises from the present to $z\sim1$ and then stays fairly flat. 

A different spin on the cosmic star formation rate has been put by 
\textcite{Lanzetta:2002}. These authors claimed that by neglecting cosmological surface brightness dimming effects, previous works have missed a significant fraction  of the ultraviolet luminosity density at high redshifts. Specifically, since the surface brightness decreases with redshift as $(1+z)^{-3}$ (because of the cosmic expansion), intrinsically faint regions of high-redshift galaxies become undetectable.

Lanzetta \textit{et al.} designated the unobscured star formation rate intensity (i.e., the intensity inferred from the \textit{observed} rest-frame UV light) by $x$, and used a distribution function $h(x)$ (defined so that $h(x)dx$ is the projected proper area per comoving volume of star formation rate intensity in the interval $x$ to $x+dx$), to estimate the ultraviolet luminosity density at high redshifts (including the surface brightness dimming effects). They found that the star formation rate density, $\dot{\rho}_s$, increases monotonically with redshift to the highest redshifts observed ($z\sim8$; although in one of the possible corrections for incompleteness $\dot{\rho}_s$ remains fairly flat above $z\sim2$). 

The low fraction of stellar mass formed by $z=3$ according to the Dickinson \emph{et~al.} results appears to contradict evidence for significant star formation occurring at still higher redshifts. One possible way out of this conundrum is that the initial mass function (IMF) at high redshifts is very top-heavy 
\cite[tilted towards massive stars;][]{Ferguson:2002}. Massive stars dominate the UV luminosity in star forming  galaxies, but their contribution to the total surviving (at lower redshifts) stellar mass is relatively low.

Most recently, \textcite{Stanway:2003} used HST's Advanced Camera for Surveys, the ground-based Sloan Digital Sky Survey and the Very Large Telescope (VLT) to determine the space density of UV-luminous starburst galaxies at $z\sim6$. They found a lower bound to the integrated, volume-averaged, global star formation rate at $z\sim6$, that was about six times less than that at $z\sim3$--4. The question of the true behavior of $\dot{\rho}_s$ above $z\sim2$ remains, therefore, presently somewhat unresolved.

Fortunately, a more definitive answer may come in the near future, through a combination of planned observations with HST and with the Space Infrared Telescope Facility (SIRTF; currently scheduled to be launched in April 2003). The Great Observatories Origins Deep Survey (GOODS; Principal Investigator M.~Dickinson) will produce a very deep image of two fields (the HDF-N  and the southern deep field observed with the Chandra Observatory) with SIRTF at 3.6--24~$\mu$m, and will thereby produce a much more complete census of stellar mass at high redshifts. At the same time, observations (of the same fields) with Hubble's Advanced Camera for Surveys (Principal Investigator M.~Giavalisco; the observations are being carried out as these lines are being written) will determine the star formation rates, sizes, and morphologies of galaxies. The combined observations will allow for the first time for a determination of the evolving mass assembly distribution $f(M$, $\dot{M}$, $t)$ (where $M$ denotes the stellar mass and $\dot{M}$ the star formation rate). Given the fact that under pure luminosity evolution, $M=\int \dot{M} dt$, a comparison between the \textit{observed} evolution of the ($M$, $\dot{M}$) phase space with time and the evolution obtained from \textit{direct integration} (of $\dot{M}$, to produce $M$), will allow, in principle, for an identification of the role of mergers and interactions. Even when the expected observational uncertainties are taken into account, there is no doubt that the planned GOODS observations will yield a huge step forward in the understanding of the \textit{assembly} of present-day galaxies, and their \textit{morphological} evolution (the emergence of the ``Hubble Sequence''). Furthermore, the planned Hubble Ultra Deep Field with the Advanced Camera for Surveys (currently scheduled for July--August 2003) could extend the redshift coverage unambiguously to $z\sim6$, close to the tail of the tentative second \textit{reionization epoch} \cite{Fan:2001,Becker:2001} of the universe 
\cite[the first reionization having tentatively occurred at $z\sim20^{+10}_{-9}$;][]{Bennett:2003}.
Such a study could therefore produce results that are not merely incremental in our understanding of the cosmic star formation history, galaxy evolution, and the ionization history of the universe, and that can be used to place meaningful constraints on theoretical models \cite[e.g.,][]{Somerville:2003}.

\section{THE HUBBLE CONSTANT}

\subsection{A brief background}

Ever since Edwin Hubble's pioneering measurements in the 1920s 
\cite[and to some extent even before that, since Vesto Slipher's measurements, that started in 1912; e.g.,][]{Slipher:1917}, we knew that we live in an expanding universe 
\cite{Hubble:1929,Hubble:1931}.
In the standard big bang theory the universe expands uniformly, with the recession velocity being related to the distance through the Hubble law, $v=H_0d$. More generally, based on the ``Cosmological Principle'' (the assumption that the universe is homogeneous and isotropic on large scales), the expansion is governed by the Friedmann equation in the context of general relativity
\begin{equation}
\left(\frac{\dot{R}}{R}\right)^2\equiv H^2 =
\frac{8\pi G\rho_M}{3}-
\frac{kc^2}{R^2}+
\frac{\Lambda c^2}{3}~~.
\end{equation}
Here $R(t)$ is the scale factor, $H=\dot{R}/R$ measures expansion rate (with $H_0$, the ``Hubble Constant,'' giving the rate at present), $\rho_M$ is the mass density, $k$ is the curvature parameter and $\Lambda$ is Einstein's cosmological constant (which represents the energy density of the vacuum). Commonly, the density of matter and that associated with the vacuum are represented by the density parameters (at present) $\Omega_M=8\pi G\rho_M/3H^2_0$ and $\Omega_\Lambda =\Lambda c^2/3H^2_0$, with which the Friedmann equation can be expressed as 
$\displaystyle\frac{kc^2}{R_o^2}=H^2_0(\Omega_M+\Omega_\Lambda -1)$. The Hubble Constant is thus the key parameter in determining the age of the universe (with $\Omega_M$ and $\Omega_\Lambda$ also playing a role). Similarly, physical processes such as the growth of structure and the nucleosynthesis of light elements (H, D, $^3$He, $^4$He, Li), as well as critical epochs in the Universe's history, such as the transition from a radiation-dominated to a matter-dominated universe, depend on the cosmic expansion rate and thereby on the value of $H_0$. It should therefore come as no surprise that the determination of the value of the Hubble Constant became a major observational goal for the past eight decades.

The first value for the Hubble Constant may have actually been derived by 
\textcite{Lemaitre:1927}, who, on the basis of Slipher's radial velocity measurements and Hubble's mean absolute magnitude for galaxies (``nebulae'') obtained $H_0=526$~km~s$^{-1}$ Mpc$^{-1}$. The next set of values by \textcite{Hubble:1929} and Hubble and Humason (who produced a velocity-distance relation up to $V\sim20000$~km~s$^{-1}$ and obtained $H_0=559$~km~s$^{-1}$ Mpc$^{-1}$ in 1931), were all around 500~km~s$^{-1}$ Mpc$^{-1}$, with an uncertainty stated rather naively as ``of the area of ten percent.'' About twenty years passed before 
\textcite{Baade:1954} revised the distance to nearby galaxies, recognizing that Hubble confused two classes of ``standard candles'' (Population~I Cepheids and Population~II W~Virginis stars), thereby reducing the value of $H_0$ by about a factor two 
\cite[a revision suggested also by][]{Behr:1951}. The value of the Hubble Constant first reached the range of values accepted today through the work of Alan 
\textcite{Sandage:1958}. Sandage demonstrated that Hubble mistakenly identified H~II regions as bright stars, and he [Sandage] was able to revise the value further to $H_0\simeq75$~km~s$^{-1}$ 
Mpc$^{-1}$ (recognizing that the uncertainty could still be by a factor~2).

In the three decades that followed, published values of the Hubble Constant varied by about a factor of two between $\sim 100$ and 50~km~s$^{-1}$ Mpc$^{-1}$. Table~1, adapted from 
\textcite{Trimble:1997}, summarizes the early history of the constant.

Generally, since redshifts (and therefore radial velocities) can be determined relatively readily (this is, of course, not true for the most distant or faintest objects), the problem of determining the Hubble Constant has always been a problem of determining accurate astronomical distances. The availability of new instrumentation, and the Hubble Space Telescope in particular, have allowed for a dramatic improvement in distance determinations.

\subsection{Distance indicators and methods}

Direct trigonometric parallaxes that use the Earth's orbit around the Sun as a baseline for triangulation can only be used to the nearest stars. Consequently, other distance indicators and methods had to be used for extragalactic distances 
\cite[see, e.g.,][for excellent reviews]{Jacoby:1992,Trimble:1997}. The most common of these employ ``standard candles,'' geometrical properties, physical properties, or various correlations to determine distances.

Standard candles are simply based on the fact that the flux of radiation decreases as an inverse square law. Objects with either a constant luminosity or whose luminosity can be related to a distance-independent measurable property (such as an oscillation period) are good standard-candle candidates. The best known and probably most reliable in this class are the Cepheid variables. Their potential as standard candles on the basis of their period-luminosity (P-L) relation was first recognized by Henrietta Leavitt in 1912, and they were used by \textcite{Hubble:1925} to determine distances to Local Group galaxies. The physical processes responsible for the P-L relation are broadly understood. Near ionization zones (in this case, primarily He$^{+}$\,\lower.2ex\hbox{$\buildrel{\rightarrow}\over{\scriptstyle\leftarrow}$} He$^{++}$), gas can absorb heat under compression and release it after maximum compression. This leads to an instability strip in the effective temperature-luminosity plane, which, in the case of Cepheids, is very narrow in temperature. The pulsation period depends on the mass and radius as
\begin{equation}
P\sim\frac{1}{(G\rho)^{\frac{1}{2}}}\sim M^{-\frac{1}{2}} R^{\frac{3}{2}}~~.
\end{equation}
The luminosity of the star (which is determined ultimately by nuclear reactions that depend on the density and temperature) is proportioned to a power of the mass, $L\sim M^k$. Since, however, we also have (for black body radiation) $L\sim R^2 T^4$, we obtain $P\sim L^{\frac{3k+2}{4}} T^{-3}$, or a 
period-luminosity-color relation. One of the ``Key Projects'' of HST has been to measure $H_0$ based on a Cepheid calibration of a number of secondary distance determination methods.

The results of this project have been described in a series of some 30 papers 
\cite[see][and references therein]{Freedman:2001}. The main goals of the project have been:  (i)~To discover Cepheids in a sample of relatively nearby galaxies (with distances 
{\raise-.5ex\hbox{$\buildrel<\over\sim$}}\,20~Mpc) and to determine distances to these galaxies. (ii)~To determine $H_0$ through several secondary distance indicators, to all of which Cepheid calibration is applied; and perhaps most importantly, (iii)~To determine the uncertainties in all the methods by comparing the distances obtained from them to Cepheid-based distances.  As a part of the project, the uncertainties in the Cepheid P-L relation itself (and its dependence on other factors such as metallicity) have been investigated.

The Key Project used Cepheid calibration to 31 galaxies, of which 18 have been observed and analyzed in the context of the project, and to which archival data, and data on the nearby galaxies M31, M33, IC~1613, NGC~300, and NGC~2403 have been added.

The Key Project used the following secondary methods based on Cepheid distances:  Type~Ia Supernovae, the Tully-Fisher Relation, the Fundamental Plane for elliptical galaxies, Surface Brightness Fluctuations and Type~II Supernovae. Let me describe very briefly the physical basis for each one of these methods.

\subsubsection{Type Ia supernovae}

Type Ia supernovae at peak brightness are extremely bright, with $M_B\simeq M_V\simeq-19.3 + 5\log(H_0/60)$, and they show a relatively low dispersion, $\sigma(M_B)\sim0.33$
\cite[e.g.,][]{Branch:1998}. Furthermore, there exists a relatively tight correlation between their peak luminosity and light-curve shape 
\cite[or rate of decline, with brighter supernovae declining more slowly;][]{Phillips:1993,Hamuy:1996,Riess:1996}. The homogeneity may be related to the fact that Type~Ia supernovae represent thermonuclear disruptions of mass-accreting white dwarfs, when the latter reach the Chandrasekhar limit 
\cite[e.g.,][]{Livio:2001}. The luminosity-light-curve relation may be the result of the following
\cite[e.g.,][]{Arnett:2001}. The peak luminosity of a supernova Type~Ia is proportional to the mass of $^{56}$Ni that is produced. A higher mass of $^{56}$Ni, however, also results in more heating and concomitantly a higher opacity (due mainly to UV lines). Consequently, a slower development of the light curve results.

\subsubsection{The Tully-Fisher relation}

For spiral galaxies, that are known to have flat rotation curves, there is an observationally-determined relationship 
\cite{Tully:1977} between the total luminosity and the maximum rotational velocity (both corrected for inclination effects), of the form $L\sim V^3_\mathrm{max}$ (in the $I$~band). The scatter around this relation is about $\pm0.3$~mag 
\cite[e.g.,][]{Giovanelli:1997}.

There is no precise physical understanding of the Tully-Fisher relation. Very broadly, a similar relation can be obtained from the following argument 
\cite[e.g.,][]{Eisenstein:1996}. Consider a galaxy (of mass $M$) collapsing from a spherical cloud. At epoch $t_\mathrm{coll}$, the turnaround radius $R_t$ is
\begin{equation}
R_t\sim M^{\frac{1}{3}} t^{\frac{2}{3}}_\mathrm{coll}~~~.
\end{equation}
After virialization, the energy is
\begin{equation}
E\sim M\sigma^2\sim G M^2/R_t~~~,
\end{equation}
where $\sigma$ is the velocity dispersion. Combining the above and assuming that the galaxy forms from a single collapse gives
\begin{equation}
\sigma\sim(M/t_\mathrm{coll})^{\frac{1}{3}}~~~.
\end{equation}
Therefore, if all galaxies collapse at the same epoch and the mass-to-light ratios, $M/L$, do not vary significantly, we obtain $L\sim\sigma^3$.

\subsubsection{The fundamental plane for elliptical galaxies}

Large spectrophotometric surveys conducted during the mid-1980s 
\cite[e.g.,][]{Djor:1987,Dressler:1987} revealed that, for elliptical galaxies, a tight correlation exists between the effective radius, $R_e$, the effective surface brightness, $SB_e$, and the central velocity dispersion $\sigma$, of the form (the ``fundamental plane'')
\begin{equation}
\log R_e = \alpha\log\sigma + \beta SB_e + \delta~~~.
\end{equation}
Here $R_e$ is in kpc, $\sigma$ in km~s$^{-1}$, and $SB_e$ in mag arcsec$^{-2}$.
The value of $\delta$ depends on $H_0$, since the calculation of the effective radius in kpc uses the Hubble constant.

The physical origin of the fundamental plane relation can be understood on the basis of the following simple considerations 
\cite[e.g.,][]{Treu:2001}. We can define an effective (virial) mass by 
\begin{equation}
M=\sigma^2 R_e/G~~~.
\end{equation}
Let us also assume that the mass-to-light ratio satisfies
\begin{equation}
\frac{M}{L} \sim M^{\delta}~~~.
\end{equation}
We then obtain
\begin{equation}
L\sim \sigma^{2(1-\delta)} R_e^{(1-\delta)}~~~,
\end{equation}
which reduces to the fundamental plane relation for reasonable values of $\delta$ ($\delta\sim0.25$).

\subsubsection{Surface brightness fluctuations}

The method of surface brightness fluctuations was developed by 
\textcite{Tonry:1988} and 
\textcite{Tonry:1997,Tonry:2000}. The method basically makes use of the obvious fact that the ability to resolve stars within galaxies is distance dependent. More specifically, for every region of a galaxy one can measure the average flux per pixel, $g$, and the pixel-to-pixel rms, $\sigma$. Since the flux obtained in a pixel is received from $N$ stars of average flux $\bar{f}$, we have $g=N\bar{f}$ and $\sigma=\sqrt{N}\bar{f}$. A galaxy which is twice as distant appears twice as smooth as the closer galaxy. Consequently, the average stellar flux is given by $\bar{f} = \sigma^2/g$ (and $\bar{f}$ scales as the inverse of the square of the distance).

\subsubsection{The Expanding Photosphere Method of Type II supernovae}

Type II supernovae result from the collapse of stars more massive than about 8~M$_\odot$. Generally, Type~II supernovae are fainter than Type~Ia supernovae, and they also exhibit a considerably larger range in their luminosities, making them poorer standard candles. Nevertheless, Type~II supernovae have been used as distance indicators, through an application of the Expanding Photosphere Method 
\cite[e.g.,][]{Kirshner:1974,Schmidt:1994}.

The basic idea is simple. The angular size of the photosphere is given by (for $z\ll1$)
\begin{equation}
\Theta =\frac{R}{D} =
\left( \frac{f_{\lambda}}{\zeta^2_{\lambda}\pi B_{\lambda} (T)}\right)^{\frac{1}{2}}~~~,
\end{equation}
where $T$ is the color temperature, $f_{\lambda}$ is the flux density, $B_{\lambda}(T)$ is the Planck function, and $\zeta_{\lambda}$ represents the dilution effects of scattering atmospheres (derived from model atmospheres). 

The photospheric radius is given by
\begin{equation}
R= v(t-t_o) + R_o~~,
\end{equation}
where $v$ is the expansion velocity (measured from the absorption minima of optically thin lines), and the initial radius, $R_o$, can be neglected at all but the earliest times. Combining the above yields
\begin{equation}
t = D\left(\frac{\Theta}{v} \right) + t_o~~~,
\end{equation}
making it possible to determine both the distance and the time of the explosion from a few measurements of $t$, $v$, $\Theta$.

\subsection{The results}

Table~2, adapted from 
\textcite{Freedman:2001}, lists the values of $H_0$ obtained from the different methods, based on the Cepheid distances of the Key Project. Of the five calibrated methods, it is clear that the Fundamental Plane is somewhat of an outlier. Combining these results, the $H_0$ Key Project obtained the value $H_0=72\pm3\pm7$~km~s$^{-1}$ Mpc$^{-1}$, where the first quoted error is random and the second is systematic. Using three different weighting schemes, all the results were found to be consistent with $H_0=72\pm8$~km~s$^{-1}$ Mpc$^{-1}$.

In an independent work, Allan Sandage, Gustav Tammann, Abijhit Saha, and collaborators used a Cepheid calibration of the peak brightness of Type~Ia Supernovae (using HST), and thereby determined $H_0$ directly from the Hubble diagram of the latter. This effort resulted to date in nine supernovae Ia with ``normal'' spectra (i.e., the spectra characterizing about 60\% of all Type~Ia supernovae) to which Cepheid distances are known \cite[Table~3, adapted from][]{Saha:2001}.  After corrections based on the decline rate and colors, the weighted average of the absolute magnitudes of the calibrators was fitted to a sample of 35 more distant, ``normal'' Type~Ia supernovae. After further correcting for some systematic errors, 
\textcite{Saha:2001} obtained $H_0=58.7\pm6.3$(internal)~km~s$^{-1}$ Mpc$^{-1}$ (for cosmological parameters $\Omega_M=0.3$, $\Omega_\Lambda=0.7$).

Unlike the uncertainty by a factor of two that has plagued this field for decades, therefore, the observations with HST have reduced the uncertainty in the value of the Hubble constant to about 15\%.

The values obtained by both the Key Project and the Sandage-Tammann-Saha team are also consistent with other recent measurements based on combining the Sunyaev-Zeldovich effect with x-ray flux measurements of clusters, and on time delays in gravitational lensing. The systematics in these methods are still estimated to be at the 20--25\% level, and the value obtained for the Hubble constant is 
$H_0\sim60$~km~s$^{-1}$ Mpc$^{-1}$ 
\cite[e.g.,][]{Schecter:2000,Reese:2000,Reese:2002}.

Analysis of Sunyaev-Zeldovich effect 
\cite{Sunyaev:1970} and x-ray data provides a direct method for determining distances to galaxy clusters. Clusters of galaxies are known to contain hot intracluster gas (at $kT\sim10$~keV), trapped in the clusters' potential wells. Photons from the cosmic microwave background that pass through a cluster have a finite probability (optical depth $\tau\sim0.01$) to interact with energetic electrons in the intracluster gas. The inverse Compton scattering that ensues boosts the energy of the microwave background photon, generating a small distortion (a decrement in frequencies 
{\raise-.5ex\hbox{$\buildrel<\over\sim$}}\,218~GHz and an increment above this value) in the spectrum of the microwave background. The Sunyaev-Zeldovich effect is proportional to the integral of the pressure along the line of sight ($\int n_e T_e dl$). Since the x-ray emission from the intracluster medium is proportional to a different power of the density, $\int n_e^2 \Lambda dl$ (where $\Lambda$ is the cooling function), a combination of the two measurements (given some assumptions about the cluster geometry) can be used to determine the distance to the cluster, independently of the distance scale ladder that is based on standard candles.

Finally, the Wilkinson Microwave Anisotropy Probe (WMAP) found a Hubble constant of 
$H_0=72\pm5$~km~s$^{-1}$ Mpc$^{-1}$. When the WMAP data were combined with the Key Project results, 
finer-scale cosmic microwave background experiments, and other large-scale structure and Lyman~$\alpha$ forest data, the best-fit value of the Hubble constant was $H_0=71^{+4}_{-3}$~km~s$^{-1}$ Mpc$^{-1}$ 
\cite{Spergel:2003}.

Type Ia supernovae have played a key role not only in the determination of the \textit{age} of the universe (through $H_0$), but also in the determination of the universe's \textit{geometry}, and the dynamics of the cosmic expansion. This came about through a combination of ground-based and HST observations in which both careful planning and serendipity played a part.

\section{THE ACCELERATING UNIVERSE}

In 1998, two teams of astronomers, working independently, presented evidence that the expansion of the universe is \emph{accelerating} 
\cite{Riess:1998,Perlmutter:1999}. This evidence was based primarily on the faintness (by about $\sim0.25$~mag) of distant Type~Ia supernovae (SNe~Ia), compared to their expected brightness in a universe decelerating under its own gravity.

The first suggestions that high-redshift Type~Ia supernovae could be used to determine the rate of cosmic deceleration came in the late 1970s 
\cite{Wagoner:1977,Colgate:1979,Tammann:1979}, but the actual discovery of an even moderate redshift ($z=0.31$) SN~Ia (SN~1988U) had to wait for a decade 
\cite{Norgaard:1989}. The samples of high-redshift supernovae started to become sufficiently large to place constraints on cosmological parameters through the efforts of the Supernova Cosmology Project, led by Saul Perlmutter, and the High-$z$ Supernova Search Team, led by Brian Schmidt. By 1998 the two teams gathered sufficient data to be able to show that the universe is characterized by a matter density paremater satisfying $\Omega_M <1$ 
\cite[i.e., that the universe is not closed by matter;][]{Perlmutter:1998,Garnavich:1998}. However, the real shocker was still to come.

\subsection{The observations}

The two supernova search teams use a similar method to find their supernovae. They take two deep images separated by about a month, subtract the first-epoch image from the second-epoch one, and search for sources above a certain threshold in the difference image. Once a candidate SN is identified, the SN type is determined by its spectrum (if that can be taken; in a few cases one has to rely on the host galaxy 
type---only SNe~Ia were found so far in ellipticals). Supernovae at relatively high redshift are then monitored photometrically to construct their light curves. An image of the host galaxy is obtained at a later time (after a year or more), and subtracted to obtain an accurate measurement of the SN brightness.

The Hubble Space Telescope has proved to be crucial especially for the highest-redshift supernovae. There, the ability to resolve and pinpoint the SN location on the host (including in cases in which the SN was found close to the galactic nucleus) was essential for a correct determination of the SN magnitudes. With samples of a few dozen SNe~Ia in hand, the two teams compared their measured distances (derived from the 
luminosity-distance relation,  $F=L/4\pi D_L^2$; where $D_L$ is the distance and $L$, $F$ are the intrinsic luminosity and observed flux, respectively) with the distances expected for the observed redshifts, for different cosmological models 
\cite[e.g.,][]{Carroll:1992}. The latter is given by
\begin{eqnarray}
D_L=&c H_0^{-1} (1+z) | 1-\Omega_M - \Omega_\Lambda |^{-1/2}
\mathrm{sinn} \left\{ | 1-\Omega_M - \Omega_\Lambda |^{1/2}     \times \right.\cr
&\left. \int^z_0 dz[(1+z)^2 (1+\Omega_M z) - z(2+z)\Omega_\Lambda ]^{-1/2} \right\} 
\end{eqnarray}
where sinn denotes sinh for $\Omega_M + \Omega_\Lambda\leq1$ and $\sin$ for $\Omega_M + \Omega_\Lambda >1$. The results for the likelihood of the cosmological parameters $\Omega_M$ and $\Omega_\Lambda$ are shown in Fig.~(28). As can be seen from the figure, the results favor values of $\Omega_M\simeq0.3$, $\Omega_\Lambda\simeq0.7$ and a \textit{negative} deceleration parameter ($a$ is the scale factor, $\Omega_0$ is the sum of today's energy densities, and $w\equiv P_\Lambda/\rho_\Lambda$ characterizes the dark energy ``equation of state''; see section VIII~D)
\begin{equation}
q_0\equiv \frac{-\ddot{a}(t_0)a(t_0)}{\dot{a}^2(t_0)} = 
\frac{\Omega_0}{2} + \frac{3}{2}w \Omega_\Lambda < 0~~,
\end{equation}
corresponding to an accelerating universe.

\subsection{Alternative interpretations}

Clearly, the interpretation of an accelerating cosmic expansion cannot go unchallenged
\cite[see also][]{Riess:2000,Turner:2000}. The two most serious challenges to acceleration are, in principle, evolution/progenitors and dust extinction. 

The possibility that SNe~Ia undergo some type of cosmic evolution cannot be easily dismissed. After all, almost everything else (e.g., the host galaxies themselves, the metallicity) evolves. The inference of acceleration, on the other hand, assumes that the local and high-redshift supernovae are drawn from the same statistical sample. I should explain that by evolution, I do not mean that somehow the processes governing supernova explosions are necessarily changing with redshift, but rather that when observing the 
high-redshift universe we are sampling younger galaxies, and consequently the population from which the supernovae are drawn is typically younger.

Since SNe~Ia represent thermonuclear disruptions of mass-accreting white dwarfs 
\cite{Livio:2001,Hillebrandt:2000}, changes in the C/O ratio in the white dwarfs (as a result of lower metallicities in the distant past), for example, could produce some inhomogeneity in SNe~Ia light curves 
\cite{Umeda:1999}. However, observations tend to show that even if such an effect exists, it is insignificant. In fact, existing observations of local galaxies 
\cite[e.g.,][]{vandenBergh:1994,Cappellaro:1997,Riess:1999} already span a wider range of metallicities and host properties than that expected on the average between the sample of galaxies at $z\sim0$ and $z\sim0.5$. 

To my knowledge, there is only one known evolutionary effect that is physically meaningful, that can mimic accelerated expansion. This is the effect of the metallicity on the density at the point of central carbon ignition (the trigger of SNe~Ia). Generally, a lower metallicity (as expected at high-$z$) will result in a lower central density 
\cite{Nomoto:1997}. This is because a lower metallicity results in a lower abundance of $^{21}$Ne, that is responsible for much of the neutrino cooling (via the so-called local URCA shell process, involving the $^{21}$Ne-$^{21}$F pair). A lower metallicity therefore reduces the cooling and leads to an earlier ignition. Due to the lower white dwarf binding energy, the light curve exhibits a more rapid development, and a lower implied maximum brightness. However, as I noted above, the local (low-$z$) sample of galaxies does not appear to exhibit any significant metallicity-dependent effect.

Dust extinction is another natural candidate for making the distant supernovae appear dimmer. In fact, the observed decrease in the brightness ($\sim0.25$~mag) of the distant sample would require only a $\sim25$\% increase in the extinction. Both teams used colors to correct for dust extinction (like in sunsets, ordinary dust also reddens the light). Furthermore, observations from the optical to the infrared of SN~1999Q (at $z=0.46$) showed that a large extinction to this supernova is highly improbable 
\cite{Riessetal:2000}, making the dust hypothesis untenable.

Another potential uncertainty is associated with the progenitors of SNe~Ia. The progenitor systems of SNe~Ia are not known without doubt 
\cite{Livio:2001}. In particular, the two leading progenitor-system candidates are double white dwarf systems 
\cite[that merge to produce the explosion;][]{Webbink:1984,Iben:1984} and systems in which the white dwarf accretes from a normal companion 
\cite{Whelan:1973,Nomoto:1982}.  The possibility therefore exists, in principle, that the local and distant samples are dominated by different progenitors, with one class producing somewhat dimmer supernovae. This possibility should certainly be considered, especially in view of the fact that 
\textcite{Li:2001} find a relatively high ($\sim40$\%) fraction of peculiar SNe~Ia among the local sample.

\textcite{Yungelson:2000} have shown that while it is possible, in principle, that one class of progenitors (e.g., double white dwarf systems) dominates the local sample and another (a white dwarf with a normal companion) the high-$z$ one, such a transition is not likely, because of the following reason. If such a transition were to occur, one would expect that around the transition point (at $z\sim1$), SNe~Ia would be composed of an equal mix of the two classes.  This is inconsistent with the observations, that show that the high-$z$ sample is actually extremely homogeneous 
\cite{Li:2001}. Consequently, it is highly unlikely that the interpretation of an accelerating universe is an artifact of the existence of two classes of progenitors \cite[see][for a more complete discussion]{Livio:2001}.

Other effects, such as those resulting from gravitational lensing, will require some further consideration once larger databases of high-redshift supernovae become available. Generally, it can be expected that most lines of sight to detected supernovae do not pass through large mass concentrations. Consequently, light paths are being bent \textit{out} of the line of sight---resulting in deamplification. In rare cases however, the line of sight may cross significant matter concentrations resulting in strong amplification. The picture that emerges, therefore, is that of a brightness distribution in which the peak is shifted toward a lower value, but with a tail of bright objects 
\cite[e.g.,][]{Holz:1998}.  Effects of this type can be averaged out once deep data from many lines of sight become available, perhaps with the proposed wide-field imager, the Super Nova/Acceleration Probe (SNAP). 

\subsection{Supernova 1997ff---serendipity and careful planning}

In the spirit of ``extraordinary claims require extraordinary proof,'' the magnitude of the discovery of an accelerating universe requires an extremely careful elimination of potential contaminating astrophysical effects. A pervasive screen of ``gray'' dust, for example, that can dim the light while leaving little imprint on the spectral energy distribution, has been suggested as an alternative explanation for the unexpected faintness of high-$z$ SNe~Ia 
\cite{Aguirre:1999}. While gray dust, e.g., quasispherical grains larger then 0.1~$\mu$m, could be made to reproduce some of the observations, some measurements do disfavor a 30\% visual opacity due to gray dust at the $\sim2.5\sigma$ level 
\cite{Riessetal:2000}. Nevertheless, both the possibility of gray dust and, in particular, the 
ever-existing challenge of luminosity evolution, did leave the case for an accelerating universe somewhat short of compelling by 2001. The strongest evidence so far against alternatives to an accelerating universe came from SN~1997ff.

In 1997, Ron Gilliland and Mark Phillips reobserved the Hubble Deep Field North with HST, with the goal of detecting high-$z$ supernovae 
\cite{Gilliland:1999}. They discovered two supernovae: SN~1997fg, at a redshift of $z=0.95$ was found in a 
late-type galaxy, while SN~1997ff was found to be hosted by an elliptical galaxy. The initial, photometrically determined redshift, placed the latter galaxy at $z\sim0.95$--1.32 with considerable uncertainty. Serendipitously, the Guaranteed Time Observer (GTO) program of 
\textcite{Thompson:1999} included imaging the host of SN~1977ff with the near infrared camera, NICMOS, on board HST. One of these exposures, amazingly enough, was actually taken within hours of the discovery of the SN. Furthermore, six months after the GTO program, 
\textcite[][in preparation; GO 7817]{Dickinson:2003} reobserved the field with NICMOS. The combination of all the HST and ground-based observations (in the $U$, $B$, $V$, $I$, $J$, $H$, and $K$ bands) allowed for an improved photometric redshift determination for the host of $z=1.65\pm0.15$ 
\cite{Budavari:2000}, making SN~1997ff the farthest known supernova to date. A probability density function based on the partial supernovae light curve in three bands also gave $z\simeq1.7$, as did a tentative determination of the redshift based on the spectrum of the host taken with the Keck telescopes 
\cite[see][for a full description]{Riess:2001}. 

The importance of detecting a SN~Ia at such a high redshift cannot be overemphasized. Even with the suggested value of the vacuum energy density of $\Omega_\Lambda\simeq0.7$, at 
$z\,{\raise-.5ex\hbox{$\buildrel>\over\sim$}}\,1$ attractive gravity would still have dominated over the cosmic repulsion, resulting in a \emph{decelerated} expansion. Consequently, SNe~Ia in the redshift range $z\sim1$--2 should appear \emph{brighter} relative to SNe in a coasting universe. On the other hand, any effect that increases monotonically with redshift, as would be expected from dust extinction and simple evolutionary effects, would predict that the SNe~Ia would be \emph{dimmer} with increasing redshift. SN~1997ff was found to be brighter by $\sim1.1$~mag (at the $>99.99$\% confidence level) than expected for an alternative source of dimming (e.g., dust or evolution) beyond $z\sim0.5$. Figure~29 shows a 
redshift-distance relation in which the points are redshift-binned data from 
\textcite{Perlmutter:1999} and 
\textcite{Riess:1998}, together with a family of curves for flat cosmological models. The transition from a decelerating to an accelerating phase occurs (if the ``dark energy'' is represented by a cosmological constant; see Section VIII~D) at $z_{tr}=(2\Omega_\Lambda/\Omega_M)^{1/3}-1$. As can be seen from the figure, the only cosmological model that is consistent with \emph{all} the data (including SN~1997ff) is one in which $\Omega_M\simeq0.35$, $\Omega_\Lambda\simeq0.65$.

Measurements of the power spectrum of the cosmic microwave background 
\cite[e.g.,][]{deBernardis:2002,Abroe:2002,Hu:2001,Netterfield:2002} provided strong evidence for a geometrically flat ($\Omega_M +\Omega_{\Lambda}\simeq1$) universe. When combined with several estimates of $\Omega_M$ from mass to light ratios, x-ray temperature of intracluster gas, numbers, and dynamics of clusters of galaxies, all giving $\Omega_M\simeq0.2$--0.3 
\cite[e.g.,][]{Carlberg:1996,Bahcall:1997,Bahcall:2000,Strauss:1995}, the inescapable conclusion was that there is a dark energy component of $\Omega_{\Lambda}\simeq0.7$---consistent with the value found from 
high-redshift supernovae. Most recently, the best fit cosmological model has been determined by the Wilkinson Microwave Anisotropy Probe (WMAP). In combination with other large-scale structure work, the WMAP results imply $\Omega_\mathrm{tot}=1.02\pm0.02$, $\Omega_\Lambda\simeq0.73$ 
\cite{Bennett:2003,Spergel:2003}.

\subsection{The nature of ``dark energy''}

It is beyond the scope of the present article to discuss dark energy in detail and the reader is referred to the excellent review by 
\textcite{Peebles:2002}. However, since this is arguably the most dramatic discovery involving HST I would like to make a few points.

In general relativity, the stress-energy tensor of the vacuum, $T^{\Lambda}_{\mu\nu}$, can be written as 
\begin{equation}
T^{\Lambda}_{\mu\nu} =\rho_{\Lambda} g_{\mu\nu}~~,
\end{equation}
where $\rho_{\Lambda}$ is constant (proportional to Einstein's cosmological constant) and $g_{\mu\nu}$ is the metric. In a Minkowski flat spacetime this can be written simply as the ``equation of state''
\begin{equation}
P_{\Lambda} = -\rho_{\Lambda}~~.
\end{equation}

More generally, if the equation of state is written as 
\cite[for example,][]{Canuto:1977,Ratra:1988,Sahni:2000,Turner:1997},
\begin{equation}
P_{\Lambda} = w\rho_{\Lambda}~~,
\end{equation}
then the dark energy density behaves (for a constant $w$) like $\rho_{\Lambda}\sim a^{-3(1+w)}$ (where $a$ is the scale factor). Since the scale factor satisfies the equation (in units in which $c=1$)
\begin{equation}
\frac{\ddot{a}}{a} = \frac{4\pi G}{3} (\rho+ 3P)~~,
\end{equation}
the dark energy contribution will result in an \textit{accelerating} universe if $w<-\frac{1}{3}$.

It is interesting to note that from a purely fluid dynamical point of view, any equation of state of the type $P=w\rho$ is in fact \emph{dynamically unstable} for any negative value of $w$ except for $w=-1$ (corresponding to a cosmological constant). This can be easily seen from perturbations on the equations for momentum and energy conservation ($<\rho>$, $<P>$ denote mean values)
\begin{equation}
\delta\dot{\rho}+(<\rho>+<P>)\nabla\cdot\vec{v} = 0
\end{equation}
\begin{equation}
(<\rho>+<P>)\dot{\vec{v}}+\left(\frac{dP}{d\rho}\right)\nabla\delta\rho=0~~.
\end{equation}
Defining the speed of sound, $\displaystyle{c^2_s=\frac{dP}{d\rho}}$, these two equations can be combined to form
\begin{equation}
\delta\ddot{\rho}=c^2_s\nabla^2\delta\rho~~,
\end{equation}
which is clearly unstable for negative $c^2_s$ (except for $w=-1$, when the combination $<\rho>+<P>$ vanishes, as expected from the fact that the vacuum is the same for all inertial observers). 

Thus, formally speaking, from a fluid-dynamical point of view only a cosmological constant yields a stable solution.

One can certainly take the dark energy to be associated with a uniform scalar field $\phi$ (a ``quintessence'' field; see below), that rolls down a potential $V(\phi)$ at a rate determined by
\begin{equation}
\ddot{\phi} + 3H\dot{\phi} + V'(\phi) = 0~~,
\end{equation}
where $\displaystyle V'(\phi) = \frac{dV}{d\phi}$.  In this case,
\begin{eqnarray}
\rho_{\phi}&= &\frac{1}{2} \dot{\phi}^2 + V(\phi)\\ \nonumber
\smallskip
P_{\phi}&= &\frac{1}{2} \dot{\phi}^2 - V(\phi)~~.
\end{eqnarray}
For a sufficiently slowly varying dynamical component, such that the kinetic energy is much smaller than the potential, $\dot{\phi}^2\ll V(\phi)$, one even obtains a field energy that mimics the cosmological constant ($P_{\phi}\simeq -\rho_{\phi}$). Generally, quintessence solutions allow for different equations of state, and even values of $w$ that are time variable. Nevertheless, I find the stability property expressed in eq.~(44) sufficiently intriguing to warrant a deeper examination of the \textit{cosmological constant} possibility, before that is abandoned in favor of other dynamical solutions. This point of view received some further support from the recent determination by WMAP that $w<-0.78$ 
\cite{Bennett:2003}.

There are two main problems with the implied energy density $\rho_V$ of the dark energy 
\cite[e.g.,][]{Weinberg:2001}. (i)~Why is its value not $\sim120$ orders of magnitude larger (as expected from fluctuations in the gravitational field up to the Planck scale)? (ii)~Why now? (Namely, why $\Omega_\Lambda\sim\Omega_M$ now, even though $\Omega_\Lambda$ may be associated with a cosmological constant, while $\Omega_M$ declined continuously from the initial singularity to its present value).

An interesting curiosity to note is that even though taking graviton energies up to Planck scale, $M_P$, misses the value of the dark energy density by $\sim10^{120}$, and taking them up to the supersymmetry scale, $M_\mathrm{SUSY}$, misses by $\sim10^{55}$, a scale of $M_V\sim(M_\mathrm{SUSY}/M_P)M_\mathrm{SUSY}$ actually does give the right order of magnitude! While I am not aware at present of any theory that produces this scale in a natural way \cite[although see, e.g.,][]{Arkani:2000}, I find this coincidence worth following up. 

Much of the efforts to resolve the above two problems revolved not around a cosmological constant, but rather around the behavior of quintessence fields. In particular, attention has concentrated on ``tracker'' solutions, in which the final value of the quintessence energy density is independent of fine-tuning of the initial conditions \cite[e.g.,][]{Zlatev:1998,Albrecht:2000}. For example if one takes a potential of the form
\begin{equation}
V(\phi)=\phi^{-\alpha}M^{4+\alpha}~~,
\end{equation}
where $\alpha>0$ and $M$ is an adjustable constant, and the field is initially much smaller than the Planck mass, then for $\rho_M\gg V(\phi)$, $\dot{\phi}^2$, we find that $\rho_M$ decreases initially faster 
($\sim t^{-2}$) than $\phi(t)(\sim t^{-2\alpha/(2+\alpha)})$.

Eventually, however, a transition to a $\rho_\phi$-dominated universe occurs (and $\rho_\phi$ decreases as $t^{-2/(4+\alpha)}$). In other words, the quintessence answer to the question: why is the dark energy density so small? is simply:  because the universe is very old. Nevertheless, simple potentials of the form (47) do not offer a clear solution to the ``why now?'' problem. In fact, to have $\rho_\phi\sim\rho_M$ (and of order of the critical density) at the present time requires fine-tuning the parameters so that 
\cite[for example,][]{Weinberg:2001}
\begin{equation}
M^{4+\alpha}\simeq(8\pi G)^{-1-\alpha/2} H_0^2~~,
\end{equation}
with no simple explanation as to why this equality should hold.

In order to overcome this requirement for fine-tuning, some versions of the quintessence models choose potentials in which the universe has periodically been accelerating in the past (the dark energy has periodically dominated the energy density; e.g., Dodelson, Koplinghat \& Stewart 2000).
\textcite{Dodelson:2000} have shown, for example, that with a potential of the form
\begin{equation}
V(\phi)=V_o \exp \left(-\lambda\phi\sqrt{8\pi G}\right)
\left[1+A\sin\left(\nu\phi\sqrt{8\pi G}\right)\right]~~,
\end{equation}
one can obtain solutions in which the dark energy density tracks the ambient energy density of the universe and satisfies observational constraints.

A different approach to the problems associated with the dark energy density has been through anthropic considerations 
\cite[e.g.,][]{Weinberg:2001,Vilenkin:1995,Kallosh:2002}. The key assumption in this class of models is that \textit{some} constants of nature, in particular the cosmological constant, and possibly the density contrast at the time of recombination, $\sigma_\mathrm{rec}$, are in fact random variables, whose range of values and a~priori probabilities are nevertheless determined by the laws of physics. In this picture, some values of the constants which are allowed in principle, may be incompatible with the very existence of observers. Assuming a \textit{principle of mediocrity}---that we should expect to find ourselves in a universe \textit{typical} of those that allow the emergence of intelligent life---\textcite{Garriga:2000} were able to show that the ``why so small?'' and ``why now?'' questions find a natural explanation. While I personally believe that anthropic considerations should only be used as a last resort, there is no denial of the fact that most versions of ``eternal inflation,'' the notion that once inflation starts it never stops, unavoidably produce an ensemble of ``pocket'' universes 
\cite{Guth:1981,Vilenkin:1983,Linde:1986,Steinhardt:1983}. Once such an infinite ensemble is believed to exist, the problem of defining probabilities on it, and the concept of ``mediocrity'' become real 
\cite[e.g.,][]{Linde:1995,Vilenkin:1998}.

In spite of the apparent ``success'' of the anthropic argumentation in solving the two main problems associated with dark energy, the search for a fundamental explanation is not, and should not, be abandoned. An interesting, if speculative, new direction is provided by alternative theories of gravity.  For example, models in which ordinary particles are localized on a three-dimensional surface (3-brane) embedded in infinite volume extradimensions to which gravity can spread, have been developed \cite{Deffayet:2002}. In a particular version of these models, the Friedmann equation (eq.~24) is replaced by
\begin{equation}
H^2 + \frac{k}{a^2} =
\left( 
\sqrt{\frac{\rho}{3M_p^2} + \frac{1}{4r_c^2}}
+\epsilon \frac{1}{2r_c^2}
\right)^2~~,
\end{equation}
where $\epsilon=\pm1$, and $r_c$ represents a crossover scale. At distances shorter than $r_c$ (which can be of astronomical size, e.g., $r_c\sim c H_0^{-1}$), observers on the brane discover the familiar Newtonian gravity. At large cosmological distances, however, the force-law of gravity becomes 
five-dimensional (as gravity spreads into the extra dimensions) and weaker. The dynamics are governed by whether $\rho/M_p^2$ is larger or smaller than $1/r_c^2$. While highly speculative at this stage, some aspects of these alternative theories of gravity can be experimentally tested \cite[e.g., by lunar ranging experiments;][]{Dvali:2002}. Also, the relative lack of power on large scales (in particular, absence of any correlated signal on scales larger than 60~degrees) found by the Wilkinson Microwave Anisotropy Probe \cite{Spergel:2003}, may indicate a breakdown of canonical gravity on large scales (or, more speculatively, a finite universe).

Luckily, observations that are already being performed with HST, and observations with the proposed Super Nova/Acceleration Probe (SNAP) will provide some crucial information on the nature of dark energy. In particular, if the equation of state parameter $w$ is constant, then a transition from a decelerating phase to an accelerating phase occurs at
\begin{equation}
z_\mathrm{tr} = [(1+3w)(\Omega_M-1)/\Omega_M]^{-1/3w} -1~~.
\end{equation}
For example, for $\Omega_M=0.27$ \cite{Bennett:2003} and a cosmological constant ($w= -1$), this gives $z_\mathrm{tr} \simeq0.76$. Equation~(51) shows that if $\Omega_M$ can be determined independently to within a few percent (a goal that can perhaps be achieved by a conbination of cosmic microwave background and large scale structure measurements), then observations of a large sample of Type~Ia supernovae in the redshift range 0.8--2 can place very meaningful constraints on the dark energy equation of state. The situation is more ambiguous if $w$ is time-dependent, but even then, a sufficiently large sample of SNe~Ia will give useful information. At the time of this writing, a search for SNe~Ia conducted in the GOODS fields has already yielded 10~SNe~Ia, with six of them in the redshift range $z\sim1$--2 (Principal Investigator: Adam Riess). This, together with a few current ground-based searches at lower redshifts, is an excellent first step. The survey proposed by SNAP, of 15~square degrees, could yield light curves and spectra for over 2000 SNe~Ia at redshifts up to $z\sim1.7$. With this sample in hand, one could reach the elimination or at least control of the sources of systematic uncertainties needed to discriminate between different dark-energy models.

\section{EPILOGUE}

The Hubble Space Telescope, together with many other space-based and ground-based observatories, have given us a much clearer picture than we ever had before of the universe we live in. We now know not only that other planetary systems are quite common, but that the cosmic expansion is accelerating. We know the age of the universe, its geometry, and its composition. Observations of stars start to reach the level of detail previously characterizing only solar physics. A general picture of the formation of structure in the universe, and many clues for how galaxies form and evolve, start to emerge. We know that supermassive black holes reside at the centers of most galaxies. This is, however, far from marking the end of HST's contributions. The recent installation of the Advanced Camera for Surveys, and the two future (currently planned for 2004) instruments, the Cosmic Origins Spectrograph (COS) and the Wide Field Camera~3 (equipped with an infrared channel), promise that the coming seven years will be at least as exciting as the first thirteen were.

\begin{acknowledgments}
This review has benefited from discussions with more colleagues than I can practically thank here. I would like, however, to extend special thanks to John Bally, Stefano Casertano, Mark Dickinson, Gia Dvali, Harry Ferguson, Andy Fruchter, Mauro Giavalisco, Ron Gilliland, Andy Ingersoll, Ken Lanzetta, Nino Panagia, Saul Perelmutter, Adam Riess, Kailash Sahu, Rachel Somerville, Massimo Stiavelli, Roeland van der Marel, and Alex Vilinkin for very helpful discussions.
\end{acknowledgments}

\bibliographystyle{apsrmp}

\begin{thebibliography}{42}
\expandafter\ifx\csname natexlab\endcsname\relax\def\natexlab#1{#1}\fi
\expandafter\ifx\csname bibnamefont\endcsname\relax
  \def\bibnamefont#1{#1}\fi
\expandafter\ifx\csname bibfnamefont\endcsname\relax
  \def\bibfnamefont#1{#1}\fi
\expandafter\ifx\csname citenamefont\endcsname\relax
  \def\citenamefont#1{#1}\fi
\expandafter\ifx\csname url\endcsname\relax
  \def\url#1{\texttt{#1}}\fi
\expandafter\ifx\csname urlprefix\endcsname\relax\def\urlprefix{URL }\fi
\providecommand{\bibinfo}[2]{#2}
\providecommand{\eprint}[2][]{\url{#2}}

\bibitem[{\citenamefont{Abraham} \emph{et~al.}(1999a)\citenamefont{Abraham \emph{et~al.}}}] {Abraham:1999a}
\bibinfo{author}{\bibnamefont{Abraham}, \bibfnamefont{R.~G.}},
\bibinfo{author}{\bibfnamefont{R.~S.} \bibnamefont{Ellis}}, 
\bibinfo{author}{\bibfnamefont{A.~C.} \bibnamefont{Fabian}}, 
\bibinfo{author}{\bibfnamefont{N.~R.} \bibnamefont{Tanvir}}, and
\bibinfo{author}{\bibfnamefont{K.}~\bibnamefont{Glazebrook}}, \bibinfo{year}{1999a}, \bibinfo{journal}{Mon.\ Not.~R.\ Astron.\ Soc.} \textbf{\bibinfo{volume}{303}}, \bibinfo{pages}{641}.

\bibitem[{\citenamefont{Abraham} \emph{et~al.}(1999b)\citenamefont{Abraham \emph{et~al.}}}] {Abraham:1999b}
\bibinfo{author}{\bibnamefont{Abraham}, \bibfnamefont{R.~G.}},
\bibinfo{author}{\bibfnamefont{M.~R.} \bibnamefont{Merrifield}},
\bibinfo{author}{\bibfnamefont{R.~S.} \bibnamefont{Ellis}}, 
\bibinfo{author}{\bibfnamefont{N.~R.} \bibnamefont{Tanvir}}, and
\bibinfo{author}{\bibfnamefont{J.}~\bibnamefont{Brinchman}}, \bibinfo{year}{1999b}, \bibinfo{journal}{Mon.\ Not.~R.\ Astron.\ Soc.} \textbf{\bibinfo{volume}{308}}, \bibinfo{pages}{569}.

\bibitem[{\citenamefont{Abraham} \emph{et~al.}(1996)\citenamefont{Abraham \emph{et~al.}}}] {Abraham:1996}
\bibinfo{author}{\bibnamefont{Abraham}, \bibfnamefont{R.~G.}},
\bibinfo{author}{\bibfnamefont{N.~R.} \bibnamefont{Tanvir}}, 
\bibinfo{author}{\bibfnamefont{B.~X.} \bibnamefont{Santiago}}, 
\bibinfo{author}{\bibfnamefont{R.~S.} \bibnamefont{Ellis}}, 
\bibinfo{author}{\bibfnamefont{K.}~\bibnamefont{Glazebrook}}, and
\bibinfo{author}{\bibfnamefont{S.}~\bibnamefont{van den Bergh}}, \bibinfo{year}{1996}, \bibinfo{journal}{Mon.\ Not.~R.\ Astron.\ Soc.} \textbf{\bibinfo{volume}{279}}, \bibinfo{pages}{L47}.

\bibitem[{\citenamefont{Abroe} \emph{et~al.}(2002)\citenamefont{Abroe \emph{et~al.}}}] {Abroe:2002}
\bibinfo{author}{\bibnamefont{Abroe}, \bibfnamefont{M.~E.}},
\bibinfo{author}{\bibfnamefont{A.}~\bibnamefont{Balbi}}, 
\bibinfo{author}{\bibfnamefont{J.}~\bibnamefont{Borrill, \emph{et~al.}}}, 
\bibinfo{year}{2002}, \bibinfo{journal}{Mon.\ Not.~R.\ Astron.\ Soc.} \textbf{\bibinfo{volume}{334}}, \bibinfo{pages}{11}.

\bibitem[\citenamefont{Adams, Graff and Richstone}(2001)]{Adams:2001}
\bibinfo{author}{\bibnamefont{Adams}, \bibfnamefont{F.~C.}},
\bibinfo{author}{\bibfnamefont{D.~S.} \bibnamefont{Graff}}, and
\bibinfo{author}{\bibfnamefont{D.~D.} \bibnamefont{Richstone}}, \bibinfo{year}{2001}, \bibinfo{journal}{Astrophys.~J.\ Lett.} \textbf{\bibinfo{volume}{551}}, \bibinfo{pages}{L31}.

\bibitem[{\citenamefont{Aguirre}(1999)\citenamefont{Aguirre}}]{Aguirre:1999}
\bibinfo{author}{\bibnamefont{Aguirre}, \bibfnamefont{A.}}, \bibinfo{year}{1999}, \bibinfo{journal}{Astrophys.~J.} \textbf{\bibinfo{volume}{525}}, \bibinfo{pages}{583}.

\bibitem[{\citenamefont{Albrecht and Skordis}(2000)\citenamefont{Albrecht and Skordis}}]{Albrecht:2000}
\bibinfo{author}{\bibnamefont{Albrecht},~\bibfnamefont{A.}} and 
\bibinfo{author}{\bibfnamefont{C.}~\bibnamefont{Skordis}}, \bibinfo{year}{2002}, \bibinfo{journal}{Phys.\ Rev.\ Lett.} \textbf{\bibinfo{volume}{84}}, \bibinfo{pages}{2076}.

\bibitem[{\citenamefont{Andersen} \emph{et~al.}(2002)\citenamefont{Andersen, Hjorth, and Gorosabel}}] {Andersen:2002}
\bibinfo{author}{\bibnamefont{Andersen}, \bibfnamefont{M.~I.}},
\bibinfo{author}{\bibfnamefont{J.}~\bibnamefont{Hjorth}}, and
\bibinfo{author}{\bibfnamefont{J.}~\bibnamefont{Gorosabel}},
\bibinfo{year}{2002}, \bibinfo{journal}{Astron.\ Astrophys.} \bibinfo{pages}{(in press)}.

\bibitem[{\citenamefont{Arkani-Hamed} \emph{et~al.}(2000)\citenamefont{Arkani-Hamed \emph{et~al.}}}] {Arkani:2000}
\bibinfo{author}{\bibnamefont{Arkani-Hamed}, \bibfnamefont{N.}},
\bibinfo{author}{\bibfnamefont{L.~J.} \bibnamefont{Hall}}, 
\bibinfo{author}{\bibfnamefont{C.}~\bibnamefont{Colda}}, and
\bibinfo{author}{\bibfnamefont{H.}~\bibnamefont{Murayama}},
\bibinfo{year}{2000}, \bibinfo{journal}{Phys.\ Rev.\ Lett.} \textbf{\bibinfo{volume}{85}}, 
\bibinfo{pages}{4434}.

\bibitem[{\citenamefont{Armitage}(2000)\citenamefont{Armitage}}]{Armitage:2000}
\bibinfo{author}{\bibnamefont{Armitage}, \bibfnamefont{P.~J.}}, \bibinfo{year}{2000}, \bibinfo{journal}{Astron.\ Astrophys.} \textbf{\bibinfo{volume}{362}}, \bibinfo{pages}{968}.

\bibitem[{\citenamefont{Armitage} \emph{et~al.}(2002)\citenamefont{Armitage \emph{et~al.}}}] {Armitage:2002}
\bibinfo{author}{\bibnamefont{Armitage}, \bibfnamefont{P.~J.}},
\bibinfo{author}{\bibfnamefont{M.}~\bibnamefont{Livio}}, 
\bibinfo{author}{\bibfnamefont{S.~H.} \bibnamefont{Lubow}}, and
\bibinfo{author}{\bibfnamefont{J.~E.} \bibnamefont{Pringle}}, \bibinfo{year}{2002}, \bibinfo{journal}{Mon.\ Not.~R.\ Astron.\ Soc.} \textbf{\bibinfo{volume}{334}}, \bibinfo{pages}{248}.

\bibitem[{\citenamefont{Arnett}(2001)\citenamefont{Arnett}}]{Arnett:2001}
\bibinfo{author}{\bibnamefont{Arnett}, \bibfnamefont{D.}}, \bibinfo{year}{2001}, 
in \emph{\bibinfo{booktitle}{Supernovae and Gamma-Ray Bursts}}, edited by
  \bibinfo{editor}{\bibfnamefont{M.}~\bibnamefont{Livio}},
\bibinfo{editor}{\bibfnamefont{N.}~\bibnamefont{Panagia}}, and
\bibinfo{editor}{\bibfnamefont{K.}~\bibnamefont{Sahu}}
  (\bibinfo{publisher}{Cambridge University, Cambridge, England}), p.~\bibinfo{pages}{250}.

\bibitem[{\citenamefont{Baade}(1954)\citenamefont{Baade}}]{Baade:1954}
\bibinfo{author}{\bibnamefont{Baade}, \bibfnamefont{W.}}, \bibinfo{year}{1954}, \bibinfo{journal}{IAU Trans.} \textbf{\bibinfo{volume}{8}}, \bibinfo{pages}{397}.

\bibitem[{\citenamefont{Bahcall} \emph{et~al.}(1997)\citenamefont{Bahcall \emph{et~al.}}}] {Bahcalletal:1997}
\bibinfo{author}{\bibnamefont{Bahcall}, \bibfnamefont{J.~N.}},
\bibinfo{author}{\bibfnamefont{S.}~\bibnamefont{Kirhakos}}, 
\bibinfo{author}{\bibfnamefont{D.~H.} \bibnamefont{Saxe}}, and
\bibinfo{author}{\bibfnamefont{D.~P.} \bibnamefont{Schneider}}, \bibinfo{year}{1997}, \bibinfo{journal}{Astrophys.~J.} \textbf{\bibinfo{volume}{479}}, \bibinfo{pages}{642}.

\bibitem[{\citenamefont{Bahcall, Kirhakos, and Schneider}(1994)\citenamefont{Bahcall \emph{et~al.}}}] {Bahcall:1994}
\bibinfo{author}{\bibnamefont{Bahcall}, \bibfnamefont{J.~N.}},
\bibinfo{author}{\bibfnamefont{S.}~\bibnamefont{Kirhakos}}, and
\bibinfo{author}{\bibfnamefont{D.~P.} \bibnamefont{Schneider}}, \bibinfo{year}{1994}, \bibinfo{journal}{Astrophys.~J.\ Lett.} \textbf{\bibinfo{volume}{435}}, \bibinfo{pages}{L11}.

\bibitem[{\citenamefont{Bahcall} \emph{et~al.}(2000)\citenamefont{Bahcall \emph{et~al.}}}] {Bahcall:2000}
\bibinfo{author}{\bibnamefont{Bahcall}, \bibfnamefont{N.~A.}},
\bibinfo{author}{\bibfnamefont{R.}~\bibnamefont{Cen}}, 
\bibinfo{author}{\bibfnamefont{R.}~\bibnamefont{Dav\'e}}, 
\bibinfo{author}{\bibfnamefont{J.~P.} \bibnamefont{Ostriker}}, and
\bibinfo{author}{\bibfnamefont{Q.}~\bibnamefont{Yu}}, \bibinfo{year}{2000}, \bibinfo{journal}{Astrophys.~J.} \textbf{\bibinfo{volume}{541}}, \bibinfo{pages}{1}.

\bibitem[{\citenamefont{Bahcall, Fan, and Cen}(1997)\citenamefont{Bahcall Fan, and Cen}}] {Bahcall:1997}
\bibinfo{author}{\bibnamefont{Bahcall}, \bibfnamefont{N.~A.}},
\bibinfo{author}{\bibfnamefont{X.}~\bibnamefont{Fan}}, and
\bibinfo{author}{\bibfnamefont{R.}~\bibnamefont{Cen}}, \bibinfo{year}{1997}, \bibinfo{journal}{Astrophys.~J.\ Lett.} \textbf{\bibinfo{volume}{485}}, \bibinfo{pages}{L53}.

\bibitem[{\citenamefont{Balick}(1987)\citenamefont{Balick}}]{Balick:1987}
\bibinfo{author}{\bibnamefont{Balick}, \bibfnamefont{B.}}, \bibinfo{year}{1987}, \bibinfo{journal}{Astron.\ Astrophys.} \textbf{\bibinfo{volume}{94}}, \bibinfo{pages}{671}.

\bibitem[{\citenamefont{Balick and Frank}(2002)\citenamefont{Balick and Frank}}]{Balick:2002}
\bibinfo{author}{\bibnamefont{Balick},~\bibfnamefont{B.}} and 
\bibinfo{author}{\bibfnamefont{A.}~\bibnamefont{Frank}}, \bibinfo{year}{2002}, \bibinfo{journal}{Annu.\ Rev.\ Astron.\ Astrophys.} \textbf{\bibinfo{volume}{40}}, \bibinfo{pages}{439}.

\bibitem[{\citenamefont{Bally}(2003)}]{Bally:2003}
\bibinfo{author}{\bibnamefont{Bally}, \bibfnamefont{J.}},
\bibinfo{year}{2003}, in \emph{\bibinfo{booktitle}{A Decade of HST Science}}, edited by 
\bibinfo{editor}{\bibfnamefont{Mario} \bibnamefont{Livio}},
\bibinfo{editor}{\bibfnamefont{Keith} \bibnamefont{Noll}}, and
\bibinfo{editor}{\bibfnamefont{Massimo} \bibnamefont{Stiavelli}},
\bibinfo{publisher}{(Cambridge University, Cambridge, England)}, p.~\bibinfo{pages}{44}.

\bibitem[{\citenamefont{Bally, O'Dell and McCaughrean}(2000)\citenamefont{Bally \emph{et~al.}}}] {Bally:2000}
\bibinfo{author}{\bibnamefont{Bally}, \bibfnamefont{J.}},
\bibinfo{author}{\bibfnamefont{C.~R.} \bibnamefont{O'Dell}}, and
\bibinfo{author}{\bibfnamefont{M.~J.} \bibnamefont{McCaughrean}}, \bibinfo{year}{2000}, \bibinfo{journal}{Astron.~J.} \textbf{\bibinfo{volume}{119}}, \bibinfo{pages}{2919}.

\bibitem[{\citenamefont{Baumgardt} \emph{et~al.}(2002)\citenamefont{Baumgart \emph{et~al.}}}] {Baumgardt:2002}
\bibinfo{author}{\bibnamefont{Baumgardt}, \bibfnamefont{H.}},
\bibinfo{author}{\bibfnamefont{P.}~\bibnamefont{Hut}}, 
\bibinfo{author}{\bibfnamefont{J.}~\bibnamefont{Makino}}, 
\bibinfo{author}{\bibfnamefont{S.}~\bibnamefont{McMillan}}, and 
\bibinfo{author}{\bibfnamefont{S.}~\bibnamefont{Portegies Zwart}}, \bibinfo{year}{2002}, 
\eprint{astro-ph/0210133}.

\bibitem[{\citenamefont{Becker} \emph{et~al.}(2001)\citenamefont{Becker \emph{et~al.}}}] {Becker:2001}
\bibinfo{author}{\bibnamefont{Becker}, \bibfnamefont{R.~H.}}, 
\bibinfo{author}{\bibfnamefont{X.}~\bibnamefont{Fan}}, 
\bibinfo{author}{\bibfnamefont{R.~L.} \bibnamefont{White, \emph{et~al.}}}, \bibinfo{year}{2001}, \bibinfo{journal}{Astron.~J.}, \textbf{\bibinfo{volume}{122}}, \bibinfo{pages}{2850}.

\bibitem[{\citenamefont{Begelman and Li}(1992)\citenamefont{Begelman and Li}}]{Begelman:1992}
\bibinfo{author}{\bibnamefont{Begelman}, \bibfnamefont{M.}} and 
\bibinfo{author}{\bibfnamefont{Z.~Y.}~\bibnamefont{Li}}, \bibinfo{year}{1992}, \bibinfo{journal}{Astrophys.~J.} \textbf{\bibinfo{volume}{397}}, \bibinfo{pages}{187}.

\bibitem[{\citenamefont{Behr}(1951)\citenamefont{Behr}}]{Behr:1951}
\bibinfo{author}{\bibnamefont{Behr}, \bibfnamefont{A.}}, \bibinfo{year}{1951}, \bibinfo{journal}{Astr.\ Nachr.} \textbf{\bibinfo{volume}{279}}, \bibinfo{pages}{97}.

\bibitem[{\citenamefont{Benitez} \emph{et~al.}(1999)\citenamefont{Benitez \emph{et~al.}}}] {Benitez:1999}
\bibinfo{author}{\bibnamefont{Benitez}, \bibfnamefont{N.}},
\bibinfo{author}{\bibfnamefont{T.}~\bibnamefont{Broadhurst}}, 
\bibinfo{author}{\bibfnamefont{R.}~\bibnamefont{Bouwens}}, 
\bibinfo{author}{\bibfnamefont{J.}~\bibnamefont{Silk}}, and
\bibinfo{author}{\bibfnamefont{P.}~\bibnamefont{Rosati}}, \bibinfo{year}{1999}, \bibinfo{journal}{Astrophys.~J.} \textbf{\bibinfo{volume}{515}}, \bibinfo{pages}{65}.

\bibitem[{\citenamefont{Bennett} \emph{et~al.}(2003)\citenamefont{Bennett \emph{et~al.}}}] {Bennett:2003}
\bibinfo{author}{\bibnamefont{Bennett}, \bibfnamefont{C.~L.}},
\bibinfo{author}{\bibfnamefont{M.}~\bibnamefont{Halpern}}, 
\bibinfo{author}{\bibfnamefont{G.}~\bibnamefont{Hinshaw, \emph{et~al.}}}, 
\bibinfo{year}{2003}, \bibinfo{journal}{Astrophys.~J.} (submitted). 

\bibitem[\citenamefont{Bershady, Jangren, and Conselice}(2000)]{Bershady:2000}
\bibinfo{author}{\bibnamefont{Bershady}, \bibfnamefont{M.~A.}}, 
\bibinfo{author}{\bibfnamefont{A.}~\bibnamefont{Jangren}}, and
\bibinfo{author}{\bibfnamefont{C.~J.} \bibnamefont{Conselice}}, \bibinfo{year}{2000}, \bibinfo{journal}{Astron.~J.}, \textbf{\bibinfo{volume}{119}}, \bibinfo{pages}{2645}.

\bibitem[{\citenamefont{Bertoldi and McKee}(1990)\citenamefont{Bertoldi and McKee}}]{Bertoldi:1990}
\bibinfo{author}{\bibnamefont{Bertoldi}, \bibfnamefont{F.}} and 
\bibinfo{author}{\bibfnamefont{C.~F.}~\bibnamefont{McKee}}, \bibinfo{year}{1990}, \bibinfo{journal}{Astrophys.~J.} \textbf{\bibinfo{volume}{354}}, \bibinfo{pages}{529}.

\bibitem[{\citenamefont{Binney and Mamon}(1982)\citenamefont{Binney and Mamon}}]{Binney:1982}
\bibinfo{author}{\bibnamefont{Binney}, \bibfnamefont{J.}} and 
\bibinfo{author}{\bibfnamefont{G.~A.} \bibnamefont{Mamon}}, \bibinfo{year}{1982}, \bibinfo{journal}{Mon.\ Not.~R.\ Astron.\ Soc.} \textbf{\bibinfo{volume}{200}}, \bibinfo{pages}{361}.

\bibitem[{\citenamefont{Binney and Tremaine}(1987)\citenamefont{Binney and Tremaine}}]{Binney:1987}
\bibinfo{author}{\bibnamefont{Binney}, \bibfnamefont{J.}} and 
\bibinfo{author}{\bibfnamefont{S.}~\bibnamefont{Tremaine}}, \bibinfo{year}{1987}, 
\emph{\bibinfo{booktitle}{Galactic Dynamics}}, 
\bibinfo{publisher}{(Princeton University, Princeton)}, p.~\bibinfo{pages}{428}.

\bibitem[\citenamefont{Biretta, Sparks, and Macchetto}(1999)]{Biretta:1999}
\bibinfo{author}{\bibnamefont{Biretta}, \bibfnamefont{J.~A.}},
\bibinfo{author}{\bibfnamefont{W.~B.}~\bibnamefont{Sparks}}, and
\bibinfo{author}{\bibfnamefont{F.}~\bibnamefont{Macchetto}}, \bibinfo{year}{1999}, \bibinfo{journal}{Astrophys.~J.} \textbf{\bibinfo{volume}{520}}, \bibinfo{pages}{621}.

\bibitem[{\citenamefont{Bjorkman and Cassinelli}(1993)\citenamefont{Bjorkman and Cassinelli}}]{Bjorkman:1993}
\bibinfo{author}{\bibnamefont{Bjorkman}, \bibfnamefont{J.~E.}} and 
\bibinfo{author}{\bibfnamefont{J.~P.}~\bibnamefont{Cassinelli}}, \bibinfo{year}{1993}, \bibinfo{journal}{Astrophys.~J.} \textbf{\bibinfo{volume}{409}}, \bibinfo{pages}{429}.

\bibitem[{\citenamefont{Blandford and Znajek}(1977)\citenamefont{Blandford and Znajek}}]{Blandford:1977}
\bibinfo{author}{\bibnamefont{Blandford}, \bibfnamefont{R.}} and 
\bibinfo{author}{\bibfnamefont{R.}~\bibnamefont{Znajek}}, \bibinfo{year}{1977}, \bibinfo{journal}{Mon. Not.~R.\ Astron.\ Soc.} \textbf{\bibinfo{volume}{179}}, \bibinfo{pages}{433}.

\bibitem[{\citenamefont{Blandford and Payne}(1982)\citenamefont{Blandford and Payne}}]{Blandford:1982}
\bibinfo{author}{\bibnamefont{Blandford}, \bibfnamefont{R.~D.}} and 
\bibinfo{author}{\bibfnamefont{D.~G.}~\bibnamefont{Payne}}, \bibinfo{year}{1982}, \bibinfo{journal}{Mon.\ Not.~R.\ Astron.\ Soc.} \textbf{\bibinfo{volume}{199}}, \bibinfo{pages}{883}.

\bibitem[{\citenamefont{Bloom} \emph{et~al.}(2002)\citenamefont{Bloom \emph{et~al.}}}] {Bloom:2002}
\bibinfo{author}{\bibnamefont{Bloom}, \bibfnamefont{J.~S.}},
\bibinfo{author}{\bibfnamefont{S.~R.} \bibnamefont{Kulkarni}}, 
\bibinfo{author}{\bibfnamefont{P.~A.} \bibnamefont{Price, \emph{et~al.}}}, \bibinfo{year}{2002}, \bibinfo{journal}{Astrophys.~J.\ Lett.} \textbf{\bibinfo{volume}{572}}, \bibinfo{pages}{L45}.

\bibitem[{\citenamefont{Bond} \emph{et~al.}(1996)\citenamefont{Bond \emph{et~al.}}}] {Bond:1996}
\bibinfo{author}{\bibnamefont{Bond}, \bibfnamefont{H.~E.}},
\bibinfo{author}{\bibfnamefont{R.}~\bibnamefont{Ciardullo}}, 
\bibinfo{author}{\bibfnamefont{L.~K.} \bibnamefont{Fullton}}, and
\bibinfo{author}{\bibfnamefont{K.~G.} \bibnamefont{Schaefer}}, \bibinfo{year}{1996} (unpublished).

\bibitem[{\citenamefont{Boroson, Oke and Green}(1982)\citenamefont{Boroson \emph{et~al.}}}]{Boroson:1982}
\bibinfo{author}{\bibnamefont{Boroson}, \bibfnamefont{T.~A.}} 
\bibinfo{author}{\bibfnamefont{J.~B.} \bibnamefont{Oke}}, and 
\bibinfo{author}{\bibfnamefont{R.~F.} \bibnamefont{Green}}, \bibinfo{year}{1982}, \bibinfo{journal}{Astrophys.~J.} \textbf{\bibinfo{volume}{263}}, \bibinfo{pages}{32}.

\bibitem[{\citenamefont{Boss}(2000)\citenamefont{Boss}}]{Boss:2000}
\bibinfo{author}{\bibnamefont{Boss}, \bibfnamefont{A.~P.}}, \bibinfo{year}{2000}, \bibinfo{journal}{Astrophys.~J.\ Lett.} \textbf{\bibinfo{volume}{536}}, \bibinfo{pages}{L101}.

\bibitem[{\citenamefont{Bouwens, Cayon, and Silk}(1997)\citenamefont{Bouwens \emph{et~al.}}}] {Bouwens:1997}
\bibinfo{author}{\bibnamefont{Bouwens}, \bibfnamefont{R.~J.}},
\bibinfo{author}{\bibfnamefont{L.}~\bibnamefont{Cayon}}, and
\bibinfo{author}{\bibfnamefont{J.}~\bibnamefont{Silk}}, \bibinfo{year}{1997}, \bibinfo{journal}{Astrophys.~J.\ Lett.} \textbf{\bibinfo{volume}{489}}, \bibinfo{pages}{L21}.

\bibitem[{\citenamefont{Branch}(1998)\citenamefont{Branch}}]{Branch:1998}
\bibinfo{author}{\bibnamefont{Branch}, \bibfnamefont{D.}}, \bibinfo{year}{1998}, \bibinfo{journal}{Annu.\ Rev.\ Astron.\ Astrophys.} \textbf{\bibinfo{volume}{36}}, \bibinfo{pages}{17}.

\bibitem[{\citenamefont{Brinchmann} \emph{et~al.}(1998)\citenamefont{Brinchmann \emph{et~al.}}}] {Brinchmann:1998}
\bibinfo{author}{\bibnamefont{Brinchmann}, \bibfnamefont{J.}},
\bibinfo{author}{\bibfnamefont{R.}~\bibnamefont{Abraham}}, and
\bibinfo{author}{\bibfnamefont{D.}~\bibnamefont{Schade, \emph{et~al.}}}, \bibinfo{year}{1998}, \bibinfo{journal}{Astrophys.~J.} \textbf{\bibinfo{volume}{499}}, \bibinfo{pages}{112}.

\bibitem[{\citenamefont{Brown} \emph{et~al.}(2001)\citenamefont{Brown \emph{et~al.}}}]{Brown:2001}
\bibinfo{author}{\bibnamefont{Brown}, \bibfnamefont{T.~M.}}, 
\bibinfo{author}{\bibfnamefont{D.}~\bibnamefont{Charbonneau}}, 
\bibinfo{author}{\bibfnamefont{R.~L.} \bibnamefont{Gilliland}},
\bibinfo{author}{\bibfnamefont{R.~W.} \bibnamefont{Noyes}}, and
\bibinfo{author}{\bibfnamefont{A.}~\bibnamefont{Burrows}}, \bibinfo{year}{2001}, \bibinfo{journal}{Astrophys.~J.} \textbf{\bibinfo{volume}{552}}, \bibinfo{pages}{699}.

\bibitem[{\citenamefont{Budav\'ari} \emph{et~al.}(2000)\citenamefont{Budav\'ari 
\emph{et al.}}}]{Budavari:2000}
\bibinfo{author}{\bibnamefont{Budav\'ari}, \bibfnamefont{T.}}, 
\bibinfo{author}{\bibfnamefont{A.~S.} \bibnamefont{Szalay}}, 
\bibinfo{author}{\bibfnamefont{A.~J.} \bibnamefont{Connolly}}, 
\bibinfo{author}{\bibfnamefont{I.}~\bibnamefont{Casbai}}, and
\bibinfo{author}{\bibfnamefont{M.}~\bibnamefont{Dickinson}}, \bibinfo{year}{2000}, 
\bibinfo{journal}{Astron.~J.}, \textbf{\bibinfo{volume}{120}}, \bibinfo{pages}{1588}.

\bibitem[{\citenamefont{Burrows} \emph{et~al.}(2000)\citenamefont{Burrows, Marley, and Sharp}}] {Burrows:2000}
\bibinfo{author}{\bibnamefont{Burrows}, \bibfnamefont{A.}},
\bibinfo{author}{\bibfnamefont{M.~S.} \bibnamefont{Marley}}, and
\bibinfo{author}{\bibfnamefont{C.~M.} \bibnamefont{Sharp}}, \bibinfo{year}{2000}, \bibinfo{journal}{Astrophys.~J.} \textbf{\bibinfo{volume}{531}}, \bibinfo{pages}{438}.

\bibitem[{\citenamefont{Burrows} \emph{et~al.}(1995)\citenamefont{Burrows \emph{et~al.}}}] {Burrows:1995}
\bibinfo{author}{\bibnamefont{Burrows}, \bibfnamefont{C.~J.}},
\bibinfo{author}{\bibfnamefont{J.}~\bibnamefont{Krist}}, 
\bibinfo{author}{\bibfnamefont{J.~J.} \bibnamefont{Hester, \emph{et~al.}}}, \bibinfo{year}{1995}, \bibinfo{journal}{Astrophys.~J.} \textbf{\bibinfo{volume}{452}}, \bibinfo{pages}{680}.

\bibitem[{\citenamefont{Burrows} \emph{et~al.}(1996)\citenamefont{Burrows \emph{et~al.}}}] {Burrows:1996}
\bibinfo{author}{\bibnamefont{Burrows}, \bibfnamefont{C.~J.}},
\bibinfo{author}{\bibfnamefont{K.~R.} \bibnamefont{Stapelfeldt}}, 
\bibinfo{author}{\bibfnamefont{A.~M.} \bibnamefont{Watson, \emph{et~al.}}}, \bibinfo{year}{1996}, \bibinfo{journal}{Astrophys.~J.} \textbf{\bibinfo{volume}{473}}, \bibinfo{pages}{437}.

\bibitem[{\citenamefont{Butler} \emph{et~al.}(2000)\citenamefont{Butler \emph{et~al.}}}] {Butler:2000}
\bibinfo{author}{\bibnamefont{Butler}, \bibfnamefont{R.~P.}},
\bibinfo{author}{\bibfnamefont{S.~S.} \bibnamefont{Vogt}},
\bibinfo{author}{\bibfnamefont{G.~W.} \bibnamefont{Marcy}},
\bibinfo{author}{\bibfnamefont{D.~A.} \bibnamefont{Fischer}},
\bibinfo{author}{\bibfnamefont{G.~W.} \bibnamefont{Henry}}, and
\bibinfo{author}{\bibfnamefont{K.}~\bibnamefont{Apps}}, \bibinfo{year}{2000}, \bibinfo{journal}{Astrophys.~J.} \textbf{\bibinfo{volume}{545}}, \bibinfo{pages}{504}.

\bibitem[{\citenamefont{Calzetti and Heckman}(1999)\citenamefont{Calzetti and Heckman}}]{Calzetti:1999}
\bibinfo{author}{\bibnamefont{Calzetti}, \bibfnamefont{D.}} and 
\bibinfo{author}{\bibfnamefont{T.~M.}~\bibnamefont{Heckman}}, \bibinfo{year}{1999}, \bibinfo{journal}{Astrophys.~J.} \textbf{\bibinfo{volume}{519}}, \bibinfo{pages}{27}.

\bibitem[{\citenamefont{Calzetti, Kinney, and Storchi-Bergmann}(1994)\citenamefont{Calzetti, Kinney, and Storchi-Bergman}}] {Calzetti:1994}
\bibinfo{author}{\bibnamefont{Calzetti}, \bibfnamefont{D.}},
\bibinfo{author}{\bibfnamefont{A.~L.} \bibnamefont{Kinney}}, and
\bibinfo{author}{\bibfnamefont{T.}~\bibnamefont{Storchi-Bergmann}}, \bibinfo{year}{1994}, \bibinfo{journal}{Astrophys.~J.} \textbf{\bibinfo{volume}{429}}, \bibinfo{pages}{582}.

\bibitem[{\citenamefont{Camenzind}(1990)\citenamefont{Camenzind}}]{Camenzind:1990}
\bibinfo{author}{\bibnamefont{Camenzind}, \bibfnamefont{M.}}, \bibinfo{year}{1990}, 
in \emph{\bibinfo{booktitle}{Reviews of Modern Astronomy~3}}, edited by 
\bibinfo{editor}{\bibfnamefont{G.}~\bibnamefont{Klare}},
\bibinfo{publisher}{(Springer, Berlin)}, p.~\bibinfo{pages}{259}.

\bibitem[{\citenamefont{Canuto} \emph{et~al.}(1977)\citenamefont{Canuto \emph{et~al.}}}] {Canuto:1977}
\bibinfo{author}{\bibnamefont{Canuto}, \bibfnamefont{V.}},
\bibinfo{author}{\bibfnamefont{P.~J.} \bibnamefont{Adams}},
\bibinfo{author}{\bibfnamefont{S.-H.} \bibnamefont{Hsieh}}, and
\bibinfo{author}{\bibfnamefont{E.}~\bibnamefont{Tsiang}}, \bibinfo{year}{1977}, \bibinfo{journal}{Phys.\ Rev.~D} \textbf{\bibinfo{volume}{16}}, \bibinfo{pages}{1643}.

\bibitem[{\citenamefont{Cappellaro} \emph{et~al.}(1997)\citenamefont{Cappellaro, Turatto, Tsvelkov,  Bartunov, Pollas, Evans, and Hamuy}}] {Cappellaro:1997}
\bibinfo{author}{\bibnamefont{Cappellaro},~\bibfnamefont{E.}},
\bibinfo{author}{\bibfnamefont{M.}~\bibnamefont{Turatto}},
\bibinfo{author}{\bibfnamefont{D.~Yu.} \bibnamefont{Tsvelkov}}, 
\bibinfo{author}{\bibfnamefont{O.~S.} \bibnamefont{Bartunov}},
\bibinfo{author}{\bibfnamefont{C.}~\bibnamefont{Pollas}},
\bibinfo{author}{\bibfnamefont{R.}~\bibnamefont{Evans}}, and
\bibinfo{author}{\bibfnamefont{M.}~\bibnamefont{Hamuy}}, \bibinfo{year}{1997}, \bibinfo{journal}{Astron.\ Astrophys.} \textbf{\bibinfo{volume}{322}}, \bibinfo{pages}{431}.

\bibitem[{\citenamefont{Carlberg} \emph{et~al.}(1996)\citenamefont{Carlberg \emph{et~al.}}}] {Carlberg:1996}
\bibinfo{author}{\bibnamefont{Carlberg}, \bibfnamefont{R.~G.}},
\bibinfo{author}{\bibfnamefont{H.~K.~C.} \bibnamefont{Yee}},
\bibinfo{author}{\bibfnamefont{E.}~\bibnamefont{Ellingson}},
\bibinfo{author}{\bibfnamefont{R.}~\bibnamefont{Abraham}},
\bibinfo{author}{\bibfnamefont{P.}~\bibnamefont{Grabel}},
\bibinfo{author}{\bibfnamefont{S.}~\bibnamefont{Morris}}, and
\bibinfo{author}{\bibfnamefont{C.~J.} \bibnamefont{Pritchet}}, \bibinfo{year}{1996}, \bibinfo{journal}{Astrophys.~J.} \textbf{\bibinfo{volume}{462}}, \bibinfo{pages}{32}.

\bibitem[{\citenamefont{Carlson} \emph{et~al.}(1995)\citenamefont{Carlson \emph{et~al.}}}] {Carlson:1995}
\bibinfo{author}{\bibnamefont{Carlson}, \bibfnamefont{R.}},
\bibinfo{author}{\bibfnamefont{P.}~\bibnamefont{Weissman}}, 
\bibinfo{author}{\bibfnamefont{M.}~\bibnamefont{Segura, \emph{et~al.}}},
\bibinfo{year}{1995}, \bibinfo{journal}{Geophys.\ Res.\ Lett.} \textbf{\bibinfo{volume}{22}}, \bibinfo{pages}{L1557}.

\bibitem[\citenamefont{Carroll, Press, and Turner}(1992)]{Carroll:1992}
\bibinfo{author}{\bibnamefont{Carroll}, \bibfnamefont{S.~M.}}, 
\bibinfo{author}{\bibfnamefont{W.~H.}~\bibnamefont{Press}}, and
\bibinfo{author}{\bibfnamefont{E.~L.}~\bibnamefont{Turner}}, \bibinfo{year}{1992}, \bibinfo{journal}{Annu.\ Rev.\ Astron.\ Astrophys.}, \textbf{\bibinfo{volume}{30}}, \bibinfo{pages}{499}.

\bibitem[{\citenamefont{Charbonneau} \emph{et~al.}(2000)\citenamefont{Charbonneau, Brown, Latham, and Mayor}}] {Charbonneau:2000}
\bibinfo{author}{\bibnamefont{Charbonneau}, \bibfnamefont{D.}},
\bibinfo{author}{\bibfnamefont{T.~M.}~\bibnamefont{Brown}},
\bibinfo{author}{\bibfnamefont{D.~W.}~\bibnamefont{Latham}}, and
\bibinfo{author}{\bibfnamefont{M.}~\bibnamefont{Mayor}}, \bibinfo{year}{2000}, \bibinfo{journal}{Astrophys.~J.\ Lett.} \textbf{\bibinfo{volume}{529}}, \bibinfo{pages}{L41}.

\bibitem[{\citenamefont{Charbonneau} \emph{et~al.}(2002)\citenamefont{Charbonneau, Brown, Noyes, and Gilliland}}] {Charbonneau:2002}
\bibinfo{author}{\bibnamefont{Charbonneau}, \bibfnamefont{D.}},
\bibinfo{author}{\bibfnamefont{T.~M.} \bibnamefont{Brown}},
\bibinfo{author}{\bibfnamefont{R.~W.} \bibnamefont{Noyes}}, and
\bibinfo{author}{\bibfnamefont{R.~L.} \bibnamefont{Gilliland}}, \bibinfo{year}{2002}, \bibinfo{journal}{Astrophys.~J.} \textbf{\bibinfo{volume}{568}}, \bibinfo{pages}{377}.

\bibitem[{\citenamefont{Chevalier and Luo}(1994)\citenamefont{Chevalier and Leo}}]{Chevalier:1994}
\bibinfo{author}{\bibnamefont{Chevalier}, \bibfnamefont{R.~A.}} and 
\bibinfo{author}{\bibfnamefont{D.}~\bibnamefont{Luo}}, \bibinfo{year}{1994}, \bibinfo{journal}{Astrophys.~J.} \textbf{\bibinfo{volume}{435}}, \bibinfo{pages}{815}.

\bibitem[{\citenamefont{Churchwell} \emph{et~al.}(1987)\citenamefont{Churchwell \emph{et~al.}}}] {Churchwell:1987}
\bibinfo{author}{\bibnamefont{Churchwell}, \bibfnamefont{E.}},
\bibinfo{author}{\bibfnamefont{M.}~\bibnamefont{Feli}}, 
\bibinfo{author}{\bibfnamefont{D.~O.~S.} \bibnamefont{Wood}}, and
\bibinfo{author}{\bibfnamefont{M.}~\bibnamefont{Massi}},
\bibinfo{year}{1987}, \bibinfo{journal}{Astrophys.~J.} \textbf{\bibinfo{volume}{321}}, 
\bibinfo{pages}{516}.

\bibitem[{\citenamefont{Chyba} \emph{et~al.}(1993)\citenamefont{Chyba, Thomas, and Zahnle}}] {Chyba:1993}
\bibinfo{author}{\bibnamefont{Chyba}, \bibfnamefont{C.}},
\bibinfo{author}{\bibfnamefont{P.}~\bibnamefont{Thomas}}, and
\bibinfo{author}{\bibfnamefont{K.}~\bibnamefont{Zahnle}}, \bibinfo{year}{1993}, \bibinfo{journal}{Nature} \textbf{\bibinfo{volume}{361}}, \bibinfo{pages}{40}.

\bibitem[{\citenamefont{Colbert and Mushotzky}(1999)\citenamefont{Colbert and Mushotzky}}]{Colbert:1999}
\bibinfo{author}{\bibnamefont{Colbert}, \bibfnamefont{E.~J.~M.}} and 
\bibinfo{author}{\bibfnamefont{R.~F.} \bibnamefont{Mushotzky}}, \bibinfo{year}{1999}, \bibinfo{journal}{Astrophys.~J.} \textbf{\bibinfo{volume}{519}}, \bibinfo{pages}{89}.

\bibitem[{\citenamefont{Cole} \emph{et~al.}(2000)\citenamefont{Cole \emph{et~al.}}}] {Cole:2000}
\bibinfo{author}{\bibnamefont{Cole}, \bibfnamefont{S.}},
\bibinfo{author}{\bibfnamefont{C.~G.} \bibnamefont{Lacey}}, 
\bibinfo{author}{\bibfnamefont{C.~M.} \bibnamefont{Baugh}}, and
\bibinfo{author}{\bibnamefont{C.~S.} \bibnamefont{Frenk}}, \bibinfo{year}{2000}, \bibinfo{journal}{Mon.\ Not.~R.\ Astron.\ Soc.} \textbf{\bibinfo{volume}{319}}, \bibinfo{pages}{168}.

\bibitem[{\citenamefont{Colgate}(1979)\citenamefont{Colgate}}]{Colgate:1979}
\bibinfo{author}{\bibnamefont{Colgate}, \bibfnamefont{S.}}, \bibinfo{year}{1979}, \bibinfo{journal}{Astrophys.~J.} \textbf{\bibinfo{volume}{232}}, \bibinfo{pages}{404}.

\bibitem[{\citenamefont{Conselice}(1997)\citenamefont{Conselice}}]{Conselice:1997}
\bibinfo{author}{\bibnamefont{Conselice}, \bibfnamefont{C.~J.}}, \bibinfo{year}{1997}, \bibinfo{journal}{Publ.\ Astron.\ Soc.\ Pac.} \textbf{\bibinfo{volume}{109}}, \bibinfo{pages}{1251}.

\bibitem[{\citenamefont{Conselice}(2001)\citenamefont{Conselice}}]{Conselice:2001}
\bibinfo{author}{\bibnamefont{Conselice}, \bibfnamefont{C.~J.}}, \bibinfo{year}{2001}, 
in \emph{\bibinfo{booktitle}{Deep Fields}}, edited by
\bibinfo{editor}{\bibfnamefont{S.}~\bibnamefont{Cristiani}}, 
\bibinfo{editor}{\bibfnamefont{A.}~\bibnamefont{Renzini}}, and
\bibinfo{editor}{\bibfnamefont{R.~E.} \bibnamefont{Williams}},
(\bibinfo{publisher}{Springer, Berlin}), p.~\bibinfo{pages}{91}.

\bibitem[\citenamefont{Conselice, Bershady, and Jangern}(2000)]{Conselice:2000}
\bibinfo{author}{\bibnamefont{Conselice}, \bibfnamefont{C.~J.}}, 
\bibinfo{author}{\bibfnamefont{M.~A.} \bibnamefont{Bershady}}, and
\bibinfo{author}{\bibfnamefont{A.}~\bibnamefont{Jangern}}, \bibinfo{year}{2000}, \bibinfo{journal}{Astrophys.~J.}, \textbf{\bibinfo{volume}{529}}, \bibinfo{pages}{886}.

\bibitem[{\citenamefont{Corradi} \emph{et~al.}(2001)\citenamefont{Corradi \emph{et~al.}}}] {Corradi:2001}
\bibinfo{author}{\bibnamefont{Corradi}, \bibfnamefont{R.~L.~M.}},
\bibinfo{author}{\bibfnamefont{M.}~\bibnamefont{Livio}},
\bibinfo{author}{\bibfnamefont{B.}~\bibnamefont{Balick}},
\bibinfo{author}{\bibfnamefont{U.}~\bibnamefont{Munali}}, and
\bibinfo{author}{\bibfnamefont{H.~E.} \bibnamefont{Schwarz}}, \bibinfo{year}{2001}, \bibinfo{journal}{Astrophys.~J.} \textbf{\bibinfo{volume}{553}}, \bibinfo{pages}{211}.

\bibitem[{\citenamefont{Costa} \emph{et~al.}(1997)\citenamefont{Costa \emph{et~al.}}}] {Costa:1997}
\bibinfo{author}{\bibnamefont{Costa},~\bibfnamefont{E.}},
\bibinfo{author}{\bibfnamefont{F.}~\bibnamefont{Frontera}}, and
\bibinfo{author}{\bibfnamefont{J.}~\bibnamefont{Heise}},
\bibinfo{year}{1997}, \bibinfo{journal}{Nature} \textbf{\bibinfo{volume}{387}}, \bibinfo{pages}{783}.

\bibitem[{\citenamefont{Crotts and Heathcote}(1991)\citenamefont{Crotts and Heathcote}}]{Crotts:1991}
\bibinfo{author}{\bibnamefont{Crotts}, \bibfnamefont{A.}} and 
\bibinfo{author}{\bibfnamefont{S.~R.} \bibnamefont{Heathcote}}, \bibinfo{year}{1991}, \bibinfo{journal}{Nature} \textbf{\bibinfo{volume}{350}}, \bibinfo{pages}{683}.

\bibitem[{\citenamefont{D'Amico} \emph{et~al.}(2002)\citenamefont{D'Amico \emph{et~al.}}}] {DAmico:2002}
\bibinfo{author}{\bibnamefont{D'Amico}, \bibfnamefont{N.}},
\bibinfo{author}{\bibfnamefont{A.}~\bibnamefont{Possenti}}, 
\bibinfo{author}{\bibfnamefont{L.}~\bibnamefont{Fici}}, 
\bibinfo{author}{\bibfnamefont{R.~N.} \bibnamefont{Manchester}},
\bibinfo{author}{\bibfnamefont{A.~G.} \bibnamefont{Lyne}},
\bibinfo{author}{\bibfnamefont{F.}~\bibnamefont{Camilo}}, and
\bibinfo{author}{\bibfnamefont{J.}~\bibnamefont{Sarkissian}}, \bibinfo{year}{2002}, \bibinfo{journal}{Astrophys.~J.\ Lett.} \textbf{\bibinfo{volume}{570}}, \bibinfo{pages}{L89}.

\bibitem[{\citenamefont{Damineli, Conti, and Lopes}(1997)\citenamefont{Damineli, Conti, and Lopes}}] {Damineli:1997}
\bibinfo{author}{\bibnamefont{Damineli}, \bibfnamefont{A.}},
\bibinfo{author}{\bibfnamefont{P.~S.} \bibnamefont{Conti}}, and
\bibinfo{author}{\bibfnamefont{D.~F.} \bibnamefont{Lopes}}, \bibinfo{year}{1997}, \bibinfo{journal}{New Astron.} \textbf{\bibinfo{volume}{2}}, \bibinfo{pages}{107}.

\bibitem[{\citenamefont{Damineli} \emph{et~al.}(2000)\citenamefont{Damineli \emph{et~al.}}}] {Damineli:2000}
\bibinfo{author}{\bibnamefont{Damineli}, \bibfnamefont{A.}},
\bibinfo{author}{\bibfnamefont{A.}~\bibnamefont{Kaufer}}, 
\bibinfo{author}{\bibfnamefont{B.}~\bibnamefont{Wolf}}, 
\bibinfo{author}{\bibfnamefont{O.}~\bibnamefont{Stahl}}, 
\bibinfo{author}{\bibfnamefont{D.~F.} \bibnamefont{Lopes}}, and
\bibinfo{author}{\bibfnamefont{F.~X.} \bibnamefont{de Araujo}}, \bibinfo{year}{2000}, \bibinfo{journal}{Astrophys.~J.\ Lett.} \textbf{\bibinfo{volume}{528}}, \bibinfo{pages}{L101}

\bibitem[{\citenamefont{Davidson and Humphries}(1997)\citenamefont{Davidson and Humphreys}}]{Davidson:1997}
\bibinfo{author}{\bibnamefont{Davidson}, \bibfnamefont{K.}} and 
\bibinfo{author}{\bibfnamefont{R.~M.} \bibnamefont{Humphreys}}, \bibinfo{year}{1997}, \bibinfo{journal}{Annu.\ Rev.\ Astron.\ Astrophys.} \textbf{\bibinfo{volume}{35}}, \bibinfo{pages}{1}.

\bibitem[{\citenamefont{Davies and Sigurdsson}(2001)\citenamefont{Davies and Sigurdsson}}] {Davies:2001}
\bibinfo{author}{\bibnamefont{Davies}, \bibfnamefont{M.~B.}} and
\bibinfo{author}{\bibfnamefont{S.}~\bibnamefont{Sigurdson}}, \bibinfo{year}{2001}, \bibinfo{journal}{Mon.\ Not.~R.\ Astron.\ Soc.} \textbf{\bibinfo{volume}{324}}, \bibinfo{pages}{612}.

\bibitem[{\citenamefont{de~Bernardis} \emph{et~al.}(2002)\citenamefont{de~Bernardis \emph{et~al.}}}] {deBernardis:2002}
\bibinfo{author}{\bibnamefont{de Bernardis}, \bibfnamefont{P.}},
\bibinfo{author}{\bibfnamefont{P.~A.~R.} \bibnamefont{Ade}},
\bibinfo{author}{\bibfnamefont{J.~J.} \bibnamefont{Bock, \emph{et~al.}}}, \bibinfo{year}{2002}, \bibinfo{journal}{Astrophys.~J.} \textbf{\bibinfo{volume}{564}}, \bibinfo{pages}{559}.

\bibitem[{\citenamefont{Deffayet, Dvali, and Gabadadze}(2002)}] {Deffayet:2002}
\bibinfo{author}{\bibnamefont{Deffayet}, \bibfnamefont{C.}}, 
\bibinfo{author}{\bibfnamefont{G.}~\bibnamefont{Dvali}}, and
\bibinfo{author}{\bibfnamefont{G.}~\bibnamefont{Gabadadze}}, \bibinfo{year}{2002}, 
\eprint{astro-ph/0105068}.

\bibitem[{\citenamefont{Dickinson}(1995)\citenamefont{Dickinson}}]{Dickinson:1995}
\bibinfo{author}{\bibnamefont{Dickinson}, \bibfnamefont{M.}}, \bibinfo{year}{1995}, 
in \emph{\bibinfo{booktitle}{Galaxies in the Young Universe}}, edited by
\bibinfo{editor}{\bibfnamefont{H.}~\bibnamefont{Hippelein}}, 
\bibinfo{editor}{\bibfnamefont{H.~J.} \bibnamefont{Meisenheimer}}, and
\bibinfo{editor}{\bibfnamefont{H.-J.} \bibnamefont{R\"oser}},
(\bibinfo{publisher}{Springer, Berlin}), p.~\bibinfo{pages}{144}.

\bibitem[{\citenamefont{Dickinson}(2000a)\citenamefont{Dickinson}}] {Dickinson:2000a}
\bibinfo{author}{\bibnamefont{Dickinson}, \bibfnamefont{M.}},
\bibinfo{year}{2000a}, \bibinfo{journal}{Phil.\ Trans.~R.\ Soc.\ Lond.~A} \textbf{\bibinfo{volume}{358}}, \bibinfo{pages}{2001}.

\bibitem[{\citenamefont{Dickinson}(2000b)\citenamefont{Dickinson}}]{Dickinson:2000b}
\bibinfo{author}{\bibnamefont{Dickinson}, \bibfnamefont{M.}}, \bibinfo{year}{2000b}, 
in \emph{\bibinfo{booktitle}{Building Galaxies}}, XIX Rencontre de Moriond, edited by
  \bibinfo{editor}{\bibfnamefont{F.}~\bibnamefont{Hammer, \emph{et~al.}}}
  (\bibinfo{publisher}{World Scientific, Singapore}), p.~\bibinfo{pages}{257}.

\bibitem[{\citenamefont{Dickinson} \emph{et~al.}(2003)\citenamefont{Dickenson \emph{et~al.}}}] {Dickinson:2003}
\bibinfo{author}{\bibnamefont{Dickinson}, \bibfnamefont{M.}},
\bibinfo{author}{\bibfnamefont{C.}~\bibnamefont{Papovich}}, 
\bibinfo{author}{\bibfnamefont{H.~C.} \bibnamefont{Ferguson}}, and 
\bibinfo{author}{\bibfnamefont{T.}~\bibnamefont{Budav\'ari}}, \bibinfo{year}{2003}, \bibinfo{journal}{Astrophys.~J.} (in press).

\bibitem[{\citenamefont{Djorgovski and Davis}(1987)\citenamefont{Djorgovski and Davis}}]{Djor:1987}
\bibinfo{author}{\bibnamefont{Djorgovski}, \bibfnamefont{G.}} and 
\bibinfo{author}{\bibfnamefont{M.}~\bibnamefont{Davis}}, \bibinfo{year}{1987}, \bibinfo{journal}{Astrophys.~J.} \textbf{\bibinfo{volume}{313}}, \bibinfo{pages}{59}.

\bibitem[{\citenamefont{Dobrovolskis}(1990)\citenamefont{Dobrovolskis}}]{Dobrovolskis:1990}
\bibinfo{author}{\bibnamefont{Dobrovolskis}, \bibfnamefont{A.~R.}}, \bibinfo{year}{1990}, \bibinfo{journal}{Icarus} \textbf{\bibinfo{volume}{88}}, \bibinfo{pages}{24}.

\bibitem[{\citenamefont{Dodelson} \emph{et~al.}(2000)\citenamefont{Dodelson \emph{et~al.}}}] {Dodelson:2000}
\bibinfo{author}{\bibnamefont{Dodelson}, \bibfnamefont{S.}},
\bibinfo{author}{\bibfnamefont{M.}~\bibnamefont{Koplinghat}}, and
\bibinfo{author}{\bibfnamefont{E.}~\bibnamefont{Stewart}}, \bibinfo{year}{2000}, \bibinfo{journal}{Phys.\ Rev.\ Lett.} \textbf{\bibinfo{volume}{85}}, \bibinfo{pages}{5276}.

\bibitem[{\citenamefont{Dressler} \emph{et~al.}(1987)\citenamefont{Dressler \emph{et~al.}}}] {Dressler:1987}
\bibinfo{author}{\bibnamefont{Dressler}, \bibfnamefont{A.}},
\bibinfo{author}{\bibfnamefont{D.}~\bibnamefont{Lynden-Bell}}, 
\bibinfo{author}{\bibfnamefont{D.}~\bibnamefont{Burstein}}, 
\bibinfo{author}{\bibfnamefont{R.~L.} \bibnamefont{Davies}}, 
\bibinfo{author}{\bibfnamefont{S.~M.} \bibnamefont{Faber}}, 
\bibinfo{author}{\bibfnamefont{R.}~\bibnamefont{Terlevich}}, and
\bibinfo{author}{\bibfnamefont{G.}~\bibnamefont{Wegner}}, \bibinfo{year}{1987}, \bibinfo{journal}{Astrophys.~J.} \textbf{\bibinfo{volume}{313}}, \bibinfo{pages}{42}.

\bibitem[{\citenamefont{Dressler} \emph{et~al.}(1994)\citenamefont{Dressler \emph{et~al.}}}] {Dressler:1994}
\bibinfo{author}{\bibnamefont{Dressler}, \bibfnamefont{A.}},
\bibinfo{author}{\bibfnamefont{A.~J.} \bibnamefont{Oemler}}, 
\bibinfo{author}{\bibfnamefont{W.~B.} \bibnamefont{Sparks}}, and 
\bibinfo{author}{\bibfnamefont{R.~A.} \bibnamefont{Lucas}}, \bibinfo{year}{1994}, \bibinfo{journal}{Astrophys.~J.\ Lett.} \textbf{\bibinfo{volume}{435}}, \bibinfo{pages}{L23}.

\bibitem[{\citenamefont{Dressler and Richstone}(1988)\citenamefont{Dressler and Richstone}}]{Dressler:1988}
\bibinfo{author}{\bibnamefont{Dressler}, \bibfnamefont{A.}} and 
\bibinfo{author}{\bibfnamefont{D.~O.} \bibnamefont{Richstone}}, \bibinfo{year}{1988}, \bibinfo{journal}{Aastrophys.~J.} \textbf{\bibinfo{volume}{324}}, \bibinfo{pages}{701}.

\bibitem[{\citenamefont{Dressler and Richstone}(1990)\citenamefont{Dressler and Richstone}}]{Dressler:1990}
\bibinfo{author}{\bibnamefont{Dressler}, \bibfnamefont{A.}} and 
\bibinfo{author}{\bibfnamefont{D.~O.} \bibnamefont{Richstone}}, \bibinfo{year}{1990}, \bibinfo{journal}{Astrophys.~J.} \textbf{\bibinfo{volume}{348}}, \bibinfo{pages}{120}.

\bibitem[{\citenamefont{Dull} \emph{et~al.}(2002)\citenamefont{Dull \emph{et~al.}}}] {Dull:2002}
\bibinfo{author}{\bibnamefont{Dull}, \bibfnamefont{J.~D.}},
\bibinfo{author}{\bibfnamefont{H.~N.} \bibnamefont{Cohn}}, 
\bibinfo{author}{\bibfnamefont{P.~M.} \bibnamefont{Lugger}}, 
\bibinfo{author}{\bibfnamefont{B.~W.} \bibnamefont{Murphy}}, 
\bibinfo{author}{\bibfnamefont{P.~O.} \bibnamefont{Seitzer}}, 
\bibinfo{author}{\bibfnamefont{P.~J.} \bibnamefont{Callanan}}, 
\bibinfo{author}{\bibfnamefont{R.~G.~M.} \bibnamefont{Rutten}}, and
\bibinfo{author}{\bibfnamefont{P.~A.} \bibnamefont{Charles}}, \bibinfo{year}{2002}, 
\eprint{astro-ph/0210588}.

\bibitem[\citenamefont{Dvali, Gruzinov, and Zaldarriaga}(2002)]{Dvali:2002}
\bibinfo{author}{\bibnamefont{Dvali}, \bibfnamefont{G.}}, 
\bibinfo{author}{\bibfnamefont{A.}~\bibnamefont{Gruzinov}}, and
\bibinfo{author}{\bibfnamefont{M.}~\bibnamefont{Zaldarriaga}}, \bibinfo{year}{2002}, 
\eprint{hep-ph/0212069}.

\bibitem[{\citenamefont{Dwarkadas} \emph{et~al.}(1996)\citenamefont{Dwarkadas, Chevalier, and Blondin}}]{Dwarkadas:1996}
\bibinfo{author}{\bibnamefont{Dwarkadas}, \bibfnamefont{V.~V.}}, 
\bibinfo{author}{\bibfnamefont{R.~A.} \bibnamefont{Chevalier}}, and
\bibinfo{author}{\bibfnamefont{J.~M.} \bibnamefont{Blondin}}, \bibinfo{year}{1996}, \bibinfo{journal}{Astrophys.~J.}, \textbf{\bibinfo{volume}{457}}, \bibinfo{pages}{773}.

\bibitem[{\citenamefont{Ebisuzaki} \emph{et~al.}(2001)\citenamefont{Ebisuzaki \emph{et~al.}}}] {Ebisuzaki:2001}
\bibinfo{author}{\bibnamefont{Ebisuzaki}, \bibfnamefont{T.}},
\bibinfo{author}{\bibfnamefont{J.}~\bibnamefont{Makino}}, 
\bibinfo{author}{\bibfnamefont{T.~G.} \bibnamefont{Tsuru, \emph{et~al.}}}, \bibinfo{year}{2001}, \bibinfo{journal}{Astrophys.~J.\ Lett.} \textbf{\bibinfo{volume}{562}}, \bibinfo{pages}{L19}.

\bibitem[{\citenamefont{Eggen, Lynden-Bell, and Sandage}(1962)\citenamefont{Eggen \emph{et~al.}}}] {Eggen:1962}
\bibinfo{author}{\bibnamefont{Eggen}, \bibfnamefont{O.~J.}},
\bibinfo{author}{\bibfnamefont{D.}~\bibnamefont{Lynden-Bell}}, and
\bibinfo{author}{\bibfnamefont{A.}~\bibnamefont{Sandage}}, \bibinfo{year}{1962}, \bibinfo{journal}{Astrophys.~J.} \textbf{\bibinfo{volume}{136}}, \bibinfo{pages}{748}.

\bibitem[{\citenamefont{Eichler} \emph{et~al.}(1989)\citenamefont{Eichler \emph{et~al.}}}] {Eichler:1989}
\bibinfo{author}{\bibnamefont{Eichler},~\bibfnamefont{D.}},
\bibinfo{author}{\bibfnamefont{M.}~\bibnamefont{Livio}},
\bibinfo{author}{\bibfnamefont{T.}~\bibnamefont{Piran}}, and
\bibinfo{author}{\bibfnamefont{D.}~\bibnamefont{Schramm}}, \bibinfo{year}{1989}, \bibinfo{journal}{Nature} \textbf{\bibinfo{volume}{340}}, \bibinfo{pages}{126}.

\bibitem[{\citenamefont{Eisenstein and Loeb}(1996)\citenamefont{Eisenstein and Loeb}}]{Eisenstein:1996}
\bibinfo{author}{\bibnamefont{Eisenstein}, \bibfnamefont{D.~J.}}, and
\bibinfo{author}{\bibnamefont{A.}~\bibnamefont{Loeb}}, \bibinfo{year}{1996}, \bibinfo{journal}{Astrophys.~J.} \textbf{\bibinfo{volume}{459}}, \bibinfo{pages}{432}.

\bibitem[{\citenamefont{Eskridge} \emph{et~al.}(2000)\citenamefont{Eskridge \emph{et~al.}}}] {Eskridge:2000}
\bibinfo{author}{\bibnamefont{Eskridge}, \bibfnamefont{P.~B.}},
\bibinfo{author}{\bibfnamefont{J.~A.} \bibnamefont{Frogel}}, 
\bibinfo{author}{\bibfnamefont{R.~W.} \bibnamefont{Pogge, \emph{et~al.}}}, \bibinfo{year}{2000}, \bibinfo{journal}{Astron.~J.} \textbf{\bibinfo{volume}{119}}, \bibinfo{pages}{536}.

\bibitem[{\citenamefont{Fabbiano}(1989)\citenamefont{Fabbiano}}]{Fabbiano:1989}
\bibinfo{author}{\bibnamefont{Fabbiano}, \bibfnamefont{G.}}, \bibinfo{year}{1989}, \bibinfo{journal}{Annu.\ Rev.\ Astron.\ Astrophys.} \textbf{\bibinfo{volume}{27}}, \bibinfo{pages}{87}.

\bibitem[{\citenamefont{Fan \emph{et~al.}}(2001)\citenamefont{Fan \emph{et~al.}}}] {Fan:2001}
\bibinfo{author}{\bibnamefont{Fan}, \bibfnamefont{X.}}, 
\bibinfo{author}{\bibfnamefont{V.~K.} \bibnamefont{Narayanan}}, 
\bibinfo{author}{\bibfnamefont{R.~L.} \bibnamefont{Lupton, \emph{et~al.}}}, \bibinfo{year}{2001}, \bibinfo{journal}{Astron.~J.}, \textbf{\bibinfo{volume}{122}}, \bibinfo{pages}{2833}.

\bibitem[{\citenamefont{Ferguson}(1998)\citenamefont{Ferguson}}]{Ferguson:1998}
\bibinfo{author}{\bibnamefont{Ferguson}, \bibfnamefont{H.~C.}}, \bibinfo{year}{1998}, \bibinfo{journal}{Rev.\ Mod.\ Astron.} \textbf{\bibinfo{volume}{11}}, \bibinfo{pages}{83}.

\bibitem[\citenamefont{Ferguson \emph{et~al.}}(2002)\citenamefont{Ferguson \emph{et~al.}}]{Ferguson:2002}
\bibinfo{author}{\bibnamefont{Ferguson}, \bibfnamefont{H.~C.}}, 
\bibinfo{author}{\bibfnamefont{M.}~\bibnamefont{Dickinson}}, and
\bibinfo{author}{\bibfnamefont{C.}~\bibnamefont{Papovich}}, \bibinfo{year}{2002}, \bibinfo{journal}{Astrophys.~J.\ Lett.}, \textbf{\bibinfo{volume}{569}}, \bibinfo{pages}{L65}.

\bibitem[\citenamefont{Ferguson, Dickinson, and Williams}(2000)]{Ferguson:2000}
\bibinfo{author}{\bibnamefont{Ferguson}, \bibfnamefont{H.~C.}}, 
\bibinfo{author}{\bibfnamefont{M.}~\bibnamefont{Dickinson}}, and
\bibinfo{author}{\bibfnamefont{A.}~\bibnamefont{Williams}}, \bibinfo{year}{2000}, \bibinfo{journal}{Annu.\ Rev.\ Astron.\ Astrophys.}, \textbf{\bibinfo{volume}{38}}, \bibinfo{pages}{667}.

\bibitem[{\citenamefont{Ferguson and McGaugh}(1995)\citenamefont{Ferguson and McGaugh}}]{Ferguson:1995}
\bibinfo{author}{\bibnamefont{Ferguson}, \bibfnamefont{H.~C.}}, and
\bibinfo{author}{\bibnamefont{S.~S.}~\bibnamefont{McGaugh}}, \bibinfo{year}{1995}, \bibinfo{journal}{Astrophys.~J.} \textbf{\bibinfo{volume}{440}}, \bibinfo{pages}{470}.

\bibitem[{\citenamefont{Ferrarese and Merritt}(2000)\citenamefont{Ferrarese and Merritt}}]{Ferrarese:2000}
\bibinfo{author}{\bibnamefont{Ferrarese}, \bibfnamefont{L.}} and 
\bibinfo{author}{\bibfnamefont{D.}~\bibnamefont{Merritt}}, \bibinfo{year}{2000}, \bibinfo{journal}{Astrophys.~J.\ Lett.} \textbf{\bibinfo{volume}{539}}, \bibinfo{pages}{L9}.

\bibitem[{\citenamefont{Field and Ferrara}(1995)\citenamefont{Field and Ferrara}}] {Field:1995}
\bibinfo{author}{\bibnamefont{Field}, \bibfnamefont{G.}} and
\bibinfo{author}{\bibfnamefont{A.}~\bibnamefont{Ferrara}}, \bibinfo{year}{1995}, \bibinfo{journal}{Astrophys.~J.} \textbf{\bibinfo{volume}{438}}, \bibinfo{pages}{957}.

\bibitem[{\citenamefont{Filippenko and Sargent}(1986)\citenamefont{Filippenko and Sargent}}] {Filippenko:1986}
\bibinfo{author}{\bibnamefont{Filippenko}, \bibfnamefont{A.~V.}} and
\bibinfo{author}{\bibfnamefont{W.~L.~W.} \bibnamefont{Sargent}}, \bibinfo{year}{1986}, 
\bibinfo{journal}{Astron.~J.} \textbf{\bibinfo{volume}{91}}, \bibinfo{pages}{691}.

\bibitem[{\citenamefont{Fishman}(2001)\citenamefont{Fishman}}]{Fishman:2001}
\bibinfo{author}{\bibnamefont{Fishman}, \bibfnamefont{G.~J.}}, \bibinfo{year}{2001}, 
in \emph{\bibinfo{booktitle}{Supernovae and Gamma-Ray Bursts}}, edited by
  \bibinfo{editor}{\bibfnamefont{M.}~\bibnamefont{Livio}},
\bibinfo{editor}{\bibfnamefont{N.}~\bibnamefont{Panagia}}, and
\bibinfo{editor}{\bibfnamefont{K.}~\bibnamefont{Sahu}}
  (\bibinfo{publisher}{Cambridge University, Cambridge, England}), p.~\bibinfo{pages}{9}.

\bibitem[{\citenamefont{Ford} \emph{et~al.}(1994)\citenamefont{Ford \emph{et~al.}}}] {Ford:1994}
\bibinfo{author}{\bibnamefont{Ford}, \bibfnamefont{H.~C}},
\bibinfo{author}{\bibfnamefont{R.~J.} \bibnamefont{Harms}}, 
\bibinfo{author}{\bibfnamefont{Z.~I.} \bibnamefont{Tsvetanov}}, 
\bibinfo{author}{\bibfnamefont{G.~F.} \bibnamefont{Hartig}}, 
\bibinfo{author}{\bibfnamefont{L.~L.} \bibnamefont{Dressel, \emph{et~al.}}}, \bibinfo{year}{1994}, \bibinfo{journal}{Astrophys.~J.\ Lett.} \textbf{\bibinfo{volume}{435}}, \bibinfo{pages}{L27}.

\bibitem[{\citenamefont{Frail} \emph{et~al.}(2001)\citenamefont{Frail \emph{et~al.}}}] {Frail:2001}
\bibinfo{author}{\bibnamefont{Frail}, \bibfnamefont{D.~A.}},
\bibinfo{author}{\bibfnamefont{S.~R.} \bibnamefont{Kulkarni}}, 
\bibinfo{author}{\bibfnamefont{R.} \bibnamefont{Sari, \emph{et~al.}}}, \bibinfo{year}{2001}, \bibinfo{journal}{Astrophys.~J.\ Lett.} \textbf{\bibinfo{volume}{562}}, \bibinfo{pages}{L55}.

\bibitem[{\citenamefont{Frail} \emph{et~al.}(1999)}]{Frail:1999}
\bibinfo{author}{\bibnamefont{Frail}, \bibfnamefont{D.~A.}},
\bibinfo{author}{\bibfnamefont{S.~R.} \bibnamefont{Kulkarni}},
\bibinfo{author}{\bibfnamefont{M.}~\bibnamefont{Wieringa}},
\bibinfo{author}{\bibfnamefont{G.}~\bibnamefont{Taylor}}, and
\bibinfo{author}{\bibfnamefont{G.}~\bibnamefont{Moriarty-Schieven}},
\bibinfo{year}{1999}, in \emph{\bibinfo{booktitle}{Gamma-Ray Bursts}}, AIP Conf.\ Proc.~526, edited by
\bibinfo{editor}{\bibfnamefont{M.}~\bibnamefont{Kippen}}
(\bibinfo{publisher}{AIP, New York}), p.~\bibinfo{pages}{298}.

\bibitem[{\citenamefont{Franceschini} \emph{et~al.}(1998)\citenamefont{Franceschini \emph{et~al.}}}] {Franceschini:1998}
\bibinfo{author}{\bibnamefont{Franceschini}, \bibfnamefont{A.}},
\bibinfo{author}{\bibfnamefont{L.}~\bibnamefont{Silva}}, 
\bibinfo{author}{\bibfnamefont{G.}~\bibnamefont{Fasano}}, 
\bibinfo{author}{\bibfnamefont{L.}~\bibnamefont{Granato}}, 
\bibinfo{author}{\bibfnamefont{A.}~\bibnamefont{Bressan}}, 
\bibinfo{author}{\bibfnamefont{S.}~\bibnamefont{Arnouts}}, and 
\bibinfo{author}{\bibfnamefont{L.}~\bibnamefont{Danese}}, \bibinfo{year}{1998}, \bibinfo{journal}{Astrophys.~J.} \textbf{\bibinfo{volume}{506}}, \bibinfo{pages}{600}.

\bibitem[{\citenamefont{Frank} \emph{et~al.}(1993)\citenamefont{Frank \emph{et~al.}}}]{Frank:1993}
\bibinfo{author}{\bibnamefont{Frank}, \bibfnamefont{A.}}, 
\bibinfo{author}{\bibfnamefont{B.}~\bibnamefont{Balick}},
\bibinfo{author}{\bibfnamefont{V.}~\bibnamefont{Icke}}, and
\bibinfo{author}{\bibfnamefont{G.}~\bibnamefont{Mellema}}, \bibinfo{year}{1993}, \bibinfo{journal}{Astrophys.~J.\ Lett.} \textbf{\bibinfo{volume}{404}}, \bibinfo{pages}{L25}.

\bibitem[{\citenamefont{Freedman} \emph{et~al.}(2001)\citenamefont{Freedman \emph{et~al.}}}] {Freedman:2001}
\bibinfo{author}{\bibnamefont{Freedman}, \bibfnamefont{W.~L.}},
\bibinfo{author}{\bibfnamefont{B.~F.}~\bibnamefont{Madore}}, 
\bibinfo{author}{\bibfnamefont{B.~K.} \bibnamefont{Gibson, \emph{et~al.}}}, \bibinfo{year}{2001}, \bibinfo{journal}{Astrophys.~J.} \textbf{\bibinfo{volume}{553}}, \bibinfo{pages}{47}.

\bibitem[{\citenamefont{Fruchter}(2002)\citenamefont{Fruchter}}]{Fruch:2002}
\bibinfo{author}{\bibnamefont{Fruchter}, \bibfnamefont{A.}}, \bibinfo{year}{2002}, \bibinfo{journal}{APS} \textbf{\bibinfo{volume}{Apr.Y2}}, \bibinfo{pages}{004}.

\bibitem[{\citenamefont{Fruchter and Hook}(2002)\citenamefont{Fruchter and Hook}}] {Fruchter:2002}
\bibinfo{author}{\bibnamefont{Fruchter}, \bibfnamefont{A.~S.}} and
\bibinfo{author}{\bibfnamefont{R.~N.} \bibnamefont{Hook}}, \bibinfo{year}{2002}, \bibinfo{journal}{Publ.\ Astron.\ Soc.\ Pac.} \textbf{\bibinfo{volume}{114}}, \bibinfo{pages}{144}.

\bibitem[{\citenamefont{Fruchter} \emph{et~al.}(1999)\citenamefont{Fruchter \emph{et~al.}}}] {Fruchter:1999}
\bibinfo{author}{\bibnamefont{Fruchter}, \bibfnamefont{A.~S.}},
\bibinfo{author}{\bibfnamefont{S.}~\bibnamefont{Thorsett}},
\bibinfo{author}{\bibfnamefont{M.}~\bibnamefont{Metzger, \emph{et~al.}}},
\bibinfo{year}{1999}, \bibinfo{journal}{Astrophys.~J.\ Lett.} \textbf{\bibinfo{volume}{519}}, \bibinfo{pages}{L13}.

\bibitem[{\citenamefont{Galama} \emph{et~al.}(1998)\citenamefont{Galama \emph{et~al.}}}] {Galama:1998}
\bibinfo{author}{\bibnamefont{Galama}, \bibfnamefont{T.~J.}},
\bibinfo{author}{\bibfnamefont{P.~M.} \bibnamefont{Vreeswijk}}, and
\bibinfo{author}{\bibfnamefont{J.}~\bibnamefont{van Paradijs}},
\bibinfo{year}{1998}, \bibinfo{journal}{Nature} \textbf{\bibinfo{volume}{395}}, \bibinfo{pages}{670}.

\bibitem[{\citenamefont{Gallego} \emph{et~al.}(1995)\citenamefont{Gallego \emph{et~al.}}}] {Gallego:1995}
\bibinfo{author}{\bibnamefont{Gallego}, \bibfnamefont{J.}},
\bibinfo{author}{\bibfnamefont{J.}~\bibnamefont{Zamorano}}, 
\bibinfo{author}{\bibfnamefont{A.}~\bibnamefont{Aragon-Salamanca}}, and 
\bibinfo{author}{\bibfnamefont{M.}~\bibnamefont{Rego}}, \bibinfo{year}{1995}, \bibinfo{journal}{Astrophys.~J.\ Lett.} \textbf{\bibinfo{volume}{455}}, \bibinfo{pages}{L1}

\bibitem[{\citenamefont{Garcia-Segura} \emph{et~al.}(1999)\citenamefont{Garcia-Segura \emph{et~al.}}}] {Garcia:1999}
\bibinfo{author}{\bibnamefont{Garcia-Segura}, \bibfnamefont{G.}},
\bibinfo{author}{\bibfnamefont{N.}~\bibnamefont{Langer}}, 
\bibinfo{author}{\bibfnamefont{M.}~\bibnamefont{R\'ozyczka}}, and
\bibinfo{author}{\bibnamefont{J.}~\bibnamefont{Franco}}, \bibinfo{year}{1999}, \bibinfo{journal}{Astrophys.~J.} \textbf{\bibinfo{volume}{517}}, \bibinfo{pages}{767}.

\bibitem[{\citenamefont{Garnavich} \emph{et~al.}(1998)\citenamefont{Garnavich \emph{et~al.}}}] {Garnavich:1998}
\bibinfo{author}{\bibnamefont{Garnavich},~\bibfnamefont{P.~M.}},
\bibinfo{author}{\bibfnamefont{R.~P.} \bibnamefont{Kirshner}}, 
\bibinfo{author}{\bibfnamefont{P.}~\bibnamefont{Challis, \emph{et~al.}}}, \bibinfo{year}{1998}, \bibinfo{journal}{Astrophys.~J.\ Lett.} \textbf{\bibinfo{volume}{493}}, \bibinfo{pages}{L53}.

\bibitem[{\citenamefont{Garnavich} \emph{et~al.}(2003)\citenamefont{Garnavich \emph{et~al.}}}] {Garnavich:2003}
\bibinfo{author}{\bibnamefont{Garnavich},~\bibfnamefont{P.~M.}},
\bibinfo{author}{\bibfnamefont{K.~Z.} \bibnamefont{Stanek}}, 
\bibinfo{author}{\bibfnamefont{L.}~\bibnamefont{Wyrzykowski, \emph{et~al.}}}, \bibinfo{year}{2003}, \bibinfo{journal}{Astrophys.~J.} \textbf{\bibinfo{volume}{582}}, \bibinfo{pages}{924}.

\bibitem[{\citenamefont{Garriga, Livio, and Vilenkin}(2000)\citenamefont{Garriga, Livio, and Vilenkin}}] {Garriga:2000}
\bibinfo{author}{\bibnamefont{Garriga},~\bibfnamefont{J.}},
\bibinfo{author}{\bibfnamefont{M.}~\bibnamefont{Livio}}, and 
\bibinfo{author}{\bibfnamefont{A.}~\bibnamefont{Vilenkin}}, \bibinfo{year}{2000}, \bibinfo{journal}{Phys.\ Rev.~D} \textbf{\bibinfo{volume}{61}}, \bibinfo{pages}{023503}.

\bibitem[{\citenamefont{Gebhardt} \emph{et~al.}(2000)\citenamefont{Gebhardt \emph{et~al.}}}] {Gebhardt:2000}
\bibinfo{author}{\bibnamefont{Gebhardt}, \bibfnamefont{K.}},
\bibinfo{author}{\bibfnamefont{R.}~\bibnamefont{Bender}}, 
\bibinfo{author}{\bibfnamefont{G.}~\bibnamefont{Bower, \emph{et~al.}}}, \bibinfo{year}{2000}, \bibinfo{journal}{Astrophys.~J.\ Lett.} \textbf{\bibinfo{volume}{539}}, \bibinfo{pages}{L13}.

\bibitem[\citenamefont{Gebhardt, Rich, and Ho}(2002)]{Gebhardt:2002}
\bibinfo{author}{\bibnamefont{Gebhardt}, \bibfnamefont{K.}}, 
\bibinfo{author}{\bibfnamefont{R.~M.} \bibnamefont{Rich}}, and
\bibinfo{author}{\bibfnamefont{L.~C.}~\bibnamefont{Ho}}, \bibinfo{year}{2002}, \bibinfo{journal}{Astrophys.~J.\ Lett.} \textbf{\bibinfo{volume}{578}}, \bibinfo{pages}{L41}.

\bibitem[{\citenamefont{Genzel} \emph{et~al.}(2000)\citenamefont{Genzel \emph{et~al.}}}] {Genzel:2000}
\bibinfo{author}{\bibnamefont{Genzel}, \bibfnamefont{R.}},
\bibinfo{author}{\bibfnamefont{C.}~\bibnamefont{Pichon}}, 
\bibinfo{author}{\bibfnamefont{A.}~\bibnamefont{Eckart}}, 
\bibinfo{author}{\bibfnamefont{O.}~\bibnamefont{Gerhard}}, and
\bibinfo{author}{\bibfnamefont{T.}~\bibnamefont{Ott}}, \bibinfo{year}{2000}, \bibinfo{journal}{Mon.\ Not.~R.\ Astron.\ Soc.} \textbf{\bibinfo{volume}{317}}, \bibinfo{pages}{348}.

\bibitem[{\citenamefont{Gerhard}(1993)\citenamefont{Gerhard}}]{Gerhard:1993}
\bibinfo{author}{\bibnamefont{Gerhard}, \bibfnamefont{O.~E.}}, \bibinfo{year}{1993}, \bibinfo{journal}{Mon.\ Not.~R.\ Astron.\ Soc.} \textbf{\bibinfo{volume}{265}}, \bibinfo{pages}{213}.

\bibitem[{\citenamefont{Gerssen} \emph{et~al.}(2002)\citenamefont{Gerssen \emph{et~al.}}}] {Gerssen:2002}
\bibinfo{author}{\bibnamefont{Gerssen}, \bibfnamefont{J.}},
\bibinfo{author}{\bibfnamefont{R.~P.} \bibnamefont{van der Marel}}, 
\bibinfo{author}{\bibfnamefont{K.}~\bibnamefont{Gebhardt}}, 
\bibinfo{author}{\bibfnamefont{P.}~\bibnamefont{Guhathakurta}}, 
\bibinfo{author}{\bibfnamefont{R.~C.}, \bibnamefont{Peterson}}, and
\bibinfo{author}{\bibfnamefont{C.}~\bibnamefont{Pryor}}, \bibinfo{year}{2002}, \bibinfo{journal}{Astron.~J.} \textbf{\bibinfo{volume}{124}}, \bibinfo{pages}{3270}.

\bibitem[{\citenamefont{Gerssen} \emph{et~al.}(2003)\citenamefont{Gerssen \emph{et~al.}}}] {Gerssen:2003}
\bibinfo{author}{\bibnamefont{Gerssen}, \bibfnamefont{J.}},
\bibinfo{author}{\bibfnamefont{R.~P.} \bibnamefont{van der Marel}}, 
\bibinfo{author}{\bibfnamefont{K.}~\bibnamefont{Gebhardt}}, 
\bibinfo{author}{\bibfnamefont{P.}~\bibnamefont{Guhathakurta}}, 
\bibinfo{author}{\bibfnamefont{R.~C.}, \bibnamefont{Peterson}}, and
\bibinfo{author}{\bibfnamefont{C.}~\bibnamefont{Pryor}}, \bibinfo{year}{2003}, \bibinfo{journal}{Astron.~J.}, (in press).

\bibitem[{\citenamefont{Ghez} \emph{et~al.}(1998)\citenamefont{Ghez \emph{et~al.}}}] {Ghez:1998}
\bibinfo{author}{\bibnamefont{Ghez}, \bibfnamefont{A.~M.}},
\bibinfo{author}{\bibfnamefont{B.~L.} \bibnamefont{Klein}}, 
\bibinfo{author}{\bibfnamefont{M.}~\bibnamefont{Morris}}, and
\bibinfo{author}{\bibfnamefont{E.~E.} \bibnamefont{Becklin}}, \bibinfo{year}{1998}, \bibinfo{journal}{Astrophys.~J.} \textbf{\bibinfo{volume}{509}}, \bibinfo{pages}{678}.

\bibitem[{\citenamefont{Giavalisco}(2002)\citenamefont{Giavalisco}}] {Giavalisco:2002}
\bibinfo{author}{\bibnamefont{Giavalisco},~\bibfnamefont{M.}}, \bibinfo{year}{2002}, \bibinfo{journal}{Annu.\ Rev.\ Astron.\ Astrophys.} \textbf{\bibinfo{volume}{40}}, \bibinfo{pages}{579}.

\bibitem[{\citenamefont{Giavalisco} \emph{et~al.}(1996)\citenamefont{Giavalisco \emph{et~al.}}}] {Giavalisco:1996}
\bibinfo{author}{\bibnamefont{Giavalisco}, \bibfnamefont{M.}},
\bibinfo{author}{\bibfnamefont{M.}~\bibnamefont{Livio}}, 
\bibinfo{author}{\bibfnamefont{R.~C.} \bibnamefont{Bohlin}}, 
\bibinfo{author}{\bibfnamefont{F.~D.} \bibnamefont{Macchetto}}, and
\bibinfo{author}{\bibfnamefont{T.~P.} \bibnamefont{Stecher}}, \bibinfo{year}{1996}, \bibinfo{journal}{Astron.~J.} \textbf{\bibinfo{volume}{112}}, \bibinfo{pages}{369}.

\bibitem[{\citenamefont{Gilliland} \emph{et~al.}(2000)\citenamefont{Gilliland \emph{et~al.}}}] {Gilliland:2000}
\bibinfo{author}{\bibnamefont{Gilliland}, \bibfnamefont{R.~L.}},
\bibinfo{author}{\bibfnamefont{T.~M.} \bibnamefont{Brown}},
\bibinfo{author}{\bibfnamefont{P.}~\bibnamefont{Guhathakurta, \emph{et~al.}}},
\bibinfo{year}{2000}, \bibinfo{journal}{Astrophys.~J.\ Lett.} \textbf{\bibinfo{volume}{545}}, \bibinfo{pages}{L47}.

\bibitem[{\citenamefont{Gilliland, Nugent and Phillips}(1999)\citenamefont{Gilliland \emph{et~al.}}}] {Gilliland:1999}
\bibinfo{author}{\bibnamefont{Gilliland}, \bibfnamefont{R.~L.}},
\bibinfo{author}{\bibfnamefont{P.~E.} \bibnamefont{Nugent}}, and
\bibinfo{author}{\bibfnamefont{M.~M.} \bibnamefont{Phillips}},
\bibinfo{year}{1999}, \bibinfo{journal}{Astrophys.~J.} \textbf{\bibinfo{volume}{521}}, \bibinfo{pages}{30}.

\bibitem[{\citenamefont{Giovanelli} \emph{et~al.}(1997)\citenamefont{Giovanelli \emph{et~al.}}}] {Giovanelli:1997}
\bibinfo{author}{\bibnamefont{Giovanelli}, \bibfnamefont{R.}},
\bibinfo{author}{\bibfnamefont{M.~P.} \bibnamefont{Haynes}}, 
\bibinfo{author}{\bibfnamefont{N.~P.} \bibnamefont{Vogt}}, 
\bibinfo{author}{\bibfnamefont{G.}~\bibnamefont{Wegner}}, 
\bibinfo{author}{\bibfnamefont{J.~J.} \bibnamefont{Salzer}}, 
\bibinfo{author}{\bibfnamefont{L.~N.}~\bibnamefont{Da~Costa}}, and
\bibinfo{author}{\bibfnamefont{W.}~\bibnamefont{Freudling}}, \bibinfo{year}{1997}, \bibinfo{journal}{Astron.~J.} \textbf{\bibinfo{volume}{113}}, \bibinfo{pages}{22}.

\bibitem[{\citenamefont{Gonzalez}(1997)\citenamefont{Gonzalez}}]{Gonzalez:1997}
\bibinfo{author}{\bibnamefont{Gonzalez}, \bibfnamefont{G.}}, \bibinfo{year}{1997}, \bibinfo{journal}{Mon.\ Not.~R.\ Astr.\ Soc.} \textbf{\bibinfo{volume}{285}}, \bibinfo{pages}{403}.

\bibitem[{\citenamefont{Goodman}(1986)\citenamefont{Goodman}}] {Goodman:1986}
\bibinfo{author}{\bibnamefont{Goodman}, \bibfnamefont{J.}}, \bibinfo{year}{1986}, \bibinfo{journal}{Astrophys.~J.\ Lett.} \textbf{\bibinfo{volume}{308}}, \bibinfo{pages}{L47}.

\bibitem[{\citenamefont{Graham} \emph{et~al.}(1995)\citenamefont{Graham \emph{et~al.}}}]{Graham:1995}
\bibinfo{author}{\bibnamefont{Graham}, \bibfnamefont{J.}}, 
\bibinfo{author}{\bibfnamefont{I.}~\bibnamefont{DePater}},
\bibinfo{author}{\bibfnamefont{J.}~\bibnamefont{Jernigan}},
\bibinfo{author}{\bibfnamefont{M.}~\bibnamefont{Livio}}, and
\bibinfo{author}{\bibfnamefont{M.}~\bibnamefont{Brown}},
\bibinfo{year}{1995}, \bibinfo{journal}{Science} \textbf{\bibinfo{volume}{267}}, \bibinfo{pages}{1320}.

\bibitem[{\citenamefont{Greiner} \emph{et~al.}(2001)\citenamefont{Greiner \emph{et~al.}}}] {Greiner:2001}
\bibinfo{author}{\bibnamefont{Greiner}, \bibfnamefont{J.}},
\bibinfo{author}{\bibfnamefont{S.}~\bibnamefont{Klose}}, 
\bibinfo{author}{\bibfnamefont{A.}~\bibnamefont{Zeh, \emph{et~al.}}}, 
\bibinfo{year}{2001}, \bibinfo{journal}{GCN} \textbf{\bibinfo{volume}{1166}}, \bibinfo{pages}{1}.

\bibitem[{\citenamefont{Griffiths} \emph{et~al.}(1994)\citenamefont{Griffiths \emph{et~al.}}}] {Griffiths:1994}
\bibinfo{author}{\bibnamefont{Griffiths}, \bibfnamefont{R.~E.}},
\bibinfo{author}{\bibfnamefont{K.~V.} \bibnamefont{Ratnatunga}}, 
\bibinfo{author}{\bibfnamefont{L.~W.} \bibnamefont{Neuschaefer, \emph{et~al.}}}, \bibinfo{year}{1994}, \bibinfo{journal}{Astrophys.~J.} \textbf{\bibinfo{volume}{437}}, \bibinfo{pages}{67}.

\bibitem[{\citenamefont{Guth}(1981)\citenamefont{Guth}}]{Guth:1981}
\bibinfo{author}{\bibnamefont{Guth}, \bibfnamefont{A.~H.}}, \bibinfo{year}{1981}, \bibinfo{journal}{Phys.\ Rev.~D} \textbf{\bibinfo{volume}{23}}, \bibinfo{pages}{347}.

\bibitem[{\citenamefont{Haehnelt and Kauffmann}(2000)\citenamefont{Haehnelt and Kauffmann}}]{Haehnelt:2000}
\bibinfo{author}{\bibnamefont{Haehnelt}, \bibfnamefont{M.~G.}} and 
\bibinfo{author}{\bibfnamefont{G.}~\bibnamefont{Kauffmann}}, \bibinfo{year}{2000}, \bibinfo{journal}{Mon.\ Not.~R.\ Astron.\ Soc.} \textbf{\bibinfo{volume}{318}}, \bibinfo{pages}{L35}.

\bibitem[{\citenamefont{Hammel} \emph{et~al.}(1995)\citenamefont{Hammel \emph{et~al.}}}] {Hammel:1995}
\bibinfo{author}{\bibnamefont{Hammel}, \bibfnamefont{H.~B.}},
\bibinfo{author}{\bibfnamefont{R.~F.} \bibnamefont{Beebe}},
\bibinfo{author}{\bibfnamefont{A.~P.} \bibnamefont{Ingersoll, \emph{et~al.}}},
\bibinfo{year}{1995}, \bibinfo{journal}{Science} \textbf{\bibinfo{volume}{267}}, \bibinfo{pages}{1288}.

\bibitem[{\citenamefont{Hamuy} \emph{et~al.}(1996)\citenamefont{Hamuy \emph{et~al.}}}] {Hamuy:1996}
\bibinfo{author}{\bibnamefont{Hamuy}, \bibfnamefont{M.}},
\bibinfo{author}{\bibfnamefont{M.~M.} \bibnamefont{Phillips}}, 
\bibinfo{author}{\bibfnamefont{N.~B.} \bibnamefont{Suntzeff}}, 
\bibinfo{author}{\bibfnamefont{R.~A.} \bibnamefont{Schommer}}, 
\bibinfo{author}{\bibfnamefont{J.}~\bibnamefont{Maza}}, and
\bibinfo{author}{\bibfnamefont{R.}~\bibnamefont{Avil\'es}}, \bibinfo{year}{1996}, \bibinfo{journal}{Astron.~J.} \textbf{\bibinfo{volume}{112}}, \bibinfo{pages}{2398}.

\bibitem[{\citenamefont{Harms} \emph{et~al.}(1994)\citenamefont{Harms \emph{et~al.}}}] {Harms:1994}
\bibinfo{author}{\bibnamefont{Harms}, \bibfnamefont{R.~J.}},
\bibinfo{author}{\bibfnamefont{H.~C.} \bibnamefont{Ford}}, 
\bibinfo{author}{\bibfnamefont{Z.~I.} \bibnamefont{Tsvetanov, \emph{et~al.}}}, \bibinfo{year}{1994}, \bibinfo{journal}{Astrophys.~J.\ Lett.} \textbf{\bibinfo{volume}{435}}, \bibinfo{pages}{L35}.

\bibitem[{\citenamefont{Henney and O'Dell}(1999)\citenamefont{Henny and O'Dell}}]{Henney:1999}
\bibinfo{author}{\bibnamefont{Henney}, \bibfnamefont{W.~J.}} and 
\bibinfo{author}{\bibfnamefont{C.~R.} \bibnamefont{O'Dell}}, \bibinfo{year}{1999}, \bibinfo{journal}{Astron.~J.} \textbf{\bibinfo{volume}{118}}, \bibinfo{pages}{2350}.

\bibitem[{\citenamefont{Henry} \emph{et~al.}(2000)\citenamefont{Henry \emph{et~al.}}}] {Henry:2000}
\bibinfo{author}{\bibnamefont{Henry}, \bibfnamefont{G.~W.}},
\bibinfo{author}{\bibfnamefont{G.~W.} \bibnamefont{Marcy}},
\bibinfo{author}{\bibfnamefont{R.~P.} \bibnamefont{Butler}}, and
\bibinfo{author}{\bibfnamefont{S.~S.} \bibnamefont{Vogt}}, \bibinfo{year}{2000}, \bibinfo{journal}{Astrophys.~J.\ Lett.} \textbf{\bibinfo{volume}{529}}, \bibinfo{pages}{L41}.

\bibitem[{\citenamefont{Hillebrand} \emph{et~al.}(1998)\citenamefont{Hillebrand \emph{et~al.}}}] {Hillebrand:1998}
\bibinfo{author}{\bibnamefont{Hillebrand}, \bibfnamefont{L.~A.}},
\bibinfo{author}{\bibfnamefont{S.~E.} \bibnamefont{Strom}},
\bibinfo{author}{\bibfnamefont{N.}~\bibnamefont{Calvet}}, 
\bibinfo{author}{\bibfnamefont{K.~M.} \bibnamefont{Merrill}}, 
\bibinfo{author}{\bibfnamefont{I.}~\bibnamefont{Gatley}}, 
\bibinfo{author}{\bibfnamefont{R.~B.} \bibnamefont{Makidon}}, 
\bibinfo{author}{\bibfnamefont{M.~R.} \bibnamefont{Meyer}}, and
\bibinfo{author}{\bibfnamefont{M.~F.} \bibnamefont{Skrutskie}}, \bibinfo{year}{1998}, \bibinfo{journal}{Astron.~J.} \textbf{\bibinfo{volume}{116}}, \bibinfo{pages}{1816}.

\bibitem[{\citenamefont{Hillebrandt and Niemeyer}(2000)\citenamefont{Hillebrandt and Niemeyer}}]{Hillebrandt:2000}
\bibinfo{author}{\bibnamefont{Hillebrandt},~\bibfnamefont{W.}} and 
\bibinfo{author}{\bibfnamefont{J.~C.} \bibnamefont{Niemeyer}}, \bibinfo{year}{2000}, \bibinfo{journal}{Annu.\ Rev.\ Astron.\ Astrophys.} \textbf{\bibinfo{volume}{38}}, \bibinfo{pages}{191}.

\bibitem[{\citenamefont{Hillier} \emph{et~al.}(2001)\citenamefont{Hillier \emph{et~al.}}}] {Hillier:2001}
\bibinfo{author}{\bibnamefont{Hillier}, \bibfnamefont{D.~J.}},
\bibinfo{author}{\bibfnamefont{K.}~\bibnamefont{Davidson}}, 
\bibinfo{author}{\bibfnamefont{K.}~\bibnamefont{Ishibashi}}, and
\bibinfo{author}{\bibfnamefont{T.}~\bibnamefont{Gull}}, \bibinfo{year}{2001}, \bibinfo{journal}{Astrophys.~J.} \textbf{\bibinfo{volume}{553}}, \bibinfo{pages}{837}.

\bibitem[{\citenamefont{Holz}(1998)\citenamefont{Holz}}]{Holz:1998}
\bibinfo{author}{\bibnamefont{Holz}, \bibfnamefont{D.~E.}}, \bibinfo{year}{1998}, \bibinfo{journal}{Astrophys.~J.\ Lett.} \textbf{\bibinfo{volume}{506}}, \bibinfo{pages}{L1}.

\bibitem[\citenamefont{Hopkins, Connolly, and Szalay}(2000)]{Hopkins:2000}
\bibinfo{author}{\bibnamefont{Hopkins}, \bibfnamefont{A.~M.}}, 
\bibinfo{author}{\bibfnamefont{A.~J.}~\bibnamefont{Connolly}}, and
\bibinfo{author}{\bibfnamefont{A.~S.}~\bibnamefont{Szalay}}, \bibinfo{year}{2000}, \bibinfo{journal}{Astron.~J.}, \textbf{\bibinfo{volume}{120}}, \bibinfo{pages}{2843}.

\bibitem[{\citenamefont{Howell} \emph{et~al.}(2000)\citenamefont{Howell, Guhathakurta, and Gilliland}}]{Howell:2000}
\bibinfo{author}{\bibnamefont{Howell}, \bibfnamefont{J.~H.}}, 
\bibinfo{author}{\bibfnamefont{R.}~\bibnamefont{Guhathakurta}}, and
\bibinfo{author}{\bibfnamefont{R.~L.} \bibnamefont{Gilliland}}, \bibinfo{year}{2000}, \bibinfo{journal}{Publ.\ Astron.\ Soc.\ Pac.} \textbf{\bibinfo{volume}{112}}, \bibinfo{pages}{775}.

\bibitem[{\citenamefont{Hu} \emph{et~al.}(2001)\citenamefont{Hu \emph{et~al.}}}]{Hu:2001}
\bibinfo{author}{\bibnamefont{Hu}, \bibfnamefont{W.}}, 
\bibinfo{author}{\bibfnamefont{M.}~\bibnamefont{Fukugita}}, 
\bibinfo{author}{\bibfnamefont{M.}~\bibnamefont{Zaldarriaga}}, and
\bibinfo{author}{\bibfnamefont{M.} \bibnamefont{Tegmark}}, \bibinfo{year}{2000}, \bibinfo{journal}{Astrophys.~J.} \textbf{\bibinfo{volume}{549}}, \bibinfo{pages}{669}.

\bibitem[{\citenamefont{Hubble}(1929)\citenamefont{Hubble}}]{Hubble:1929}
\bibinfo{author}{\bibnamefont{Hubble}, \bibfnamefont{E.}}, \bibinfo{year}{1929}, \bibinfo{journal}{Proc.\ Nat.\ Acad.\ Sci.} \textbf{\bibinfo{volume}{15}}, \bibinfo{pages}{168}.

\bibitem[{\citenamefont{Hubble and Humason}(1931)\citenamefont{Hubble and Humason}}]{Hubble:1931}
\bibinfo{author}{\bibnamefont{Hubble}, \bibfnamefont{E.}} and
\bibinfo{author}{\bibfnamefont{M.~L.} \bibnamefont{Humason}}, \bibinfo{year}{1931}, \bibinfo{journal}{Astrophys.~J.} \textbf{\bibinfo{volume}{74}}, \bibinfo{pages}{43}.

\bibitem[{\citenamefont{Hubble}(1925)\citenamefont{Hubble}}]{Hubble:1925}
\bibinfo{author}{\bibnamefont{Hubble}, \bibfnamefont{E.~P.}}, \bibinfo{year}{1925}, \bibinfo{journal}{Astrophys.~J.} \textbf{\bibinfo{volume}{62}}, \bibinfo{pages}{409}.

\bibitem[{\citenamefont{Hubble}(1926)\citenamefont{Hubble}}]{Hubble:1926}
\bibinfo{author}{\bibnamefont{Hubble}, \bibfnamefont{E.~P.}}, \bibinfo{year}{1926}, \bibinfo{journal}{Astrophys.~J.} \textbf{\bibinfo{volume}{64}}, \bibinfo{pages}{321}.

\bibitem[{\citenamefont{Hutchings and Campbell}(1983)\citenamefont{Hutchings and Campbell}}]{Hutchings:1983}
\bibinfo{author}{\bibnamefont{Hutchings}, \bibfnamefont{J.~B.}} and 
\bibinfo{author}{\bibfnamefont{B.}~\bibnamefont{Campbell}}, \bibinfo{year}{1983}, \bibinfo{journal}{Nature} \textbf{\bibinfo{volume}{303}}, \bibinfo{pages}{584}.

\bibitem[{\citenamefont{Hutchings, Janson and Neff}(1989)\citenamefont{Hutchings \emph{et~al.}}}]{Hutchings:1989}
\bibinfo{author}{\bibnamefont{Hutchings}, \bibfnamefont{J.~B.}} and 
\bibinfo{author}{\bibfnamefont{T.}~\bibnamefont{Janson}}, and 
\bibinfo{author}{\bibfnamefont{S.~G.} \bibnamefont{Neff}}, \bibinfo{year}{1989}, \bibinfo{journal}{Astrophys.~J.} \textbf{\bibinfo{volume}{342}}, \bibinfo{pages}{660}.

\bibitem[{\citenamefont{Iben}(1985)\citenamefont{Iben}}]{Iben:1985}
\bibinfo{author}{\bibnamefont{Iben}, \bibfnamefont{I.\ Jr.}}, \bibinfo{year}{1985}, \bibinfo{journal}{Quart.~J.\ Roy.\ Astron.\ Soc.} \textbf{\bibinfo{volume}{26}}, \bibinfo{pages}{1}.

\bibitem[{\citenamefont{Iben and Livio}(1993)\citenamefont{Iben and Livio}}]{Iben:1993}
\bibinfo{author}{\bibnamefont{Iben}, \bibfnamefont{I.\ Jr.}} and 
\bibinfo{author}{\bibfnamefont{M.}~\bibnamefont{Livio}}, \bibinfo{year}{1993}, \bibinfo{journal}{Publ.\ Astro.\ Soc.\ Pac.} \textbf{\bibinfo{volume}{105}}, \bibinfo{pages}{1373}.

\bibitem[{\citenamefont{Iben and Tutukov}(1984)\citenamefont{Iben and Tutukov}}]{Iben:1984}
\bibinfo{author}{\bibnamefont{Iben}, \bibfnamefont{I.\ Jr.}} and 
\bibinfo{author}{\bibfnamefont{A.~V.} \bibnamefont{Tutukov}}, \bibinfo{year}{1993}, \bibinfo{journal}{Astrophys.~J.} \textbf{\bibinfo{volume}{418}}, \bibinfo{pages}{343}.

\bibitem[{\citenamefont{Icke} \emph{et~al.}(1992)\citenamefont{Icke, ???, and ???}}]{Icke:1992}
\bibinfo{author}{\bibnamefont{Icke}, \bibfnamefont{V.}}, 
\bibinfo{author}{\bibfnamefont{G.}~\bibnamefont{Mellema}},
\bibinfo{author}{\bibfnamefont{B.}~\bibnamefont{Balick}}, 
\bibinfo{author}{\bibfnamefont{F.}~\bibnamefont{Eulderink}}, and 
\bibinfo{author}{\bibfnamefont{A.} \bibnamefont{Frank}}, \bibinfo{year}{1992}, \bibinfo{journal}{Nature}, \textbf{\bibinfo{volume}{355}}, \bibinfo{pages}{524}.

\bibitem[{\citenamefont{Im} \emph{et~al.}(1999)\citenamefont{Im \emph{et~al.}}}] {Im:1999}
\bibinfo{author}{\bibnamefont{Im}, \bibfnamefont{M.}},
\bibinfo{author}{\bibfnamefont{R.~E.} \bibnamefont{Griffiths}}, 
\bibinfo{author}{\bibfnamefont{A.}~\bibnamefont{Naim, \emph{et~al.}}}, \bibinfo{year}{1999}, \bibinfo{journal}{Astrophys.~J.} \textbf{\bibinfo{volume}{510}}, \bibinfo{pages}{82}.

\bibitem[{\citenamefont{Ingersoll and Kanamori}(1996)\citenamefont{Ingersoll and Kanamori}}]{Ingersoll:1996}
\bibinfo{author}{\bibnamefont{Ingersoll}, \bibfnamefont{A.~P.}} and
\bibinfo{author}\bibfnamefont{H.}~{\bibnamefont{Kanamori}}, \bibinfo{year}{1996}, 
in \emph{\bibinfo{booktitle}{The Collision of Comet Shoemaker-Levy~9 and Jupiter}}, edited by
\bibinfo{editor}{\bibfnamefont{K.~S.} \bibnamefont{Noll}},
\bibinfo{editor}{\bibfnamefont{H.~A.} \bibnamefont{Weaver}}, and
\bibinfo{editor}{\bibfnamefont{P.~D.} \bibnamefont{Feldman}},
(\bibinfo{publisher}{Cambridge University, Cambridge, England}), p.~\bibinfo{pages}{329}.

\bibitem[{\citenamefont{Ishibashi} \emph{et~al.}(1999)\citenamefont{Ishibashi \emph{et~al.}}}] {Ishibashi:1999}
\bibinfo{author}{\bibnamefont{Ishibashi}, \bibfnamefont{K.}},
\bibinfo{author}{\bibfnamefont{M.~F.} \bibnamefont{Corcoran}}, 
\bibinfo{author}{\bibfnamefont{K.}~\bibnamefont{Davidson}}, 
\bibinfo{author}{\bibfnamefont{J.~H.} \bibnamefont{Swank}}, 
\bibinfo{author}{\bibfnamefont{R.}~\bibnamefont{Petre}}, 
\bibinfo{author}{\bibfnamefont{S.~A.} \bibnamefont{Drake}}, 
\bibinfo{author}{\bibfnamefont{A.}~\bibnamefont{Damineli}}, and
\bibinfo{author}{\bibfnamefont{S.}~\bibnamefont{White}}, \bibinfo{year}{1999}, \bibinfo{journal}{Astrophys.~J.} \textbf{\bibinfo{volume}{524}}, \bibinfo{pages}{983}

\bibitem[{\citenamefont{Iwamoto} \emph{et~al.}(1998)\citenamefont{Iwamoto \emph{et~al.}}}] {Iwamoto:1998}
\bibinfo{author}{\bibnamefont{Iwamoto},~\bibfnamefont{K.}},
\bibinfo{author}{\bibfnamefont{P.~A.} \bibnamefont{Mazzali}}, and
\bibinfo{author}{\bibfnamefont{K.}~\bibnamefont{Nomoto}}, 
\bibinfo{year}{1998}, \bibinfo{journal}{Nature} \textbf{\bibinfo{volume}{395}}, \bibinfo{pages}{672}.

\bibitem[{\citenamefont{Jacoby} \emph{et~al.}(1992)\citenamefont{Jacoby \emph{et~al.}}}] {Jacoby:1992}
\bibinfo{author}{\bibnamefont{Jacoby}, \bibfnamefont{G.~H.}},
\bibinfo{author}{\bibfnamefont{D.}~\bibnamefont{Branch}}, 
\bibinfo{author}{\bibfnamefont{R.}~\bibnamefont{Ciardullo}}, 
\bibinfo{author}{\bibfnamefont{R.~L.} \bibnamefont{Davies}}, 
\bibinfo{author}{\bibfnamefont{W.~H.} \bibnamefont{Harris}}, 
\bibinfo{author}{\bibfnamefont{M.~J.} \bibnamefont{Pierce}}, 
\bibinfo{author}{\bibfnamefont{C.~J.} \bibnamefont{Pritchett}}, 
\bibinfo{author}{\bibfnamefont{J.~L.} \bibnamefont{Tonry}}, and
\bibinfo{author}{\bibfnamefont{D.~L.} \bibnamefont{Welch}}, \bibinfo{year}{1992}, \bibinfo{journal}{Publ.\ Astron.\ Soc.\ Pac.} \textbf{\bibinfo{volume}{104}}, \bibinfo{pages}{599}.

\bibitem[{\citenamefont{Jeltema} \emph{et~al.}(2002)\citenamefont{Jeltema \emph{et~al.}}}] {Jeltema:2002}
\bibinfo{author}{\bibnamefont{Jeltema}, \bibfnamefont{J.~E.}},
\bibinfo{author}{\bibfnamefont{C.~R.} \bibnamefont{Canizares}}, 
\bibinfo{author}{\bibfnamefont{D.~A.} \bibnamefont{Buote}}, and
\bibinfo{author}{\bibfnamefont{G.~P.} \bibnamefont{Garmire}}, \bibinfo{year}{2002}, \bibinfo{journal}{Astrophys.~J.} (submitted).

\bibitem[{\citenamefont{Johnstone} \emph{et~al.}(1998)\citenamefont{Johnstone, Hollenbach, and Bally}}]{Johnstone:1998}
\bibinfo{author}{\bibnamefont{Johnstone}, \bibfnamefont{D.}}, 
\bibinfo{author}{\bibfnamefont{D.}~\bibnamefont{Hollenbach}}, and
\bibinfo{author}{\bibfnamefont{J.}~\bibnamefont{Bally}}, \bibinfo{year}{1998}, \bibinfo{journal}{Astrophys.~J.}, \textbf{\bibinfo{volume}{499}}, \bibinfo{pages}{758}.

\bibitem[{\citenamefont{Kahn}(1982)\citenamefont{Kahn}}]{Kahn:1982}
\bibinfo{author}{\bibnamefont{Kahn}, \bibfnamefont{F.~D.}}, \bibinfo{year}{1982},  
in \emph{\bibinfo{booktitle}{Planetary Nebulae}}, IAU Symp.~103, edited by
\bibinfo{editor}{\bibfnamefont{D.~R.}~\bibnamefont{Flower}}
(\bibinfo{publisher}{Reidel, Dordrecht}), p.~\bibinfo{pages}{305}.

\bibitem[{\citenamefont{Kalas} \emph{et~al.}(2000)\citenamefont{Kalas \emph{et~al.}}}] {Kalas:2000}
\bibinfo{author}{\bibnamefont{Kalas}, \bibfnamefont{P.}},
\bibinfo{author}{\bibfnamefont{J.}~\bibnamefont{Larwood}},
\bibinfo{author}{\bibfnamefont{B.~A.} \bibnamefont{Smith}}, and
\bibinfo{author}{\bibfnamefont{A.}~\bibnamefont{Schultz}}, \bibinfo{year}{2000}, \bibinfo{journal}{Astrophys.~J.\ Lett.} \textbf{\bibinfo{volume}{530}}, \bibinfo{pages}{L133}.

\bibitem[{\citenamefont{Kallosh and Linde}(2002)\citenamefont{Kallosh and Linde}}]{Kallosh:2002}
\bibinfo{author}{\bibnamefont{Kallosh}, \bibfnamefont{R.}} and 
\bibinfo{author}{\bibfnamefont{A.}~\bibnamefont{Linde}}, \bibinfo{year}{2002}, \eprint{hep-th/0208157}.

\bibitem[{\citenamefont{Kauffmann} \emph{et~al.}(1999)\citenamefont{Kauffmann \emph{et~al.}}}] {Kauffmann:1999}
\bibinfo{author}{\bibnamefont{Kauffmann}, \bibfnamefont{G.}},
\bibinfo{author}{\bibfnamefont{J.}~\bibnamefont{Colberg}}, 
\bibinfo{author}{\bibfnamefont{A.}~\bibnamefont{Diaferio}}, and
\bibinfo{author}{\bibfnamefont{S.~D.~M.} \bibnamefont{White}}, \bibinfo{year}{1999}, \bibinfo{journal}{Mon.\ Not.~R.\ Astron.\ Soc.} \textbf{\bibinfo{volume}{303}}, \bibinfo{pages}{188}.

\bibitem[{\citenamefont{Kauffmann, White, and Guideroni}(1993)\citenamefont{Kauffmann \emph{et~al.}}}] {Kauffmann:1993}
\bibinfo{author}{\bibnamefont{Kauffmann}, \bibfnamefont{G.}},
\bibinfo{author}{\bibfnamefont{S.~D.~M.} \bibnamefont{White}}, and
\bibinfo{author}{\bibfnamefont{B.}~\bibnamefont{Guideroni}}, \bibinfo{year}{1993}, \bibinfo{journal}{Mon.\ Not.~R.\ Astron.\ Soc.} \textbf{\bibinfo{volume}{264}}, \bibinfo{pages}{201}.

\bibitem[{\citenamefont{King} \emph{et~al.}(2001)\citenamefont{King \emph{et~al.}}}] {King:2001}
\bibinfo{author}{\bibnamefont{King}, \bibfnamefont{A.~R.}},
\bibinfo{author}{\bibfnamefont{M.~B.} \bibnamefont{Davies}}, 
\bibinfo{author}{\bibfnamefont{M.~J.} \bibnamefont{Ward}}, 
\bibinfo{author}{\bibfnamefont{G.}~\bibnamefont{Fabbiano}}, and
\bibinfo{author}{\bibfnamefont{M.}~\bibnamefont{Elvis}}, \bibinfo{year}{2001}, \bibinfo{journal}{Astrophys.~J.\ Lett.} \textbf{\bibinfo{volume}{552}}, \bibinfo{pages}{L109}.

\bibitem[{\citenamefont{Kirhakos} \emph{et~al.}(1999)\citenamefont{Kirhakos \emph{et~al.}}}] {Kirhakos:1999}
\bibinfo{author}{\bibnamefont{Kirhakos}, \bibfnamefont{S.}},
\bibinfo{author}{\bibfnamefont{J.~N.} \bibnamefont{Bahcall}}, 
\bibinfo{author}{\bibfnamefont{D.~P.} \bibnamefont{Schneider}}, and
\bibinfo{author}{\bibfnamefont{J.}~\bibnamefont{Kristian}}, 
\bibinfo{year}{1999}, \bibinfo{journal}{Astrophys.~J.} \textbf{\bibinfo{volume}{520}}, \bibinfo{pages}{67}.

\bibitem[{\citenamefont{Kirshner and Kwan}(1974)\citenamefont{Kirshner and Kwan}}]{Kirshner:1974}
\bibinfo{author}{\bibnamefont{Kirshner}, \bibfnamefont{R.~P.}} and 
\bibinfo{author}{\bibfnamefont{J.}~\bibnamefont{Kwan}}, \bibinfo{year}{1974}, \bibinfo{journal}{Astrophys.~J.} \textbf{\bibinfo{volume}{193}}, \bibinfo{pages}{27}.

\bibitem[{\citenamefont{K\"{o}nigl}(1989)\citenamefont{K\"{o}nigl}}] {Konigl:1989}
\bibinfo{author}{\bibnamefont{K\"{o}nigl}, \bibfnamefont{A.}}, \bibinfo{year}{1989}, \bibinfo{journal}{Astrophys.~J.} \textbf{\bibinfo{volume}{342}}, \bibinfo{pages}{208}.

\bibitem[{\citenamefont{Kormendy}(1988)\citenamefont{Kormendy}}]{Kormendy:1988}
\bibinfo{author}{\bibnamefont{Kormendy}, \bibfnamefont{J.}}, \bibinfo{year}{1988}, \bibinfo{journal}{Astrophys.~J.} \textbf{\bibinfo{volume}{325}}, \bibinfo{pages}{128}.

\bibitem[{\citenamefont{Kormendy and Richstone}(1995)\citenamefont{Kormendy and Richstone}}]{Kormendy:1995}
\bibinfo{author}{\bibnamefont{Kormendy}, \bibfnamefont{J.}} and 
\bibinfo{author}{\bibfnamefont{D.}~\bibnamefont{Richstone}}, \bibinfo{year}{1995}, \bibinfo{journal}{Annu.\ Rev.\ Astron.\ Astrophys.} \textbf{\bibinfo{volume}{33}}, \bibinfo{pages}{581}.

\bibitem[{\citenamefont{Kouveliotou} \emph{et~al.}(1993)\citenamefont{Kouveliotou \emph{et~al.}}}] {Kouveliotou:1993}
\bibinfo{author}{\bibnamefont{Kouveliotou},~\bibfnamefont{C.}},
\bibinfo{author}{\bibfnamefont{C.~A.} \bibnamefont{Meegan}},
\bibinfo{author}{\bibfnamefont{G.~J.} \bibnamefont{Fishman}},
\bibinfo{author}{\bibfnamefont{N.~P.} \bibnamefont{Bhat}}, 
\bibinfo{author}{\bibfnamefont{M.~S.} \bibnamefont{Briggs}}, 
\bibinfo{author}{\bibfnamefont{T.~M.} \bibnamefont{Koshut}}, 
\bibinfo{author}{\bibfnamefont{W.~S.} \bibnamefont{Paciesas}}, and 
\bibinfo{author}{\bibfnamefont{G.~N.} \bibnamefont{Pendleton}}, \bibinfo{year}{1993}, \bibinfo{journal}{Astrophys.~J.\ Lett.} \textbf{\bibinfo{volume}{413}}, \bibinfo{pages}{L101}.

\bibitem[{\citenamefont{Kristian}(1973)\citenamefont{Kristian}}]{Kristian:1973}
\bibinfo{author}{\bibnamefont{Kristian}, \bibfnamefont{J.}}, \bibinfo{year}{1973}, \bibinfo{journal}{Astrophys.~J.\ Lett.} \textbf{\bibinfo{volume}{179}}, \bibinfo{pages}{L61}.

\bibitem[{\citenamefont{Krolik}(1999a)\citenamefont{Krolik}}]{Krolik:1999a}
\bibinfo{author}{\bibnamefont{Krolik}, \bibfnamefont{J.~H.}}, \bibinfo{year}{1999a}, \bibinfo{journal}{Astrophys.~J.\ Lett.} \textbf{\bibinfo{volume}{515}}, \bibinfo{pages}{L73}.

\bibitem[{\citenamefont{Krolik}(1999b)\citenamefont{Krolik}}]{Krolik:1999b}
\bibinfo{author}{\bibnamefont{Krolik}, \bibfnamefont{J.~H.}}, 
\bibinfo{year}{1999b}, \emph{\bibinfo{booktitle}{Active Galactic Nuclei: From the Central Black Hole to the Galactic Environment}}, (\bibinfo{publisher}{Princeton University Press, Princeton}).

\bibitem[{\citenamefont{Kukula \emph{et~al.}}(1996)\citenamefont{Kukula \emph{et~al.}}}]{Kukula:1996}
\bibinfo{author}{\bibnamefont{Kukula}, \bibfnamefont{M.~J.}}, 
\bibinfo{author}{\bibfnamefont{J.~S.} \bibnamefont{Dunlop}}, 
\bibinfo{author}{\bibfnamefont{D.~H.} \bibnamefont{Hughes}}, 
\bibinfo{author}{\bibfnamefont{G.}~\bibnamefont{Taylor}}, and
\bibinfo{author}{\bibfnamefont{T.}~\bibnamefont{Boroson}}, 
\bibinfo{year}{1996},  in \emph{\bibinfo{booktitle}{Quasar Hosts}}, edited by
  \bibinfo{editor}{\bibfnamefont{D.}~\bibnamefont{Clements}} and
  \bibinfo{editor}{\bibfnamefont{I.}~\bibnamefont{Perez-Fournon}}, 
  (\bibinfo{publisher}{Springer, Berlin}), p.~\bibinfo{pages}{177}.

\bibitem[{\citenamefont{Kulkarni} \emph{et~al.}(1999)\citenamefont{Kulkarni \emph{et~al.}}}] {Kulkarni:1999}
\bibinfo{author}{\bibnamefont{Kulkarni},~\bibfnamefont{S.~R.}},
\bibinfo{author}{\bibfnamefont{S.~G.} \bibnamefont{Djorgovski}},
\bibinfo{author}{\bibfnamefont{S.~C.} \bibnamefont{Odewahn, \emph{et~al.}}},
\bibinfo{year}{1999}, \bibinfo{journal}{Nature} \textbf{\bibinfo{volume}{398}}, \bibinfo{pages}{389}.

\bibitem[{\citenamefont{Kwok}(1982)\citenamefont{Kwok}}]{Kwok:1982}
\bibinfo{author}{\bibnamefont{Kwok}, \bibfnamefont{S.}}, \bibinfo{year}{1982}, \bibinfo{journal}{Astrophys.~J.} \textbf{\bibinfo{volume}{258}}, \bibinfo{pages}{280}.

\bibitem[{\citenamefont{Lada} \emph{et~al.}(2000)\citenamefont{Lada \emph{et~al.}}}] {Lada:2000}
\bibinfo{author}{\bibnamefont{Lada}, \bibfnamefont{C.~J.}},
\bibinfo{author}{\bibfnamefont{A.~A.} \bibnamefont{Muench}}, 
\bibinfo{author}{\bibfnamefont{K.~E.} \bibnamefont{Haisch, \emph{et~al.}}}, 
\bibinfo{year}{2000}, \bibinfo{journal}{Astron.~J.} \textbf{\bibinfo{volume}{120}}, \bibinfo{pages}{3162}.

\bibitem[{\citenamefont{Lanzetta} \emph{et~al.}(1995)\citenamefont{Lanzetta \emph{et~al.}}}] {Lanzetta:1995}
\bibinfo{author}{\bibnamefont{Lanzetta}, \bibfnamefont{K.~M.}},
\bibinfo{author}{\bibfnamefont{D.~V.} \bibnamefont{Bowen}}, 
\bibinfo{author}{\bibfnamefont{D.}~\bibnamefont{Tytler}}, and 
\bibinfo{author}{\bibfnamefont{J.~K.} \bibnamefont{Webb}}, \bibinfo{year}{1995}, \bibinfo{journal}{Astrophys.~J.} \textbf{\bibinfo{volume}{442}}, \bibinfo{pages}{538}.

\bibitem[{\citenamefont{Lanzetta} \emph{et~al.}(2002)\citenamefont{Lanzetta \emph{et~al.}}}] {Lanzetta:2002}
\bibinfo{author}{\bibnamefont{Lanzetta}, \bibfnamefont{K.~M.}},
\bibinfo{author}{\bibfnamefont{N.}~\bibnamefont{Yahata}}, 
\bibinfo{author}{\bibfnamefont{S.}~\bibnamefont{Pascarelle}}, 
\bibinfo{author}{\bibfnamefont{H.-W.} \bibnamefont{Chen}}, and 
\bibinfo{author}{\bibfnamefont{A.}~\bibnamefont{Fern\'andez-Soto}}, \bibinfo{year}{2002}, \bibinfo{journal}{Astrophys.~J.} \textbf{\bibinfo{volume}{570}}, \bibinfo{pages}{492}.

\bibitem[{\citenamefont{Lauer} \emph{et~al.}(1992)\citenamefont{Lauer \emph{et~al.}}}] {Lauer:1992}
\bibinfo{author}{\bibnamefont{Lauer}, \bibfnamefont{T.~R.}},
\bibinfo{author}{\bibfnamefont{S.~M.} \bibnamefont{Faber}}, 
\bibinfo{author}{\bibfnamefont{C.~R.} \bibnamefont{Lynds, \emph{et~al.}}}, \bibinfo{year}{1992}, \bibinfo{journal}{Astron.~J.} \textbf{\bibinfo{volume}{103}}, \bibinfo{pages}{703}.

\bibitem[{\citenamefont{Laughlin}(2000)\citenamefont{Laughlin}}]{Laughlin:2000}
\bibinfo{author}{\bibnamefont{Laughlin}, \bibfnamefont{G.}}, \bibinfo{year}{2000}, \bibinfo{journal}{Astrophys.~J.} \textbf{\bibinfo{volume}{545}}, \bibinfo{pages}{1064}.

\bibitem[{\citenamefont{Lema\^{\i}tre}(1927)\citenamefont{Lema\^{\i}tre}}]{Lemaitre:1927}
\bibinfo{author}{\bibnamefont{Lema\^{\i}tre}, \bibfnamefont{G.}}, \bibinfo{year}{1927}, \bibinfo{journal}{Ann.\ Acad.\ Sci.\ Bruxelles} \textbf{\bibinfo{volume}{47A}}, \bibinfo{pages}{49}.

\bibitem[{\citenamefont{Li} \emph{et~al.}(2001)\citenamefont{Li \emph{et~al.}}}] {Li:2001}
\bibinfo{author}{\bibnamefont{Li}, \bibfnamefont{W.}},
\bibinfo{author}{\bibfnamefont{A.~V.} \bibnamefont{Filippenko}}, 
\bibinfo{author}{\bibfnamefont{R.~R.} \bibnamefont{Treffers}}, 
\bibinfo{author}{\bibfnamefont{A.~G.} \bibnamefont{Riess}}, 
\bibinfo{author}{\bibfnamefont{J.}~\bibnamefont{Hu}}, and 
\bibinfo{author}{\bibfnamefont{Y.}~\bibnamefont{Qiu}}, \bibinfo{year}{2001}, \bibinfo{journal}{Astrophys.~J.} \textbf{\bibinfo{volume}{546}}, \bibinfo{pages}{734}.

\bibitem[{\citenamefont{Lilly} \emph{et~al.}(1996)\citenamefont{Lilly \emph{et~al.}}}] {Lilly:1996}
\bibinfo{author}{\bibnamefont{Lilly}, \bibfnamefont{S.~J.}},
\bibinfo{author}{\bibfnamefont{O.}~\bibnamefont{Le F\'evre}}, 
\bibinfo{author}{\bibfnamefont{F.}~\bibnamefont{Hammer}}, and
\bibinfo{author}{\bibfnamefont{D.}~\bibnamefont{Crampton}}, \bibinfo{year}{1996}, \bibinfo{journal}{Astrophys.~J.\ Lett.} \textbf{\bibinfo{volume}{460}}, \bibinfo{pages}{L1}.

\bibitem[\citenamefont{Lin, Bodenheimer, and Richardson}(1996)]{Lin:1996}
\bibinfo{author}{\bibnamefont{Lin}, \bibfnamefont{D.~N.~C.}}, 
\bibinfo{author}{\bibfnamefont{P.}~\bibnamefont{Bodenheimer}}, and
\bibinfo{author}{\bibfnamefont{D.~C.}~\bibnamefont{Richardson}}, \bibinfo{year}{1996}, \bibinfo{journal}{Nature}, \textbf{\bibinfo{volume}{380}}, \bibinfo{pages}{606}.

\bibitem[{\citenamefont{Linde}(1986)\citenamefont{Linde}}]{Linde:1986}
\bibinfo{author}{\bibnamefont{Linde}, \bibfnamefont{A.}}, \bibinfo{year}{1986}, \bibinfo{journal}{Phys.\ Rev.\ Lett.} \textbf{\bibinfo{volume}{A1}}, \bibinfo{pages}{81}.

\bibitem[\citenamefont{Linde, Linde, and Mezhlumian}(1995)]{Linde:1995}
\bibinfo{author}{\bibnamefont{Linde}, \bibfnamefont{A.}}, 
\bibinfo{author}{\bibfnamefont{D.}~\bibnamefont{Linde}}, and
\bibinfo{author}{\bibfnamefont{A.}~\bibnamefont{Mezhlumian}}, \bibinfo{year}{1995}, \bibinfo{journal}{Phys.\ Lett.}, \textbf{\bibinfo{volume}{B345}}, \bibinfo{pages}{203}.

\bibitem[\citenamefont{Lira, Johnson, and Lawrence}(2002)]{Lira:2002}
\bibinfo{author}{\bibnamefont{Lira}, \bibfnamefont{P.}}, 
\bibinfo{author}{\bibfnamefont{R.}~\bibnamefont{Johnson}}, and
\bibinfo{author}{\bibfnamefont{A.}~\bibnamefont{Lawrence}}, \bibinfo{year}{2002}, \bibinfo{journal}{Mon.\ Not.~R.\ Astron.\ Soc.} (submitted).

\bibitem[{\citenamefont{Lissauer}(1987)\citenamefont{Lissauer}}]{Lissauer:1987}
\bibinfo{author}{\bibnamefont{Lissauer}, \bibfnamefont{J.~J.}}, \bibinfo{year}{1987}, \bibinfo{journal}{Icarus} \textbf{\bibinfo{volume}{69}}, \bibinfo{pages}{249}.

\bibitem[{\citenamefont{Livio}(1994)\citenamefont{Livio}}]{Livio:1994}
\bibinfo{author}{\bibnamefont{Livio}, \bibfnamefont{M.}}, \bibinfo{year}{1994}, in \emph{\bibinfo{booktitle}{Circumstellar Media in the Late Stages of Stellar Evolution}}, edited by
  \bibinfo{editor}{\bibfnamefont{R.}~\bibnamefont{Clegg}} \emph{et~al.} 
  (\bibinfo{publisher}{Cambridge University, Cambridge, England}), p.~\bibinfo{pages}{35}.

\bibitem[{\citenamefont{Livio}(1996)}]{Livio:1996}
\bibinfo{author}{\bibnamefont{Livio}, \bibfnamefont{M.}},
  \bibinfo{year}{1996}, in \emph{\bibinfo{booktitle}{Evolutionary Processes in Binary Stars}}, edited by
  \bibinfo{editor}{\bibfnamefont{R.~A.~M.~J.} \bibnamefont{Wijers}},
  \bibinfo{editor}{\bibfnamefont{M.~B.} \bibnamefont{Davies}}, and
  \bibinfo{editor}{\bibfnamefont{C.~A.} \bibnamefont{Tout}}
  (\bibinfo{publisher}{Kluwer, Dordrecht}), p.~\bibinfo{pages}{141}.

\bibitem[{\citenamefont{Livio}(2000)\citenamefont{Livio}}]{Livio:2000}
\bibinfo{author}{\bibnamefont{Livio}, \bibfnamefont{M.}}, \bibinfo{year}{2000}, in \emph{\bibinfo{booktitle}{Cosmic Explosions}}, AIP Conference Proceedings No.~522, edited by
  \bibinfo{editor}{\bibfnamefont{S.~S.} \bibnamefont{Holt}} and
  \bibinfo{editor}{\bibfnamefont{W.~W.} \bibnamefont{Zhang}}
  (\bibinfo{publisher}{AIP, Melville}), p.~\bibinfo{pages}{275}.

\bibitem[{\citenamefont{Livio}(2001)\citenamefont{Livio}}]{Livio:2001}
\bibinfo{author}{\bibnamefont{Livio}, \bibfnamefont{M.}}, \bibinfo{year}{2001},  in \emph{\bibinfo{booktitle}{Supernovae and Gamma-Ray Bursts}}, edited by
  \bibinfo{editor}{\bibfnamefont{M.}~\bibnamefont{Livio}},
  \bibinfo{editor}{\bibfnamefont{N.}~\bibnamefont{Panagia}}, and
  \bibinfo{editor}{\bibfnamefont{K.}~\bibnamefont{Sahu}}
  (\bibinfo{publisher}{Cambridge University, Cambridge, England}), p.~\bibinfo{pages}{334}.

\bibitem[\citenamefont{Livio, Fall, and Madau}(1998)]{Livio:1998}
\bibinfo{author}{\bibnamefont{Livio}, \bibfnamefont{Mario}}, 
\bibinfo{author}{\bibfnamefont{S.\ Michael}~\bibnamefont{Fall}}, and
\bibinfo{author}{\bibfnamefont{Piero}~\bibnamefont{Madau}}, \bibinfo{year}{1998}, \emph{\bibinfo{title}{The Hubble Deep Field}} (\bibinfo{publisher}{Cambridge University, Cambridge, England}).

\bibitem[{\citenamefont{Livio and Soker}(2001)\citenamefont{Livio and Soker}}] {LivioSoker:2001}
\bibinfo{author}{\bibnamefont{Livio}, \bibfnamefont{M.}} and
\bibinfo{author}{\bibfnamefont{N.}~\bibnamefont{Soker}}, \bibinfo{year}{2001}, \bibinfo{journal}{Astrophys.~J.} \textbf{\bibinfo{volume}{552}}, \bibinfo{pages}{685}.

\bibitem[{\citenamefont{Livio and Waxman}(2000)\citenamefont{Livio and Waxman}}] {LivioWaxman:2000}
\bibinfo{author}{\bibnamefont{Livio}, \bibfnamefont{M.}} and
\bibinfo{author}{\bibfnamefont{E.}~\bibnamefont{Waxman}}, \bibinfo{year}{2000}, \bibinfo{journal}{Astrophys.~J.} \textbf{\bibinfo{volume}{538}}, \bibinfo{pages}{187}.

\bibitem[{\citenamefont{Luo, McCray, and Slavin}(1994)\citenamefont{Luo \emph{et~al.}}}] {Luo:1994}
\bibinfo{author}{\bibnamefont{Luo}, \bibfnamefont{D.}},
\bibinfo{author}{\bibfnamefont{R.}~\bibnamefont{McCray}}, and
\bibinfo{author}{\bibfnamefont{J.}~\bibnamefont{Slavin}}, \bibinfo{year}{1994}, \bibinfo{journal}{Astrophys.~J.} \textbf{\bibinfo{volume}{430}}, \bibinfo{pages}{264}.

\bibitem[{\citenamefont{Lynden-Bell}(1969)\citenamefont{Lynden-Bell}}]{Lynden:1969}
\bibinfo{author}{\bibnamefont{Lynden-Bell}, \bibfnamefont{D.}}, \bibinfo{year}{1969}, \bibinfo{journal}{Nature} \textbf{\bibinfo{volume}{223}}, \bibinfo{pages}{690}.

\bibitem[{\citenamefont{Macchetto} \emph{et~al.}(1997)\citenamefont{Macchetto \emph{et~al.}}}] {Macchetto:1997}
\bibinfo{author}{\bibnamefont{Macchetto}, \bibfnamefont{F.~D.}},
\bibinfo{author}{\bibfnamefont{A.}~\bibnamefont{Marconi}}, 
\bibinfo{author}{\bibfnamefont{D.~J.} \bibnamefont{Axon}}, 
\bibinfo{author}{\bibfnamefont{A.}~\bibnamefont{Capetti}}, 
\bibinfo{author}{\bibfnamefont{W.}~\bibnamefont{Sparks}}, and 
\bibinfo{author}{\bibfnamefont{P.}~\bibnamefont{Crane}}, \bibinfo{year}{1997}, \bibinfo{journal}{Astrophys.~J.} \textbf{\bibinfo{volume}{489}}, \bibinfo{pages}{579}.

\bibitem[{\citenamefont{MacLow}(1996)}]{MacLow:1996}
\bibinfo{author}{\bibnamefont{MacLow}, \bibfnamefont{M.-M.}},
\bibinfo{year}{1996}, in \emph{\bibinfo{booktitle}{The Collision of Comet 
Shoemaker-Levy~9 and Jupiter}}, edited by 
\bibinfo{editor}{\bibfnamefont{K.~S.}~\bibnamefont{Noll}},
\bibinfo{editor}{\bibfnamefont{H.~A.}~\bibnamefont{Weaver}}, and
\bibinfo{editor}{\bibfnamefont{P.~D.}~\bibnamefont{Feldman}},
\bibinfo{publisher}{(Cambridge University, New York)}, p.~\bibinfo{pages}{157}.

\bibitem[{\citenamefont{Madau} \emph{et~al.}(1996)\citenamefont{Madau \emph{et~al.}}}] {Madau:1996}
\bibinfo{author}{\bibnamefont{Madau}, \bibfnamefont{P.}},
\bibinfo{author}{\bibfnamefont{H.~C.} \bibnamefont{Ferguson}}, 
\bibinfo{author}{\bibfnamefont{M.~E.} \bibnamefont{Dickinson}}, 
\bibinfo{author}{\bibfnamefont{M.}~\bibnamefont{Giavalisco}}, 
\bibinfo{author}{\bibfnamefont{C.~C.} \bibnamefont{Steidel}}, and
\bibinfo{author}{\bibfnamefont{A.}~\bibnamefont{Fruchter}}, \bibinfo{year}{1996}, \bibinfo{journal}{Mon.\ Not.~R.\ Astron.\ Soc.} \textbf{\bibinfo{volume}{283}}, \bibinfo{pages}{1388}.

\bibitem[{\citenamefont{Magorrian} \emph{et~al.}(1998)\citenamefont{Magorrian \emph{et~al.}}}] {Magorrian:1998}
\bibinfo{author}{\bibnamefont{Magorrian}, \bibfnamefont{J.}},
\bibinfo{author}{\bibfnamefont{S.}~\bibnamefont{Tremaine}}, 
\bibinfo{author}{\bibfnamefont{D.}~\bibnamefont{Richstone, \emph{et~al}.}}, \bibinfo{year}{1998}, \bibinfo{journal}{Astron.~J.} \textbf{\bibinfo{volume}{115}}, \bibinfo{pages}{2285}.

\bibitem[{\citenamefont{Marcy} \emph{et~al.}(2000)}]{Marcy:2000}
\bibinfo{author}{\bibnamefont{Marcy}, \bibfnamefont{G.~W.}},
\bibinfo{author}{\bibfnamefont{W.~D.}~\bibnamefont{Cochran}}, and
\bibinfo{author}{\bibfnamefont{M.}~\bibnamefont{Mayor}}, 
\bibinfo{year}{2000}, in \emph{\bibinfo{booktitle}{Protostars and Planets IV}}, edited by 
\bibinfo{editor}{\bibfnamefont{V.}~\bibnamefont{Mannings}},
\bibinfo{editor}{\bibfnamefont{A.~P.}~\bibnamefont{Boss}}, and
\bibinfo{editor}{\bibfnamefont{S.}~\bibnamefont{Russell}},
\bibinfo{publisher}{(Univ.\ Arizona, Tucson)}, p.~\bibinfo{pages}{1285}.

\bibitem[{\citenamefont{Mayer} \emph{et~al.}(2002)\citenamefont{Mayer \emph{et~al.}}}]{Mayer:2002}
\bibinfo{author}{\bibnamefont{Mayer}, \bibfnamefont{L.}},
\bibinfo{author}{\bibfnamefont{T.}~\bibnamefont{Quinn}}, \bibinfo{author}{\bibfnamefont{J.}~\bibnamefont{Wadsley}}, and
\bibinfo{author}{\bibfnamefont{J.}~\bibnamefont{Stadel}}, \bibinfo{year}{2002}, \bibinfo{journal}{Science} \textbf{\bibinfo{volume}{298}}, \bibinfo{pages}{1756}.

\bibitem[{\citenamefont{Mayor and Queloz}(1995)\citenamefont{Mayor and Queloz}}] {Mayor:1995}
\bibinfo{author}{\bibnamefont{Mayor}, \bibfnamefont{M.}} and
\bibinfo{author}{\bibfnamefont{D.}~\bibnamefont{Queloz}}, \bibinfo{year}{1995}, \bibinfo{journal}{Nature} \textbf{\bibinfo{volume}{378}}, \bibinfo{pages}{355}.

\bibitem[{\citenamefont{Mazeh} \emph{et~al.}(2000)\citenamefont{Mazeh \emph{et~al.}}}]{Mazeh:2000}
\bibinfo{author}{\bibnamefont{Mazeh}, \bibfnamefont{T.}}, 
\bibinfo{author}{\bibfnamefont{D.}~\bibnamefont{Naef}}, 
\bibinfo{author}{\bibfnamefont{G.}~\bibnamefont{Torres, \emph{et al.}}}, 
\bibinfo{year}{2000}, \bibinfo{journal}{Astrophys.~J.\ Lett.}, \textbf{\bibinfo{volume}{532}}, \bibinfo{pages}{L55}.

\bibitem[{\citenamefont{McCarthy} \emph{et~al.}(2001)\citenamefont{McCarthy \emph{et~al.}}}] {McCarthy:2001}
\bibinfo{author}{\bibnamefont{McCarthy}, \bibfnamefont{P.~J.}},
\bibinfo{author}{\bibfnamefont{R.~G.} \bibnamefont{Carlberg}}, 
\bibinfo{author}{\bibfnamefont{H.-W.} \bibnamefont{Chen, \emph{et~al.}}}, \bibinfo{year}{2001}, \bibinfo{journal}{Astrophys.~J.\ Lett.} \textbf{\bibinfo{volume}{560}}, \bibinfo{pages}{L131}.

\bibitem[{\citenamefont{McCray}(2003)}]{McCray:2003}
\bibinfo{author}{\bibnamefont{McCray}, \bibfnamefont{R.}},
\bibinfo{year}{2003}, in \emph{\bibinfo{booktitle}{A Decade of HST Science}}, edited by 
\bibinfo{editor}{\bibfnamefont{Mario} \bibnamefont{Livio}},
\bibinfo{editor}{\bibfnamefont{Keith} \bibnamefont{Noll}}, and
\bibinfo{editor}{\bibfnamefont{Massimo} \bibnamefont{Stiavelli}},
\bibinfo{publisher}{(Cambridge University, Cambridge, England)}, p.~\bibinfo{pages}{64}.

\bibitem[{\citenamefont{Meaburn}(1988)\citenamefont{Meaburn}}]{Meaburn:1988}
\bibinfo{author}{\bibnamefont{Meaburn}, \bibfnamefont{J.}}, \bibinfo{year}{1988}, \bibinfo{journal}{Mon.\ Not.~R.\ Astron.\ Soc.} \textbf{\bibinfo{volume}{233}}, \bibinfo{pages}{791}.

\bibitem[{\citenamefont{Mellema}(1995)\citenamefont{Mellema}}]{Mellema:1995}
\bibinfo{author}{\bibnamefont{Mellema}, \bibfnamefont{G.}}, \bibinfo{year}{1995}, \bibinfo{journal}{Mon.\ Not.~R.\ Astron.\ Soc.} \textbf{\bibinfo{volume}{277}}, \bibinfo{pages}{173}.

\bibitem[{\citenamefont{M\'esz\'aros}(2002)\citenamefont{M\'esz\'aros}}] {Meszaros:2002}
\bibinfo{author}{\bibnamefont{M\'esz\'aros},~\bibfnamefont{P.}}, \bibinfo{year}{2002}, \bibinfo{journal}{Annu.\ Rev.\ Astron.\ Astrophys.} \textbf{\bibinfo{volume}{40}}, \bibinfo{pages}{137}.

\bibitem[{\citenamefont{M\'esz\'aros and Rees}(1997)\citenamefont{M\'esz\'aros and Rees}}] {Meszaros:1997}
\bibinfo{author}{\bibnamefont{M\'esz\'aros}, \bibfnamefont{P.}} and
\bibinfo{author}{\bibfnamefont{M.~J.} \bibnamefont{Rees}}, \bibinfo{year}{1997}, \bibinfo{journal}{Astrophys.~J.} \textbf{\bibinfo{volume}{476}}, \bibinfo{pages}{232}.

\bibitem[{\citenamefont{M\'esz\'aros and Rees}(1999)\citenamefont{M\'esz\'aros and Rees}}] {Meszaros:1999}
\bibinfo{author}{\bibnamefont{M\'esz\'aros}, \bibfnamefont{P.}} and
\bibinfo{author}{\bibfnamefont{M.~J.} \bibnamefont{Rees}}, \bibinfo{year}{1999}, \bibinfo{journal}{Mon.\ Not.~R.\ Astron.\ Soc.} \textbf{\bibinfo{volume}{306}}, \bibinfo{pages}{L39}.

\bibitem[{\citenamefont{Metcalfe} \emph{et~al.}(1996)\citenamefont{Metcalfe \emph{et~al.}}}] {Metcalfe:1996}
\bibinfo{author}{\bibnamefont{Metcalfe}, \bibfnamefont{N.}},
\bibinfo{author}{\bibfnamefont{T.}~\bibnamefont{Shanks}}, 
\bibinfo{author}{\bibfnamefont{A.}~\bibnamefont{Campos}}, 
\bibinfo{author}{\bibfnamefont{R.}~\bibnamefont{Fong}}, and
\bibinfo{author}{\bibfnamefont{J.~P.} \bibnamefont{Gardner}}, \bibinfo{year}{1996}, \bibinfo{journal}{Nature} \textbf{\bibinfo{volume}{383}}, \bibinfo{pages}{236}.

\bibitem[{\citenamefont{Metzger} \emph{et~al.}(1997)\citenamefont{Metzger \emph{et~al.}}}] {Metzger:1997}
\bibinfo{author}{\bibnamefont{Metzger}, \bibfnamefont{M.}},
\bibinfo{author}{\bibfnamefont{S.~G.} \bibnamefont{Djorgovski}},
\bibinfo{author}{\bibfnamefont{S.}~\bibnamefont{Kulkarni}},
\bibinfo{author}{\bibfnamefont{C.}~\bibnamefont{Steidel}}, and
\bibinfo{author}{\bibfnamefont{K.}~\bibnamefont{Adelberger}},
\bibinfo{year}{1997}, \bibinfo{journal}{Nature} \textbf{\bibinfo{volume}{387}}, \bibinfo{pages}{878}.

\bibitem[{\citenamefont{Meurer, Heckman, and Calzetti}(1999)\citenamefont{Meurer, Heckman, and Calzetti}}] {Meurer:1999}
\bibinfo{author}{\bibnamefont{Meurer}, \bibfnamefont{G.~R.}},
\bibinfo{author}{\bibfnamefont{T.~M.} \bibnamefont{Heckman}}, and 
\bibinfo{author}{\bibfnamefont{D.}~\bibnamefont{Calzetti}}, \bibinfo{year}{1999}, \bibinfo{journal}{Astrophys.~J.} \textbf{\bibinfo{volume}{521}}, \bibinfo{pages}{64}.

\bibitem[{\citenamefont{Moustakas and Somerville}(2002)\citenamefont{Moustakas and Somerville}}]{Moustakas:2002}
\bibinfo{author}{\bibnamefont{Moustakas}, \bibfnamefont{L.~A.}} and 
\bibinfo{author}{\bibfnamefont{R.~S.} \bibnamefont{Somerville}}, \bibinfo{year}{2002}, \bibinfo{journal}{Astrophys.~J.} \textbf{\bibinfo{volume}{577}}, \bibinfo{pages}{1}.

\bibitem[{\citenamefont{Najita}(2000)\citenamefont{Najita}}] {Najita:2000}
\bibinfo{author}{\bibnamefont{Najita}, \bibfnamefont{J.}}, \bibinfo{year}{2000}, in \emph{\bibinfo{booktitle}{Unsolved Mysteries in Stellar Evolution}}, edited by
  \bibinfo{editor}{\bibfnamefont{M.}~\bibnamefont{Livio}}  
  (\bibinfo{publisher}{Cambridge University, Cambridge, England}), p.~\bibinfo{pages}{25}.

\bibitem[{\citenamefont{Nakano}(1993)\citenamefont{Nakano}}]{Nakano:1993}
\bibinfo{author}{\bibnamefont{Nakano}, \bibfnamefont{S.}}, \bibinfo{year}{1993}, \bibinfo{journal}{IAU Circular} \textbf{\bibinfo{volume}{5800}}.

\bibitem[{\citenamefont{Netterfield} \emph{et~al.}(2002)\citenamefont{Netterfield \emph{et~al.}}}] {Netterfield:2002}
\bibinfo{author}{\bibnamefont{Netterfield}, \bibfnamefont{C.~B.}},
\bibinfo{author}{\bibfnamefont{P.~A.~R.} \bibnamefont{Ade}}, 
\bibinfo{author}{\bibfnamefont{J.~J.} \bibnamefont{Bock, \emph{et~al.}}}, 
\bibinfo{year}{2002}, \bibinfo{journal}{Astrophys.~J.} \textbf{\bibinfo{volume}{571}}, \bibinfo{pages}{604}.

\bibitem[{\citenamefont{Nicholson} \emph{et~al.}(1995)\citenamefont{Nicholson \emph{et~al.}}}] {Nicholson:1995}
\bibinfo{author}{\bibnamefont{Nicholson}, \bibfnamefont{P.}},
\bibinfo{author}{\bibfnamefont{T.}~\bibnamefont{Gierasch}}, 
\bibinfo{author}{\bibfnamefont{T.}~\bibnamefont{Hayward, \emph{et~al.}}}, 
\bibinfo{year}{1995}, \bibinfo{journal}{Geophys.\ Res.\ Lett.} \textbf{\bibinfo{volume}{22}}, \bibinfo{pages}{1613}.

\bibitem[{\citenamefont{Nomoto}(1982)\citenamefont{Nomoto}}]{Nomoto:1982}
\bibinfo{author}{\bibnamefont{Nomoto},~\bibfnamefont{K.}}, \bibinfo{year}{1982}, \bibinfo{journal}{Astrophys.~J.} \textbf{\bibinfo{volume}{253}}, \bibinfo{pages}{798}.

\bibitem[{\citenamefont{Nomoto} \emph{et~al.}(1997)\citenamefont{Nomoto \emph{et~al.}}}] {Nomoto:1997}
\bibinfo{author}{\bibnamefont{Nomoto}, \bibfnamefont{K.}},
\bibinfo{author}{\bibfnamefont{K.}~\bibnamefont{Iwamoto}},
\bibinfo{author}{\bibfnamefont{N.}~\bibnamefont{Nakasato}},
\bibinfo{author}{\bibfnamefont{F.~K.} \bibnamefont{Thielemann}},
\bibinfo{author}{\bibfnamefont{F.}~\bibnamefont{Brochwitz}},
\bibinfo{author}{\bibfnamefont{T.}~\bibnamefont{Young}},
\bibinfo{author}{\bibfnamefont{T.}~\bibnamefont{Shigeyama}},
\bibinfo{author}{\bibfnamefont{T.}~\bibnamefont{Tsufimoto}}, and
\bibinfo{author}{\bibfnamefont{Y.}~\bibnamefont{Yoshii}}, \bibinfo{year}{1997}, 
in \emph{\bibinfo{booktitle}{Thermonuclear Supernovae}}, edited by
  \bibinfo{editor}{\bibfnamefont{P.}~\bibnamefont{Ruiz-Lapuente}},
\bibinfo{editor}{\bibfnamefont{R.}~\bibnamefont{Canal}}, and
\bibinfo{editor}{\bibfnamefont{J.}~\bibnamefont{Isern}},
  (\bibinfo{publisher}{Kluwer, Dordrecht}), p.~\bibinfo{pages}{349}.

\bibitem[{\citenamefont{N{\o}rgaard-Nielsen} \emph{et~al.}(1989)\citenamefont{N{\o}rgaard-Nielsen,
  Hansen, Jorgensen, Salamanca, Ellis, and Couch}}]{Norgaard:1989}
\bibinfo{author}{\bibnamefont{N{\o}rgaard-Nielsen}, \bibfnamefont{H.~V.}},
  \bibinfo{author}{\bibfnamefont{L.}~\bibnamefont{Hansen}},
  \bibinfo{author}{\bibfnamefont{H.~E.} \bibnamefont{Jorgensen}},
  \bibinfo{author}{\bibfnamefont{A.~A.} \bibnamefont{Salamanca}}, 
  \bibinfo{author}{\bibfnamefont{R.~S.} \bibnamefont{Ellis}}, and
  \bibinfo{author}{\bibfnamefont{W.~J.} \bibnamefont{Couch}}, \bibinfo{year}{1989},
  \bibinfo{journal}{Nature} \textbf{\bibinfo{volume}{339}},
  \bibinfo{pages}{523}.

\bibitem[{\citenamefont{O'Dell}(2001)\citenamefont{O'Dell}}]{ODell:2001}
\bibinfo{author}{\bibnamefont{O'Dell}, \bibfnamefont{C.~R.}}, \bibinfo{year}{2001}, \bibinfo{journal}{Annu.\ Rev.\ Astron.\ Astrophys.} \textbf{\bibinfo{volume}{39}}, \bibinfo{pages}{99}.

\bibitem[{\citenamefont{O'Dell} \emph{et~al.}(2003)\citenamefont{O'Dell, \emph{et~al.}}}]{ODell:2003}
\bibinfo{author}{\bibnamefont{O'Dell}, \bibfnamefont{C.~R.}},
\bibinfo{author}{\bibfnamefont{B.}~\bibnamefont{Balick}}, 
\bibinfo{author}{\bibfnamefont{A.~R.} \bibnamefont{Hajian}}, 
\bibinfo{author}{\bibfnamefont{W.~J.} \bibnamefont{Henney}}, and
\bibinfo{author}{\bibfnamefont{A.}~\bibnamefont{Burkett}}, \bibinfo{year}{2003}, \bibinfo{journal}{R.~Mx.~A.~C.} \textbf{\bibinfo{volume}{15}}, \bibinfo{pages}{29}.

\bibitem[{\citenamefont{O'Dell} \emph{et~al.}(1993)\citenamefont{O'Dell, Wen and Hu}}]{ODell:1993}
\bibinfo{author}{\bibnamefont{O'Dell}, \bibfnamefont{C.~R.}},
\bibinfo{author}{\bibfnamefont{Z.}~\bibnamefont{Wen}}, and
\bibinfo{author}{\bibfnamefont{X.}~\bibnamefont{Hu}}, \bibinfo{year}{1993}, \bibinfo{journal}{Astrophys.~J.} \textbf{\bibinfo{volume}{410}}, \bibinfo{pages}{696}.

\bibitem[{\citenamefont{O'Dell and Wong}(1996)\citenamefont{O'Dell and Wong}}]{ODell:1996}
\bibinfo{author}{\bibnamefont{O'Dell}, \bibfnamefont{C.~R.}} and 
\bibinfo{author}{\bibfnamefont{S.~K.}~\bibnamefont{Wong}}, \bibinfo{year}{1996}, \bibinfo{journal}{Astron.~J.} \textbf{\bibinfo{volume}{111}}, \bibinfo{pages}{846}.

\bibitem[{\citenamefont{Ogilvie and Livio}(2001)\citenamefont{Ogilvie and Livio}}] {Ogilvie:2001}
\bibinfo{author}{\bibnamefont{Ogilvie}, \bibfnamefont{G.~I.}}, and
\bibinfo{author}{\bibfnamefont{M.}~\bibnamefont{Livio}}, \bibinfo{year}{2001}, \bibinfo{journal}{Astrophys.~J.} \textbf{\bibinfo{volume}{553}}, \bibinfo{pages}{158}.

\bibitem[{\citenamefont{Ostriker}(2000)\citenamefont{Ostriker}}]{Ostriker:2000}
\bibinfo{author}{\bibnamefont{Ostriker}, \bibfnamefont{J.~P.}}, \bibinfo{year}{2000}, \bibinfo{journal}{Phys.\ Rev.\ Lett.} \textbf{\bibinfo{volume}{84}}, \bibinfo{pages}{L5258}.

\bibitem[{\citenamefont{Owocki} \emph{et~al.}(1994)\citenamefont{Owocki \emph{et~al.}}}] {Owocki:1994}
\bibinfo{author}{\bibnamefont{Owocki}, \bibfnamefont{S.~P.}},
\bibinfo{author}{\bibfnamefont{S.~R.} \bibnamefont{Crammer}}, and
\bibinfo{author}{\bibfnamefont{J.~M.} \bibnamefont{Blondin}},
\bibinfo{year}{1994}, \bibinfo{journal}{Astrophys.~J.} \textbf{\bibinfo{volume}{424}}, \bibinfo{pages}{887}.

\bibitem[{\citenamefont{Paczynski}(1986)\citenamefont{Paczynski}}] {Paczynski:1986}
\bibinfo{author}{\bibnamefont{Paczynski}, \bibfnamefont{B.}}, \bibinfo{year}{1986}, \bibinfo{journal}{Astrophys.~J.\ Lett.} \textbf{\bibinfo{volume}{308}}, \bibinfo{pages}{L43}.

\bibitem[{\citenamefont{Paczynski}(1998)\citenamefont{Paczynski}}] {Paczynski:1998}
\bibinfo{author}{\bibnamefont{Paczynski}, \bibfnamefont{B.}}, \bibinfo{year}{1998}, \bibinfo{journal}{Astrophys.~J.\ Lett.} \textbf{\bibinfo{volume}{494}}, \bibinfo{pages}{L45}.

\bibitem[{\citenamefont{Panagia}(2002)\citenamefont{Panagia}}]{Panagia:2002}
\bibinfo{author}{\bibnamefont{Panagia}, \bibfnamefont{N.}}, \bibinfo{year}{2002}, in \emph{\bibinfo{booktitle}{Multifrequency Behavior of High Energy Cosmic Sources}}, 2001 Frascati Workshop, in press.

\bibitem[{\citenamefont{Panaitescu and Kumar}(2001)\citenamefont{Panaitescu and Kumar}}] {Panaitescu:2001}
\bibinfo{author}{\bibnamefont{Panaitescu}, \bibfnamefont{A.}}, and
\bibinfo{author}{\bibfnamefont{P.}~\bibnamefont{Kumar}}, \bibinfo{year}{2001}, \bibinfo{journal}{Astrophys.~J.} \textbf{\bibinfo{volume}{554}}, \bibinfo{pages}{667}.

\bibitem[{\citenamefont{Parks}(1991)\citenamefont{Parks}}]{Parks:1991}
\bibinfo{author}{\bibnamefont{Parks}, \bibfnamefont{G.~K.}}, \bibinfo{year}{1991}, \emph{\bibinfo{title}{Physics of Space Plasmas: An Introduction}} (\bibinfo{publisher}{Addison-Wesley, New York}).

\bibitem[{\citenamefont{Pearce} \emph{et~al.}(2001)\citenamefont{Pearce \emph{et~al.}}}] {Pearce:2001}
\bibinfo{author}{\bibnamefont{Pearce}, \bibfnamefont{F.~R.}},
\bibinfo{author}{\bibfnamefont{A.}~\bibnamefont{Jenkins}}, 
\bibinfo{author}{\bibfnamefont{C.~S.} \bibnamefont{Frenk}}, 
\bibinfo{author}{\bibfnamefont{S.~D.~M.} \bibnamefont{White}}, 
\bibinfo{author}{\bibfnamefont{P.~A.} \bibnamefont{Thomas}}, 
\bibinfo{author}{\bibfnamefont{H.~M.~P.} \bibnamefont{Conchman}}, 
\bibinfo{author}{\bibfnamefont{J.~A.} \bibnamefont{Peacock}}, and
\bibinfo{author}{\bibfnamefont{G.}~\bibnamefont{Efstathiou}}, \bibinfo{year}{2001}, \bibinfo{journal}{Mon.\ Not.~R.\ Astron.\ Soc.} \textbf{\bibinfo{volume}{326}}, \bibinfo{pages}{649}.

\bibitem[{\citenamefont{Peebles and Ratra}(2002)\citenamefont{Peebles and Ratra}}]{Peebles:2002}
\bibinfo{author}{\bibnamefont{Peebles}, \bibfnamefont{P.~J.~E.}} and 
\bibinfo{author}{\bibfnamefont{B.}~\bibnamefont{Ratra}}, \bibinfo{year}{2002}, \eprint{astro-ph/0207347}.

\bibitem[{\citenamefont{Pei and Fall}(1995)\citenamefont{Pei and Fall}}]{Pei:1995}
\bibinfo{author}{\bibnamefont{Pei}, \bibfnamefont{Y.}} and 
\bibinfo{author}{\bibfnamefont{S.~M.} \bibnamefont{Fall}}, \bibinfo{year}{1995}, \bibinfo{journal}{Astrophys.~J.} \textbf{\bibinfo{volume}{454}}, \bibinfo{pages}{69}.

\bibitem[\citenamefont{Pei, Fall, and Hauser}(1999)]{Pei:1999}
\bibinfo{author}{\bibnamefont{Pei}, \bibfnamefont{Y.}}, 
\bibinfo{author}{\bibfnamefont{S.~M.} \bibnamefont{Fall}}, and
\bibinfo{author}{\bibfnamefont{M.~G.} \bibnamefont{Hauser}}, \bibinfo{year}{1999}, \bibinfo{journal}{Astrophys.~J.}, \textbf{\bibinfo{volume}{522}}, \bibinfo{pages}{604}.

\bibitem[{\citenamefont{Perlmutter} \emph{et~al.}(1998)\citenamefont{Perlmutter \emph{et~al.}}}] {Perlmutter:1998}
\bibinfo{author}{\bibnamefont{Perlmutter}, \bibfnamefont{S.}},
\bibinfo{author}{\bibfnamefont{G.}~\bibnamefont{Aldering}},
\bibinfo{author}{\bibfnamefont{M.}~\bibnamefont{Della Valle, \emph{et~al.}}}, \bibinfo{year}{1998}, \bibinfo{journal}{Nature} \textbf{\bibinfo{volume}{391}}, \bibinfo{pages}{51}.

\bibitem[{\citenamefont{Perlmutter} \emph{et~al.}(1999)\citenamefont{Perlmutter \emph{et~al.}}}] {Perlmutter:1999}
\bibinfo{author}{\bibnamefont{Perlmutter}, \bibfnamefont{S.}},
\bibinfo{author}{\bibfnamefont{G.}~\bibnamefont{Aldering}},
\bibinfo{author}{\bibfnamefont{G.}~\bibnamefont{Goldhaber, \emph{et~al.}}}, \bibinfo{year}{1999}, \bibinfo{journal}{Astrophys.~J.} \textbf{\bibinfo{volume}{517}}, \bibinfo{pages}{565}.

\bibitem[{\citenamefont{Phillips}(1993)\citenamefont{Phillips}}]{Phillips:1993}
\bibinfo{author}{\bibnamefont{Phillips}, \bibfnamefont{M.~M.}}, \bibinfo{year}{1993}, \bibinfo{journal}{Astrophys.~J.\ Lett.} \textbf{\bibinfo{volume}{413}}, \bibinfo{pages}{L105}.

\bibitem[{\citenamefont{Podsiadlowski} \emph{et~al.}(2002)\citenamefont{Podsiadlowski, Rapport, and Han}}]{Podsiadlowski:2002}
\bibinfo{author}{\bibnamefont{Podsiadlowski}, \bibfnamefont{Ph.}}, 
\bibinfo{author}{\bibfnamefont{S.}~\bibnamefont{Rapport}}, and
\bibinfo{author}{\bibfnamefont{Z.}~\bibnamefont{Han}},
\bibinfo{year}{2002}, \bibinfo{journal}{Mon.\ Not.~R.\ Astron.\ Soc.} (submitted).

\bibitem[{\citenamefont{Pollack} \emph{et~al.}(1996)\citenamefont{Pollack \emph{et~al.}}}]{Pollack:1996}
\bibinfo{author}{\bibnamefont{Pollack}, \bibfnamefont{J.~B.}}, 
\bibinfo{author}{\bibfnamefont{O.}~\bibnamefont{Hubickys}}, 
\bibinfo{author}{\bibfnamefont{P.}~\bibnamefont{Bodenheimer}}, 
\bibinfo{author}{\bibfnamefont{J.~J.} {Lissauer}}, 
\bibinfo{author}{\bibfnamefont{M.}~{Podolak}}, and
\bibinfo{author}{\bibfnamefont{Y.}~{Greenzweig}}, \bibinfo{year}{1996}, \bibinfo{journal}{Icarus} \textbf{\bibinfo{volume}{124}}, \bibinfo{pages}{62}.

\bibitem[{\citenamefont{Prialnik}(2000)\citenamefont{Prialnik}}]{Prialnik:2000}
\bibinfo{author}{\bibnamefont{Prialnik}, \bibfnamefont{D.}}, \bibinfo{year}{2000},  
\emph{\bibinfo{booktitle}{An Introduction to the Theory of Stellar Structure and Evolution}}, 
(\bibinfo{publisher}{Cambridge University, Cambridge, England}), p.~\bibinfo{pages}{174}.

\bibitem[{\citenamefont{Price} \emph{et~al.}(2002)\citenamefont{Price \emph{et~al.}}}]{Price:2002}
\bibinfo{author}{\bibnamefont{Price}, \bibfnamefont{P.~A.}}, 
\bibinfo{author}{\bibfnamefont{E.}~\bibnamefont{Berger}}, 
\bibinfo{author}{\bibfnamefont{D.~E.} \bibnamefont{Reichert, \emph{et~al.}}}, 
\bibinfo{year}{2002}, \bibinfo{journal}{Astrophys.~J.\ Lett.} \textbf{\bibinfo{volume}{572}}, \bibinfo{pages}{L51}.

\bibitem[{\citenamefont{Pringle}(1989)\citenamefont{Pringle}}]{Pringle:1989}
\bibinfo{author}{\bibnamefont{Pringle}, \bibfnamefont{J.~E.}}, \bibinfo{year}{1989}, \bibinfo{journal}{Mon.\ Not.~R.\ Astron.\ Soc.} \textbf{\bibinfo{volume}{238}}, \bibinfo{pages}{37}.

\bibitem[{\citenamefont{Pun}(1997)\citenamefont{Pun}}] {Pun:1997}
\bibinfo{author}{\bibnamefont{Pun}, \bibfnamefont{C.~S.~J.}},
\bibinfo{year}{1997}, \eprint{http://oposite.stsci.edu//pubinfo/PR/97/03.html/}.

\bibitem[{\citenamefont{Rasio and Livio}(1996)\citenamefont{Rasio and Livio}}] {Rasio:1996}
\bibinfo{author}{\bibnamefont{Rasio}, \bibfnamefont{F.~A.}} and
\bibinfo{author}{\bibfnamefont{M.}~\bibnamefont{Livio}}, \bibinfo{year}{1996}, \bibinfo{journal}{Astrophys.~J.} \textbf{\bibinfo{volume}{471}}, \bibinfo{pages}{366}.

\bibitem[{\citenamefont{Ratra and Peebles}(1988)\citenamefont{Ratra and Peebles}}] {Ratra:1988}
\bibinfo{author}{\bibnamefont{Ratra}, \bibfnamefont{B.}} and
\bibinfo{author}{\bibfnamefont{P.~J.~E.}~\bibnamefont{Peebles}}, \bibinfo{year}{1988}, \bibinfo{journal}{Phys.\ Rev.~D} \textbf{\bibinfo{volume}{37}}, \bibinfo{pages}{3406}.

\bibitem[{\citenamefont{Rees}(1978)\citenamefont{Rees}}]{Rees:1978}
\bibinfo{author}{\bibnamefont{Rees}, \bibfnamefont{M.~J.}}, \bibinfo{year}{1978}, \bibinfo{journal}{Nature} \textbf{\bibinfo{volume}{275}}, \bibinfo{pages}{516}.

\bibitem[{\citenamefont{Rees}(1984)\citenamefont{Rees}}]{Rees:1984}
\bibinfo{author}{\bibnamefont{Rees}, \bibfnamefont{M.~J.}}, \bibinfo{year}{1984}, \bibinfo{journal}{Annu.\ Rev.\ Astron.\ Astrophys.} \textbf{\bibinfo{volume}{22}}, \bibinfo{pages}{471}.

\bibitem[{\citenamefont{Reese} \emph{et~al.}(2002)\citenamefont{Reese \emph{et~al.}}}] {Reese:2002}
\bibinfo{author}{\bibnamefont{Reese}, \bibfnamefont{E.~D.}},
\bibinfo{author}{\bibfnamefont{J.~E.} \bibnamefont{Carlstrom}},
\bibinfo{author}{\bibfnamefont{M.}~\bibnamefont{Joy}},
\bibinfo{author}{\bibfnamefont{J.~J.} \bibnamefont{Mohr}},
\bibinfo{author}{\bibfnamefont{L.}~\bibnamefont{Grego}}, and
\bibinfo{author}{\bibfnamefont{W.~L.} \bibnamefont{Holzapfel}},
\bibinfo{year}{2002}, \bibinfo{journal}{Astrophys.~J.} \textbf{\bibinfo{volume}{581}}, \bibinfo{pages}{53}.

\bibitem[{\citenamefont{Reese} \emph{et~al.}(2000)\citenamefont{Reese \emph{et~al.}}}] {Reese:2000}
\bibinfo{author}{\bibnamefont{Reese}, \bibfnamefont{E.~D.}},
\bibinfo{author}{\bibfnamefont{J.~J.} \bibnamefont{Mohr}},
\bibinfo{author}{\bibfnamefont{J.~E.} \bibnamefont{Carlstrom}},
\bibinfo{author}{\bibfnamefont{M.}~\bibnamefont{Joy}},
\bibinfo{author}{\bibfnamefont{L.}~\bibnamefont{Grego}},
\bibinfo{author}{\bibfnamefont{G.~P.} \bibnamefont{Holder}},
\bibinfo{author}{\bibfnamefont{W.~L.} \bibnamefont{Holzapfel}},
\bibinfo{author}{\bibfnamefont{J.~P.} \bibnamefont{Hughes}},
\bibinfo{author}{\bibfnamefont{S.~K.} \bibnamefont{Patel}}, and
\bibinfo{author}{\bibfnamefont{M.}~\bibnamefont{Donahue}}, \bibinfo{year}{2000}, \bibinfo{journal}{Astrophys.~J.} \textbf{\bibinfo{volume}{533}}, \bibinfo{pages}{38}.

\bibitem[{\citenamefont{Reid}(2002)\citenamefont{Reid}}]{Reid:2002}
\bibinfo{author}{\bibnamefont{Reid}, \bibfnamefont{I.~N.}}, \bibinfo{year}{2002}, \bibinfo{journal}{Publ.\ Astron.\ Soc.\ Pac.} \textbf{\bibinfo{volume}{114}}, \bibinfo{pages}{306}.

\bibitem[{\citenamefont{Reipurth}(1999)\citenamefont{Reipurth}}]{Reipurth:1999}
\bibinfo{author}{\bibnamefont{Reipurth}, \bibfnamefont{B.}}, \bibinfo{year}{1999}, 
in \emph{\bibinfo{booktitle}{A General Catalog of Herbig-Haro Objects}}, second edition,\hfil\break
http://casa.colorado.edu/hhcat/.

\bibitem[{\citenamefont{Reipurth} \emph{et~al.}(1998)\citenamefont{Reipurth, Devine and Bally}}] {Reipurth:1998}
\bibinfo{author}{\bibnamefont{Reipurth}, \bibfnamefont{B.}},
\bibinfo{author}{\bibnamefont{D.}~\bibnamefont{Devine}}, and 
\bibinfo{author}{\bibnamefont{J.}~\bibnamefont{Bally}}, \bibinfo{year}{1998}, \bibinfo{journal}{Astron.~J.} \textbf{\bibinfo{volume}{116}}, \bibinfo{pages}{1396}.

\bibitem[{\citenamefont{Reipurth and Heathcote}(1993)\citenamefont{Reipurth and Heathcoate}}] {Reipurth:1993}
\bibinfo{author}{\bibnamefont{Reipurth}, \bibfnamefont{B.}} and
\bibinfo{author}{\bibfnamefont{S.}~\bibnamefont{Heathcote}}, \bibinfo{year}{1993}, 
in \emph{\bibinfo{booktitle}{Astrophysical Jets}}, edited by
  \bibinfo{editor}{\bibfnamefont{D.}~\bibnamefont{Burgarella}},
\bibinfo{editor}{\bibfnamefont{M.}~\bibnamefont{Livio}}, and
\bibinfo{editor}{\bibfnamefont{C.~P.} \bibnamefont{O'Dea}},
  (\bibinfo{publisher}{Cambridge University, Cambridge, England}), p.~\bibinfo{pages}{35}.

\bibitem[{\citenamefont{Rhoads}(1997)\citenamefont{Rhoads}}]{Rhoads:1997}
\bibinfo{author}{\bibnamefont{Rhoads}, \bibfnamefont{J.}}, \bibinfo{year}{1997}, \bibinfo{journal}{Astrophys.~J.\ Lett.} \textbf{\bibinfo{volume}{487}}, \bibinfo{pages}{L1}.

\bibitem[{\citenamefont{Riess}(2000)\citenamefont{Riess}}] {Riess:2000}
\bibinfo{author}{\bibnamefont{Riess}, \bibfnamefont{A.~G.}}, \bibinfo{year}{2000}, 
\bibinfo{journal}{Publ.\ Astron.\ Soc.\ Pac.} \textbf{\bibinfo{volume}{112}}, \bibinfo{pages}{1284}.

\bibitem[{\citenamefont{Riess} \emph{et~al.}(1998)\citenamefont{Riess \emph{et~al.}}}] {Riess:1998}
\bibinfo{author}{\bibnamefont{Riess}, \bibfnamefont{A.~G.}},
\bibinfo{author}{\bibfnamefont{A.~V.} \bibnamefont{Filippenko}},
\bibinfo{author}{\bibfnamefont{P.}~\bibnamefont{Challis, \emph{et~al.}}}, \bibinfo{year}{1998}, \bibinfo{journal}{Astron.~J.} \textbf{\bibinfo{volume}{116}}, \bibinfo{pages}{1009}.

\bibitem[{\citenamefont{Riess} \emph{et~al.}(2000)\citenamefont{Riess \emph{et~al.}}}] {Riessetal:2000}
\bibinfo{author}{\bibnamefont{Riess}, \bibfnamefont{A.~G.}},
\bibinfo{author}{\bibfnamefont{A.~V.} \bibnamefont{Filippenko}},
\bibinfo{author}{\bibfnamefont{M.~C.} \bibnamefont{Liu, \emph{et~al.}}}, \bibinfo{year}{2000}, \bibinfo{journal}{Astrophys.~J.} \textbf{\bibinfo{volume}{526}}, \bibinfo{pages}{62}.

\bibitem[{\citenamefont{Riess} \emph{et~al.}(1999)\citenamefont{Riess \emph{et~al.}}}] {Riess:1999}
\bibinfo{author}{\bibnamefont{Riess}, \bibfnamefont{A.~G.}},
\bibinfo{author}{\bibfnamefont{R.~P.} \bibnamefont{Kirshner}},
\bibinfo{author}{\bibfnamefont{B.~P.} \bibnamefont{Schmidt, \emph{et~al.}}}, \bibinfo{year}{1999}, \bibinfo{journal}{Astron.~J.} \textbf{\bibinfo{volume}{117}}, \bibinfo{pages}{707}.

\bibitem[{\citenamefont{Riess} \emph{et~al.}(2001)\citenamefont{Riess \emph{et~al.}}}] {Riess:2001}
\bibinfo{author}{\bibnamefont{Riess}, \bibfnamefont{A.~G.}},
\bibinfo{author}{\bibfnamefont{P.~E.} \bibnamefont{Nugent}}, 
\bibinfo{author}{\bibfnamefont{R.~L.} \bibnamefont{Gilliland, \emph{et~al.}}}, \bibinfo{year}{2001}, \bibinfo{journal}{Astrophys.~J.} \textbf{\bibinfo{volume}{560}}, \bibinfo{pages}{49}.

\bibitem[{\citenamefont{Riess} \emph{et~al.}(1996)\citenamefont{Riess \emph{et~al.}}}] {Riess:1996}
\bibinfo{author}{\bibnamefont{Riess}, \bibfnamefont{A.~G.}},
\bibinfo{author}{\bibfnamefont{W.~H.} \bibnamefont{Press}}, and
\bibinfo{author}{\bibfnamefont{R.~P.} \bibnamefont{Kirshner}}, \bibinfo{year}{1996}, \bibinfo{journal}{Astrophys.~J.} \textbf{\bibinfo{volume}{473}}, \bibinfo{pages}{88}.

\bibitem[{\citenamefont{Roberts} \emph{et~al.}(2001)\citenamefont{Roberts \emph{et~al.}}}] {Roberts:2001}
\bibinfo{author}{\bibnamefont{Roberts}, \bibfnamefont{T.~P.}},
\bibinfo{author}{\bibfnamefont{M.~R.} \bibnamefont{Good}}, 
\bibinfo{author}{\bibfnamefont{M.~J.} \bibnamefont{Ward}}, 
\bibinfo{author}{\bibfnamefont{R.~S.} \bibnamefont{Warwick}}, 
\bibinfo{author}{\bibfnamefont{P.~T.} \bibnamefont{O'Brien}}, 
\bibinfo{author}{\bibfnamefont{P.}~\bibnamefont{Lira}}, and
\bibinfo{author}{\bibfnamefont{A.~D.~P.} \bibnamefont{Hands}}, \bibinfo{year}{2001}, \bibinfo{journal}{Mon.\ Not.~R.\ Astron.\ Soc.} \textbf{\bibinfo{volume}{325}}, \bibinfo{pages}{L7}.

\bibitem[{\citenamefont{Roche} \emph{et~al.}(1998)\citenamefont{Roche \emph{et~al.}}}] {Roche:1998}
\bibinfo{author}{\bibnamefont{Roche}, \bibfnamefont{N.}},
\bibinfo{author}{\bibfnamefont{K.}~\bibnamefont{Ratnatunga}}, 
\bibinfo{author}{\bibfnamefont{R.~E.} \bibnamefont{Griffiths}}, 
\bibinfo{author}{\bibfnamefont{M.}~\bibnamefont{Im}}, and
\bibinfo{author}{\bibfnamefont{A.}~\bibnamefont{Naim}}, \bibinfo{year}{1998}, \bibinfo{journal}{Mon.\ Not.~R.\ Astron.\ Soc.} \textbf{\bibinfo{volume}{293}}, \bibinfo{pages}{157}.

\bibitem[{\citenamefont{Saha} \emph{et~al.}(2001)\citenamefont{Saha \emph{et~al.}}}] {Saha:2001}
\bibinfo{author}{\bibnamefont{Saha}, \bibfnamefont{A.}},
\bibinfo{author}{\bibfnamefont{A.}~\bibnamefont{Sandage}},
\bibinfo{author}{\bibfnamefont{G.~A.} \bibnamefont{Tammann}},
\bibinfo{author}{\bibfnamefont{A.~E.} \bibnamefont{Dolphin}},
\bibinfo{author}{\bibfnamefont{J.}~\bibnamefont{Christensen}},
\bibinfo{author}{\bibfnamefont{N.}~\bibnamefont{Panagia}}, and
\bibinfo{author}{\bibfnamefont{F.~D.} \bibnamefont{Machetto}}, \bibinfo{year}{2001}, \bibinfo{journal}{Astrophys.~J.} \textbf{\bibinfo{volume}{562}}, \bibinfo{pages}{314}.

\bibitem[{\citenamefont{Sahni and Starobinsky}(2000)\citenamefont{Sahni and Starobinsky}}] {Sahni:2000}
\bibinfo{author}{\bibnamefont{Sahni}, \bibfnamefont{V.}} and
\bibinfo{author}{\bibfnamefont{A.}~\bibnamefont{Starobinsky}}, \bibinfo{year}{2000}, \bibinfo{journal}{Int.~J.\ Mod.\ Phys.~D} \textbf{\bibinfo{volume}{9}}, \bibinfo{pages}{373}.

\bibitem[{\citenamefont{Sahu} \emph{et~al.}(1997)\citenamefont{Sahu \emph{et~al.}}}] {Sahu:1997}
\bibinfo{author}{\bibnamefont{Sahu}, \bibfnamefont{K.~C.}},
\bibinfo{author}{\bibfnamefont{M.}~{Livio}}, 
\bibinfo{author}{\bibfnamefont{L.}~{Petro}}, 
\bibinfo{author}{\bibfnamefont{F.~D.} {Macchetto}}, 
\bibinfo{author}{\bibfnamefont{J.}~{van Paradijs}}, 
\bibinfo{author}{\bibfnamefont{C.}~{Kouveliotou}}, 
\bibinfo{author}{\bibfnamefont{G.~J.} {Fishman}}, 
\bibinfo{author}{\bibfnamefont{C.~A.} {Meegan}}, 
\bibinfo{author}{\bibfnamefont{P.~J.} {Groot}}, and
\bibinfo{author}{\bibfnamefont{T.~J.} {Calama}}, 
\bibinfo{year}{1997}, \bibinfo{journal}{Nature} \textbf{\bibinfo{volume}{387}}, \bibinfo{pages}{476}.

\bibitem[{\citenamefont{Salamanca} \emph{et~al.}(2002)\citenamefont{Salamanca \emph{et~al.}}}] {Salamanca:2002}
\bibinfo{author}{\bibnamefont{Salamanca},~\bibfnamefont{I.}},
\bibinfo{author}{\bibfnamefont{E.}~\bibnamefont{Rol}}, 
\bibinfo{author}{\bibfnamefont{N.}~\bibnamefont{Tanvir}}, and 
\bibinfo{author}{\bibfnamefont{L.}~\bibnamefont{Kaper}}, \bibinfo{year}{2002}, \bibinfo{journal}{GCN 1443}.

\bibitem[{\citenamefont{Salaris and Weiss}(1998)\citenamefont{Salaris and Weiss}}] {Salaris:1998}
\bibinfo{author}{\bibnamefont{Salaris}, \bibfnamefont{M.}} and
\bibinfo{author}{\bibfnamefont{A.}~\bibnamefont{Weiss}}, \bibinfo{year}{1998}, \bibinfo{journal}{Astron.\ Astrophys.} \textbf{\bibinfo{volume}{335}}, \bibinfo{pages}{943}.

\bibitem[{\citenamefont{Sandage}(1958)\citenamefont{Sandage}}]{Sandage:1958}
\bibinfo{author}{\bibnamefont{Sandage}, \bibfnamefont{A.}}, \bibinfo{year}{1958}, \bibinfo{journal}{Astrophys.~J.} \textbf{\bibinfo{volume}{127}}, \bibinfo{pages}{513}.

\bibitem[{\citenamefont{Sargent} \emph{et~al.}(1978)\citenamefont{Sargent \emph{et~al.}}}] {Sargent:1978}
\bibinfo{author}{\bibnamefont{Sargent}, \bibfnamefont{W.~L.~W.}},
\bibinfo{author}{\bibfnamefont{P.~J.} \bibnamefont{Young}}, 
\bibinfo{author}{\bibfnamefont{A.}~\bibnamefont{Boksenberg}}, 
\bibinfo{author}{\bibfnamefont{K.}~\bibnamefont{Shortridge}}, 
\bibinfo{author}{\bibfnamefont{C.~R.} \bibnamefont{Lynds}}, and
\bibinfo{author}{\bibfnamefont{F.~D.~A.}, \bibnamefont{Hartwick}}, \bibinfo{year}{1978}, \bibinfo{journal}{Astrophys.~J.} \textbf{\bibinfo{volume}{221}}, \bibinfo{pages}{731}.

\bibitem[{\citenamefont{Schecter}(2000)\citenamefont{Schecter}}] {Schecter:2000}
\bibinfo{author}{\bibnamefont{Schecter}, \bibfnamefont{P.~L.}},
\bibinfo{year}{2000}, \eprint{astro-ph/0009048}.

\bibitem[{\citenamefont{Schmidt} \emph{et~al.}(1994)\citenamefont{Schmidt \emph{et~al.}}}]{Schmidt:1994}
\bibinfo{author}{\bibnamefont{Schmidt}, \bibfnamefont{B.~P.}}, 
\bibinfo{author}{\bibfnamefont{R.~P.} \bibnamefont{Kirshner}},
\bibinfo{author}{\bibfnamefont{R.~G.} \bibnamefont{Eastman}},
\bibinfo{author}{\bibfnamefont{M.~M.} \bibnamefont{Phillips}},
\bibinfo{author}{\bibfnamefont{N.~B.} \bibnamefont{Suntzeff}},
\bibinfo{author}{\bibfnamefont{M.}~\bibnamefont{Hamuy}},
\bibinfo{author}{\bibfnamefont{J.}~\bibnamefont{Maza}}, and
\bibinfo{author}{\bibfnamefont{R.}~\bibnamefont{Avil\'es}}, \bibinfo{year}{1994}, \bibinfo{journal}{Astrophys.~J.}, \textbf{\bibinfo{volume}{432}}, \bibinfo{pages}{42}.

\bibitem[{\citenamefont{Schneider \emph{et~al.}}(1999)\citenamefont{Schneider \emph{et~al.}}}]{Schneider:1999}
\bibinfo{author}{\bibnamefont{Schneider}, \bibfnamefont{G.}}, 
\bibinfo{author}{\bibfnamefont{B.~A.} \bibnamefont{Smith}},
\bibinfo{author}{\bibfnamefont{E.~E.} \bibnamefont{Becklin}},
\bibinfo{author}{\bibfnamefont{D.~W.} \bibnamefont{Koerner}},
\bibinfo{author}{\bibfnamefont{R.}~\bibnamefont{Meier}},
\bibinfo{author}{\bibfnamefont{D.~C.} \bibnamefont{Hines}},
\bibinfo{author}{\bibfnamefont{P.~J.} \bibnamefont{Lowrance}},
\bibinfo{author}{\bibfnamefont{R.~J.} \bibnamefont{Terrile}},
\bibinfo{author}{\bibfnamefont{R.~I.} \bibnamefont{Thompson}}, and
\bibinfo{author}{\bibfnamefont{M.}~\bibnamefont{Rieke}}, \bibinfo{year}{1999}, \bibinfo{journal}{Astrophys.~J.\ Lett.} \textbf{\bibinfo{volume}{513}}, \bibinfo{pages}{L127}.

\bibitem[{\citenamefont{Seager and Sasselov}(2000)\citenamefont{Seager and Sasselov}}] {Seager:2000}
\bibinfo{author}{\bibnamefont{Seager}, \bibfnamefont{S.}} and
\bibinfo{author}{\bibfnamefont{D.~D.}~\bibnamefont{Sasselov}}, \bibinfo{year}{2000}, \bibinfo{journal}{Astrophys.~J.} \textbf{\bibinfo{volume}{537}}, \bibinfo{pages}{916}.

\bibitem[{\citenamefont{Sekanina}  \emph{et~al.}(1994)\citenamefont{Sekanina, Chodas, and Yeomans}}]{Sekanina:1994}
\bibinfo{author}{\bibnamefont{Sekanina}, \bibfnamefont{Z.}},
\bibinfo{author}\bibfnamefont{P.~W.} {\bibnamefont{Chodas}}, and
\bibinfo{author}{\bibfnamefont{D.~K.} \bibnamefont{Yeomans}}, \bibinfo{year}{1994},
\bibinfo{journal}{Astron.\ Astrophys.} \textbf{\bibinfo{volume}{304}},
\bibinfo{pages}{296}.

\bibitem[{\citenamefont{Selwood}(2000)}]{Selwood:2000}
\bibinfo{author}{\bibnamefont{Selwood}, \bibfnamefont{J.~A.}},
  \bibinfo{year}{2000}, in \emph{\bibinfo{booktitle}{Dynamics of Galaxies:  from the Early Universe to the Present}}, ASP Conf.\ Ser.~197, edited by
  \bibinfo{editor}{\bibfnamefont{F.}~\bibnamefont{Combes}},
  \bibinfo{editor}{\bibfnamefont{G.~A.}~\bibnamefont{Mamon}}, and
  \bibinfo{ediror}{\bibfnamefont{V.}~\bibnamefont{Charmandaris}}
  (\bibinfo{publisher}{ASP, San Francisco}), p.~\bibinfo{pages}{3}.

\bibitem[{\citenamefont{Shoemaker} \emph{et~al.}(1993)\citenamefont{Shoemaker, Shoemaker, and Levy}}]{Shoemaker:1993}
\bibinfo{author}{\bibnamefont{Shoemaker}, \bibfnamefont{C.~S.}},
\bibinfo{author}{\bibfnamefont{E.~M.}~\bibnamefont{Shoemaker}},  and
\bibinfo{author}{\bibfnamefont{D.}~\bibnamefont{Levy}}, \bibinfo{year}{1993},
\bibinfo{journal}{IAU Circular} \textbf{\bibinfo{volume}{5725}}.

\bibitem[{\citenamefont{Shu} \emph{et~al.}(1993)\citenamefont{Shu, Johnstone, and Hollenbach}}]{Shu:1993}
\bibinfo{author}{\bibnamefont{Shu}, \bibfnamefont{F.~H.}}, 
\bibinfo{author}{\bibfnamefont{D.}~\bibnamefont{Johnstone}}, and
\bibinfo{author}{\bibfnamefont{D.}~\bibnamefont{Hollenbach}}, 
\bibinfo{year}{1993}, \bibinfo{journal}{Icarus} \textbf{\bibinfo{volume}{106}}, \bibinfo{pages}{92}.

\bibitem[{\citenamefont{Shu} \emph{et~al.}(1994)\citenamefont{Shu \emph{et~al.}}}] {Shu:1994}
\bibinfo{author}{\bibnamefont{Shu}, \bibfnamefont{F.~H.}},
\bibinfo{author}{\bibfnamefont{J.}~\bibnamefont{Najita}}, 
\bibinfo{author}{\bibfnamefont{E.}~\bibnamefont{Ostriker}},
\bibinfo{author}{\bibfnamefont{F.}~\bibnamefont{Wilkin}},
\bibinfo{author}{\bibfnamefont{S.}~\bibnamefont{Ruden}}, and
\bibinfo{author}{\bibfnamefont{S.}~\bibnamefont{Lizano}}, \bibinfo{year}{1994}, \bibinfo{journal}{Astrophys.~J.} \textbf{\bibinfo{volume}{429}}, \bibinfo{pages}{781}.

\bibitem[{\citenamefont{Silk and Rees}(1998)\citenamefont{Silk and Rees}}] {Silk:1998}
\bibinfo{author}{\bibnamefont{Silk}, \bibfnamefont{J.}} and
\bibinfo{author}{\bibfnamefont{M.~J.}~\bibnamefont{Rees}}, \bibinfo{year}{1998}, \bibinfo{journal}{Astron.\ Astrophys.} \textbf{\bibinfo{volume}{331}}, \bibinfo{pages}{L1}.

\bibitem[{\citenamefont{Slipher}(1917)\citenamefont{Slipher}}] {Slipher:1917}
\bibinfo{author}{\bibnamefont{Slipher}, \bibfnamefont{V.~M.}},
\bibinfo{year}{1917}, \bibinfo{journal}{Proc.\ Amer.\ Phil.\ Soc.} \textbf{\bibinfo{volume}{56}}, \bibinfo{pages}{403}.

\bibitem[{\citenamefont{Soker and Livio}(1989)\citenamefont{Soker and Livio}}] {Soker:1989}
\bibinfo{author}{\bibnamefont{Soker}, \bibfnamefont{N.}} and
\bibinfo{author}{\bibfnamefont{M.}~\bibnamefont{Livio}}, \bibinfo{year}{1989}, \bibinfo{journal}{Astrophys.~J.} \textbf{\bibinfo{volume}{339}}, \bibinfo{pages}{268}.

\bibitem[{\citenamefont{Soker and Rappaport}(2001)\citenamefont{Soker and Rappaport}}] {Soker:2001}
\bibinfo{author}{\bibnamefont{Soker}, \bibfnamefont{N.}} and
\bibinfo{author}{\bibfnamefont{S.}~\bibnamefont{Rappaport}}, \bibinfo{year}{2001}, \bibinfo{journal}{Astrophys.~J.} \textbf{\bibinfo{volume}{557}}, \bibinfo{pages}{256}.

\bibitem[{\citenamefont{Somerville and Livio}(2003)\citenamefont{Somerville and Livio}}] {Somerville:2003}
\bibinfo{author}{\bibnamefont{Somerville}, \bibfnamefont{R.~S.}} and
\bibinfo{author}{\bibfnamefont{M.}~\bibnamefont{Livio}}, \bibinfo{year}{2003}, \bibinfo{journal}{Astrophys.~J.} (in press); \eprint{astro-ph/0303017}.

\bibitem[\citenamefont{Somerville, Primak, and Faber}(2001)]{Somerville:2001}
\bibinfo{author}{\bibnamefont{Somerville}, \bibfnamefont{R.~S.}}, 
\bibinfo{author}{\bibfnamefont{J.~R.} \bibnamefont{Primak}}, and
\bibinfo{author}{\bibfnamefont{S.~M.} \bibnamefont{Faber}}, \bibinfo{year}{2001}, \bibinfo{journal}{Mon.\ Not.~R.\ Astron.\ Soc.} \textbf{\bibinfo{volume}{320}}, \bibinfo{pages}{504}.

\bibitem[{\citenamefont{Speck} \emph{et~al.}(2002)\citenamefont{Speck \emph{et~al.}}}]{Speck:2002}
\bibinfo{author}{\bibnamefont{Speck}, \bibfnamefont{A.~K.}},
\bibinfo{author}{\bibfnamefont{M.}~\bibnamefont{Meixner}}, 
\bibinfo{author}{\bibfnamefont{D.}~\bibnamefont{Fong}}, 
\bibinfo{author}{\bibfnamefont{P.~R.} \bibnamefont{McCullough}}, 
\bibinfo{author}{\bibfnamefont{D.~E.} \bibnamefont{Moser}}, and
\bibinfo{author}{\bibfnamefont{T.}~\bibnamefont{Veta}}, \bibinfo{year}{2002}, \bibinfo{journal}{Astron.~J.} \textbf{\bibinfo{volume}{123}}, \bibinfo{pages}{346}.

\bibitem[{\citenamefont{Spergel and Steinhardt}(2000)\citenamefont{Spergel and Steinhardt}}] {Spergel:2000}
\bibinfo{author}{\bibnamefont{Spergel}, \bibfnamefont{D.~N.}} and
\bibinfo{author}{\bibfnamefont{P.~J.}~\bibnamefont{Steinhardt}}, \bibinfo{year}{2000}, \bibinfo{journal}{Phys.\ Rev.\ Lett.} \textbf{\bibinfo{volume}{84}}, \bibinfo{pages}{3760}.

\bibitem[{\citenamefont{Spergel} \emph{et~al.}(2003)\citenamefont{Spergel \emph{et~al.}}}] {Spergel:2003}
\bibinfo{author}{\bibnamefont{Spergel}, \bibfnamefont{D.~N.}},
\bibinfo{author}{\bibfnamefont{L.}~\bibnamefont{Verde}},
\bibinfo{author}{\bibfnamefont{H.~V.}~\bibnamefont{Peiris, \emph{et~al.}}},
\bibinfo{year}{2003}, \bibinfo{journal}{Astrophys.~J.} (in press); \eprint{astro-ph/0302209}.

\bibitem[{\citenamefont{Spruit}(1996)}]{Spruit:1996}
\bibinfo{author}{\bibnamefont{Spruit}, \bibfnamefont{H.~C.}},
  \bibinfo{year}{1996}, in \emph{\bibinfo{booktitle}{Evolutionary Processes in Binary Stars}}, edited by
  \bibinfo{editor}{\bibfnamefont{R.}~\bibnamefont{Wijers}},
  \bibinfo{editor}{\bibfnamefont{M.}~\bibnamefont{Davies}}, and
  \bibinfo{ediror}{\bibfnamefont{C.}~\bibnamefont{Tout}}
  (\bibinfo{publisher}{Cambridge University, Cambridge, England}), p.~\bibinfo{pages}{249}.

\bibitem[{\citenamefont{Stanway, Bunker, and McMahon}(2003)}] {Stanway:2003}
\bibinfo{author}{\bibnamefont{Stanway}, \bibfnamefont{E.~R.}},
\bibinfo{author}{\bibfnamefont{A.~J.} \bibnamefont{Bunker}}, and
\bibinfo{author}{\bibfnamefont{R.~G.} \bibnamefont{McMahon}}, \bibinfo{year}{2003}, \bibinfo{journal}{Mon.\ Not.~R.\ Astron.\ Soc.} (in press); \eprint{astro-ph/0302212}.

\bibitem[{\citenamefont{Steidel} \emph{et~al.}(1999)\citenamefont{Steidel \emph{et~al.}}}] {Steidel:1999}
\bibinfo{author}{\bibnamefont{Steidel}, \bibfnamefont{C.~C.}},
\bibinfo{author}{\bibfnamefont{K.~L.} \bibnamefont{Adelberger}},
\bibinfo{author}{\bibfnamefont{M.}~\bibnamefont{Giavalisco}},
\bibinfo{author}{\bibfnamefont{M.}~\bibnamefont{Dickinson}}, and 
\bibinfo{author}{\bibfnamefont{M.}~\bibnamefont{Pettini}}, \bibinfo{year}{1999}, \bibinfo{journal}{Astrophys.~J.} \textbf{\bibinfo{volume}{519}}, \bibinfo{pages}{1}.

\bibitem[{\citenamefont{Steinhardt}(1983)\citenamefont{Steinhardt}}]{Steinhardt:1983}
\bibinfo{author}{\bibnamefont{Steinhardt}, \bibfnamefont{P.~J.}}, \bibinfo{year}{1983},  in \emph{\bibinfo{booktitle}{The Very Early Universe}}, edited by
  \bibinfo{editor}{\bibfnamefont{G.~W.} \bibnamefont{Gibbons}},
  \bibinfo{editor}{\bibfnamefont{S.}~\bibnamefont{Hawking}}, and
  \bibinfo{editor}{\bibfnamefont{S.~T.~C.} \bibnamefont{Siklos}}
  (\bibinfo{publisher}{Cambridge University, Cambridge, England}), p.~\bibinfo{pages}{251}.

\bibitem[{\citenamefont{Strauss and Willick}(1995)\citenamefont{Strauss and Willick}}] {Strauss:1995}
\bibinfo{author}{\bibnamefont{Strauss}, \bibfnamefont{M.~A.}} and
\bibinfo{author}{\bibfnamefont{J.~A.} \bibnamefont{Willick}}, \bibinfo{year}{1995}, \bibinfo{journal}{Phys.\ Rep.} \textbf{\bibinfo{volume}{261}}, \bibinfo{pages}{271}.

\bibitem[{\citenamefont{Sunyaev and Zeldovich}(1970)\citenamefont{Sunyaev and Zeldovich}}] {Sunyaev:1970}
\bibinfo{author}{\bibnamefont{Sunyaev}, \bibfnamefont{R.~A.}} and
\bibinfo{author}{\bibfnamefont{Y.~B.}~\bibnamefont{Zeldovich}}, \bibinfo{year}{1970}, \bibinfo{journal}{Comments Astrophys.\ Space Phys.} \textbf{\bibinfo{volume}{2}}, \bibinfo{pages}{66}.

\bibitem[{\citenamefont{Tammann}(1979)}]{Tammann:1979}
\bibinfo{author}{\bibnamefont{Tammann}, \bibfnamefont{G.~A.}},
  \bibinfo{year}{1979}, in \emph{\bibinfo{booktitle}{ESA/ESO Workshop on Astronomical Uses of the Space Telescope}}, edited by
  \bibinfo{editor}{\bibfnamefont{F.}~\bibnamefont{Macchetto}},
\bibinfo{editor}{\bibfnamefont{F.}~\bibnamefont{Pacini}}, and
\bibinfo{editor}{\bibfnamefont{M.}~\bibnamefont{Tarenghi}}
  (\bibinfo{publisher}{ESO, Geneva}), p.~\bibinfo{pages}{229}.

\bibitem[{\citenamefont{Terman, Taam, and Hernquist}(1994)\citenamefont{Terman \emph{et~al.}}}] {Terman:1994}
\bibinfo{author}{\bibnamefont{Terman}, \bibfnamefont{J.~L.}},
\bibinfo{author}{\bibfnamefont{R.~E.} \bibnamefont{Taam}}, and
\bibinfo{author}{\bibfnamefont{L.}~\bibnamefont{Hernquist}}, \bibinfo{year}{1994}, \bibinfo{journal}{Astrophys.~J.} \textbf{\bibinfo{volume}{422}}, \bibinfo{pages}{729}.

\bibitem[{\citenamefont{Terman, Taam, and Hernquist}(1995)\citenamefont{Terman \emph{et~al.}}}] {Terman:1995}
\bibinfo{author}{\bibnamefont{Terman}, \bibfnamefont{J.~L.}},
\bibinfo{author}{\bibfnamefont{R.~E.} \bibnamefont{Taam}}, and
\bibinfo{author}{\bibfnamefont{L.}~\bibnamefont{Hernquist}}, \bibinfo{year}{1995}, \bibinfo{journal}{Astrophys.~J.} \textbf{\bibinfo{volume}{445}}, \bibinfo{pages}{367}.

\bibitem[{\citenamefont{Thompson} \emph{et~al.}(1997)\citenamefont{Thompson \emph{et~al.}}}] {Thompson:1997}
\bibinfo{author}{\bibnamefont{Thompson}, \bibfnamefont{R.}},
\bibinfo{author}{\bibfnamefont{M.}~\bibnamefont{Rieke}},
\bibinfo{author}{\bibfnamefont{G.}~\bibnamefont{Schneider}},
\bibinfo{author}{\bibfnamefont{D.}~\bibnamefont{Hines}},
\bibinfo{author}{\bibfnamefont{R.}~\bibnamefont{Sahai, \emph{et al.}}},
\bibinfo{year}{1997}, \bibinfo{journal}{ST ScI Press Release}.

\bibitem[{\citenamefont{Thompson} \emph{et~al.}(1999)\citenamefont{Thompson \emph{et~al.}}}] {Thompson:1999}
\bibinfo{author}{\bibnamefont{Thompson}, \bibfnamefont{R.~I.}},
\bibinfo{author}{\bibfnamefont{L.~J.} \bibnamefont{Storrie-Lombardi}}, 
\bibinfo{author}{\bibfnamefont{P.~J.} \bibnamefont{Weymann}}, 
\bibinfo{author}{\bibfnamefont{M.~J.} \bibnamefont{Rieke}},
\bibinfo{author}{\bibfnamefont{G.}~\bibnamefont{Schneider}},
\bibinfo{author}{\bibfnamefont{E.}~\bibnamefont{Stobie}}, and
\bibinfo{author}{\bibfnamefont{D.}~\bibnamefont{Lytle}}, \bibinfo{year}{1999}, \bibinfo{journal}{Astron.~J.} \textbf{\bibinfo{volume}{117}}, \bibinfo{pages}{17}.

\bibitem[{\citenamefont{Throop}(2000)\citenamefont{Throop}}]{Throop:2000}
\bibinfo{author}{\bibnamefont{Throop}, \bibfnamefont{H.}}, \bibinfo{year}{2000}, Ph.D.\ thesis (University of Colorado, Boulder).

\bibitem[{\citenamefont{Tinsley and Gunn}(1976)\citenamefont{Tinsley and Gunn}}] {Tinsley:1976}
\bibinfo{author}{\bibnamefont{Tinsley}, \bibfnamefont{B.~M.}} and
\bibinfo{author}{\bibfnamefont{J.}~\bibnamefont{Gunn}}, \bibinfo{year}{1976}, \bibinfo{journal}{Astrophys.~J.} \textbf{\bibinfo{volume}{302}}, \bibinfo{pages}{52}.

\bibitem[{\citenamefont{Tonry} \emph{et~al.}(1997)\citenamefont{Tonry \emph{et~al.}}}] {Tonry:1997}
\bibinfo{author}{\bibnamefont{Tonry}, \bibfnamefont{J.~L.}},
\bibinfo{author}{\bibfnamefont{J.~P.} \bibnamefont{Blakeslee}},
\bibinfo{author}{\bibfnamefont{E.~A.} \bibnamefont{Aghar}}, and
\bibinfo{author}{\bibfnamefont{A.}~\bibnamefont{Dressler}}, \bibinfo{year}{1997}, \bibinfo{journal}{Astrophys.~J.} \textbf{\bibinfo{volume}{475}}, \bibinfo{pages}{399}.

\bibitem[{\citenamefont{Tonry} \emph{et~al.}(2000)\citenamefont{Tonry \emph{et~al.}}}] {Tonry:2000}
\bibinfo{author}{\bibnamefont{Tonry}, \bibfnamefont{J.~L.}},
\bibinfo{author}{\bibfnamefont{J.~P.} \bibnamefont{Blakeslee}},
\bibinfo{author}{\bibfnamefont{E.~A.} \bibnamefont{Aghar}}, and
\bibinfo{author}{\bibfnamefont{A.}~\bibnamefont{Dressler}}, \bibinfo{year}{2000}, \bibinfo{journal}{Astrophys.~J.} \textbf{\bibinfo{volume}{530}}, \bibinfo{pages}{625}.

\bibitem[{\citenamefont{Tonry and Schneider}(1988)\citenamefont{Tonry and Schneider}}] {Tonry:1988}
\bibinfo{author}{\bibnamefont{Tonry}, \bibfnamefont{J.~L.}} and
\bibinfo{author}{\bibfnamefont{D.~P.} \bibnamefont{Schneider}}, \bibinfo{year}{1988}, \bibinfo{journal}{Astron.~J.} \textbf{\bibinfo{volume}{96}}, \bibinfo{pages}{807}.

\bibitem[{\citenamefont{Tremaine} \emph{et~al.}(2002)\citenamefont{Tremaine \emph{et~al.}}}] {Tremaine:2002}
\bibinfo{author}{\bibnamefont{Tremaine}, \bibfnamefont{S.}},
\bibinfo{author}{\bibfnamefont{K.}~\bibnamefont{Gebhardt}}, 
\bibinfo{author}{\bibfnamefont{R.}~\bibnamefont{Bender, \emph{et~al.}}}, \bibinfo{year}{2002}, \bibinfo{journal}{Astrophys.~J.} \textbf{\bibinfo{volume}{574}}, \bibinfo{pages}{740}.

\bibitem[{\citenamefont{Treu and Stiavelli}(1999)\citenamefont{Treu and Stiavelli}}] {Treu:1999}
\bibinfo{author}{\bibnamefont{Treu}, \bibfnamefont{T.}} and
\bibinfo{author}{\bibfnamefont{M.}~\bibnamefont{Stiavelli}}, \bibinfo{year}{1999}, \bibinfo{journal}{Astrophys.~J.\ Lett.} \textbf{\bibinfo{volume}{524}}, \bibinfo{pages}{L27}.

\bibitem[{\citenamefont{Treu} \emph{et~al.}(2001)\citenamefont{Treu \emph{et~al.}}}]{Treu:2001}
\bibinfo{author}{\bibnamefont{Treu}, \bibfnamefont{T.}},
\bibinfo{author}{\bibfnamefont{M.}~\bibnamefont{Stiavelli}},
\bibinfo{author}{\bibfnamefont{G.}~\bibnamefont{Bertin}},
\bibinfo{author}{\bibfnamefont{S.}~\bibfnamefont{Casertano}}, and
\bibinfo{author}{\bibfnamefont{P.}~\bibnamefont{M{\o}ller}}, \bibinfo{year}{2001}, \bibinfo{journal}{Mon.\ Not.~R.\ Astron.\ Soc.} \textbf{\bibinfo{volume}{326}}, \bibinfo{pages}{237}.

\bibitem[{\citenamefont{Trimble}(1997)\citenamefont{Trimble}}]{Trimble:1997}
\bibinfo{author}{\bibnamefont{Trimble}, \bibfnamefont{V.}}, \bibinfo{year}{1997}, \bibinfo{journal}{Space Sci.\ Rev.} \textbf{\bibinfo{volume}{79}}, \bibinfo{pages}{793}.

\bibitem[{\citenamefont{Tully and Fisher}(1977)\citenamefont{Tully and Fisher}}] {Tully:1977}
\bibinfo{author}{\bibnamefont{Tully}, \bibfnamefont{R.~B.}} and
\bibinfo{author}{\bibfnamefont{J.~R.} \bibnamefont{Fisher}}, \bibinfo{year}{1977}, \bibinfo{journal}{Astron.\ Astrophys.} \textbf{\bibinfo{volume}{54}}, \bibinfo{pages}{661}.

\bibitem[{\citenamefont{Turner}(2000)\citenamefont{Turner}}]{Turner:2000}
\bibinfo{author}{\bibnamefont{Turner}, \bibfnamefont{M.~S.}}, \bibinfo{year}{2000}, \bibinfo{journal}{Phys.\ Rep.} \textbf{\bibinfo{volume}{334}}, \bibinfo{pages}{619}.

\bibitem[{\citenamefont{Turner and White}(1997)\citenamefont{Turner and White}}] {Turner:1997}
\bibinfo{author}{\bibnamefont{Turner}, \bibfnamefont{M.~S.}} and
\bibinfo{author}{\bibfnamefont{M.}~\bibnamefont{White}}, \bibinfo{year}{1997}, \bibinfo{journal}{Phys.\ Rev.~D} \textbf{\bibinfo{volume}{56}}, \bibinfo{pages}{R4439}.

\bibitem[{\citenamefont{Umeda} \emph{et~al.}(1999)\citenamefont{Umeda \emph{et~al.}}}] {Umeda:1999}
\bibinfo{author}{\bibnamefont{Umeda},~\bibfnamefont{H.}},
\bibinfo{author}{\bibfnamefont{K.}~\bibnamefont{Nomoto}},
\bibinfo{author}{\bibfnamefont{C.}~\bibnamefont{Kobayashi}}, 
\bibinfo{author}{\bibfnamefont{I.}~\bibnamefont{Hachisu}}, and
\bibinfo{author}{\bibfnamefont{M.}~\bibnamefont{Kato}}, \bibinfo{year}{1999}, \bibinfo{journal}{Astrophys.~J.\ Lett.} \textbf{\bibinfo{volume}{522}}, \bibinfo{pages}{L43}.

\bibitem[{\citenamefont{van~den Bergh}(1994)\citenamefont{van~den Bergh}}]{vandenBergh:1994}
\bibinfo{author}{\bibnamefont{van~den Bergh},~\bibfnamefont{S.}}, \bibinfo{year}{1994}, \bibinfo{journal}{Astrophys.~J.\ Suppl.\ Ser.} \textbf{\bibinfo{volume}{92}}, \bibinfo{pages}{219}.

\bibitem[{\citenamefont{van~den Bergh} \emph{et~al.}(2002)\citenamefont{van~den Bergh \emph{et~al.}}}] {Bergh:2002}\bibinfo{author}{\bibnamefont{van~den Bergh}, \bibfnamefont{S.}},
\bibinfo{author}{\bibfnamefont{R.~G.} \bibnamefont{Abraham}}, 
\bibinfo{author}{\bibfnamefont{L.~F.} \bibnamefont{Whyte}}, 
\bibinfo{author}{\bibfnamefont{M.~R.} \bibnamefont{Merrifield}}, 
\bibinfo{author}{\bibfnamefont{P.~B.} \bibnamefont{Eskridge}}, 
\bibinfo{author}{\bibfnamefont{J.~A.} \bibnamefont{Frogel}}, and
\bibinfo{author}{\bibfnamefont{R.}~\bibnamefont{Pogge}}, \bibinfo{year}{2002}, \bibinfo{journal}{Astron.~J.} \textbf{\bibinfo{volume}{123}}, \bibinfo{pages}{2913}.

\bibitem[{\citenamefont{van~der Marel}(1994)\citenamefont{van~der Marel}}]{Marel:1994}
\bibinfo{author}{\bibnamefont{van~der Marel}, \bibfnamefont{R.}}, \bibinfo{year}{1994}, \bibinfo{journal}{Mon.\ Not.~R.\ Astron.\ Soc.} \textbf{\bibinfo{volume}{270}}, \bibinfo{pages}{271}.

\bibitem[{\citenamefont{van~der Marel} \emph{et~al.}(2002)\citenamefont{van~der Marel \emph{et~al.}}}] {Marel:2002}\bibinfo{author}{\bibnamefont{van~der Marel}, \bibfnamefont{R.~P.}},
\bibinfo{author}{\bibfnamefont{J.}~\bibnamefont{Gerssin}}, 
\bibinfo{author}{\bibfnamefont{P.}~\bibnamefont{Guhathakurta}}, 
\bibinfo{author}{\bibfnamefont{R.~C.} \bibnamefont{Peterson}}, and
\bibinfo{author}{\bibfnamefont{K.}~\bibnamefont{Gebhardt}}, \bibinfo{year}{2002}, \bibinfo{journal}{Astron.~J.} \textbf{\bibinfo{volume}{124}}, \bibinfo{pages}{3255}.

\bibitem[{\citenamefont{van Paradijs} \emph{et~al.}(1997)\citenamefont{van Paradijs \emph{et~al.}}}] {vanParadijs:1997}
\bibinfo{author}{\bibnamefont{van Paradijs},~\bibfnamefont{J.}},
\bibinfo{author}{\bibfnamefont{P.}~\bibnamefont{Groot}},
\bibinfo{author}{\bibfnamefont{T.}~\bibnamefont{Galama, \emph{et~al.}}},
\bibinfo{year}{1997}, \bibinfo{journal}{Nature} \textbf{\bibinfo{volume}{386}}, \bibinfo{pages}{686}.

\bibitem[{\citenamefont{Vidal-Madjar} \emph{et~al.}(2003)\citenamefont{Vidal-Madjar \emph{et~al.}}}] {Vidal:2003}\bibinfo{author}{\bibnamefont{Vidal-Madjar}, \bibfnamefont{A.}},
\bibinfo{author}{\bibfnamefont{A.}~\bibnamefont{Lecavelier des Etangs}}, 
\bibinfo{author}{\bibfnamefont{J.-M.} \bibnamefont{D\'esert}}, 
\bibinfo{author}{\bibfnamefont{G.~E.} \bibnamefont{Ballester}}, 
\bibinfo{author}{\bibfnamefont{R.}~\bibnamefont{Ferlet}}, 
\bibinfo{author}{\bibfnamefont{G.}~\bibnamefont{H\'ebrard}}, and
\bibinfo{author}{\bibfnamefont{M.}~\bibnamefont{Mayor}}, 
\bibinfo{year}{2003}, \bibinfo{journal}{Nature} \textbf{\bibinfo{volume}{422}}, \bibinfo{pages}{143}.

\bibitem[{\citenamefont{Vilenkin}(1983)\citenamefont{Vilenkin}}]{Vilenkin:1983}
\bibinfo{author}{\bibnamefont{Vilenkin}, \bibfnamefont{A.}}, \bibinfo{year}{1983}, \bibinfo{journal}{Phys.\ Rev.~D} \textbf{\bibinfo{volume}{27}}, \bibinfo{pages}{2848}.

\bibitem[{\citenamefont{Vilenkin}(1995)\citenamefont{Vilenkin}}]{Vilenkin:1995}
\bibinfo{author}{\bibnamefont{Vilenkin}, \bibfnamefont{A.}}, \bibinfo{year}{1995}, \bibinfo{journal}{Phys.\ Rev.\ Lett.} \textbf{\bibinfo{volume}{74}}, \bibinfo{pages}{846}.

\bibitem[{\citenamefont{Vilenkin}(1998)\citenamefont{Vilenkin}}]{Vilenkin:1998}
\bibinfo{author}{\bibnamefont{Vilenkin}, \bibfnamefont{A.}}, \bibinfo{year}{1998}, \bibinfo{journal}{Phys.\ Rev.\ Lett.} \textbf{\bibinfo{volume}{81}}, \bibinfo{pages}{5501}.

\bibitem[{\citenamefont{Vishniac}(1983)\citenamefont{Vishniac}}]{Vishniac:1983}
\bibinfo{author}{\bibnamefont{Vishniac}, \bibfnamefont{E.~T.}}, \bibinfo{year}{1983}, \bibinfo{journal}{Astrophys.~J.} \textbf{\bibinfo{volume}{274}}, \bibinfo{pages}{152}.

\bibitem[{\citenamefont{Wagoner}(1977)\citenamefont{Wagoner}}]{Wagoner:1977}
\bibinfo{author}{\bibnamefont{Wagoner}, \bibfnamefont{R.~V.}}, \bibinfo{year}{1977}, \bibinfo{journal}{Astrophys.~J.\ Lett.} \textbf{\bibinfo{volume}{214}}, \bibinfo{pages}{L5}.

\bibitem[{\citenamefont{Weaver} \emph{et~al.}(1995)\citenamefont{Weaver \emph{et~al.}}}] {Weaver:1995}
\bibinfo{author}{\bibnamefont{Weaver}, \bibfnamefont{H.~A.}},
\bibinfo{author}{\bibfnamefont{M.~F.} \bibnamefont{A'Hearn}},
\bibinfo{author}{\bibfnamefont{C.}~\bibnamefont{Arpigny, \emph{et~al.}}}, 
\bibinfo{year}{1995}, \bibinfo{journal}{Science} \textbf{\bibinfo{volume}{267}}, \bibinfo{pages}{1282}.

\bibitem[{\citenamefont{Weaver} \emph{et~al.}(1994)\citenamefont{Weaver \emph{et~al.}}}] {Weaver:1994}
\bibinfo{author}{\bibnamefont{Weaver}, \bibfnamefont{H.~A.}},
\bibinfo{author}{\bibfnamefont{P.~D.} \bibnamefont{Feldman}},
\bibinfo{author}{\bibfnamefont{M.~F.} \bibnamefont{A'Hearn, \emph{et~al.}}},
\bibinfo{year}{1994}, \bibinfo{journal}{Science} \textbf{\bibinfo{volume}{263}}, \bibinfo{pages}{787}.

\bibitem[{\citenamefont{Webbink}(1984)\citenamefont{Webbink}}]{Webbink:1984}
\bibinfo{author}{\bibnamefont{Webbink}, \bibfnamefont{R.~F.}}, \bibinfo{year}{1984}, \bibinfo{journal}{Astrophys.~J.} \textbf{\bibinfo{volume}{227}}, \bibinfo{pages}{355}.

\bibitem[{\citenamefont{Weinberg}(2001)\citenamefont{Weinberg}}] {Weinberg:2001}
\bibinfo{author}{\bibnamefont{Weinberg}, \bibfnamefont{S.}}, \bibinfo{year}{2001}, 
in \emph{\bibinfo{booktitle}{Sources and Detection of Dark Matter and Dark Energy in the Universe}}, 
edited by \bibinfo{editor}{\bibfnamefont{D.~B.} \bibnamefont{Cline}}, 
  (\bibinfo{publisher}{Springer, Berlin}), p.~\bibinfo{pages}{18}.

\bibitem[{\citenamefont{Whelan and Iben}(1973)\citenamefont{Whelan and Iben}}]{Whelan:1973}
\bibinfo{author}{\bibnamefont{Whelan}, \bibfnamefont{J.}} and 
\bibinfo{author}{\bibfnamefont{I.}~\bibnamefont{Iben Jr.}}, \bibinfo{year}{1973}, \bibinfo{journal}{Astrophys.~J.} \textbf{\bibinfo{volume}{186}}, \bibinfo{pages}{1007}.

\bibitem[{\citenamefont{Windhorst} \emph{et~al.}(1995)\citenamefont{Windhorst \emph{et~al.}}}] {Windhorst:1995}
\bibinfo{author}{\bibnamefont{Windhorst}, \bibfnamefont{R.~A.}},
\bibinfo{author}{\bibfnamefont{S.~P.} \bibnamefont{Driver}}, 
\bibinfo{author}{\bibfnamefont{E.~J.} \bibnamefont{Ostrander, \emph{et~al.}}}, \bibinfo{year}{1995}, 
in \emph{\bibinfo{booktitle}{Galaxies in the Young Universe}}, edited by
  \bibinfo{editor}{\bibfnamefont{H.}~\bibnamefont{Hippelein}},
  (\bibinfo{publisher}{Springer, Berlin}), p.~\bibinfo{pages}{265}.

\bibitem[{\citenamefont{Woosley}(1993)\citenamefont{Woosley}}] {Woosley:1993}
\bibinfo{author}{\bibnamefont{Woosley}, \bibfnamefont{S.~E.}}, \bibinfo{year}{1993}, \bibinfo{journal}{Astrophys.~J.} \textbf{\bibinfo{volume}{405}}, \bibinfo{pages}{273}.

\bibitem[{\citenamefont{Wyckoff} \emph{et~al.}(1980)\citenamefont{Wyckoff \emph{et~al.}}}] {Wyckoff:1980}
\bibinfo{author}{\bibnamefont{Wyckoff}, \bibfnamefont{S.}},
\bibinfo{author}{\bibfnamefont{T.}~\bibnamefont{Gehren}},
\bibinfo{author}{\bibfnamefont{D.~C.} \bibnamefont{Morton}}, 
\bibinfo{author}{\bibfnamefont{R.}~\bibnamefont{Albrecht}}, 
\bibinfo{author}{\bibfnamefont{P.~A.} \bibnamefont{Wehinger}}, and
\bibinfo{author}{\bibfnamefont{A.}~\bibnamefont{Boksenberg}}, \bibinfo{year}{1980}, \bibinfo{journal}{Astrophys.~J.} \textbf{\bibinfo{volume}{242}}, \bibinfo{pages}{59}.

\bibitem[{\citenamefont{Yeomans and Chodas}(1993)\citenamefont{Yeomans and Chodas}}]{Yeomans:1993}
\bibinfo{author}{\bibnamefont{Yeomans}, \bibfnamefont{D.~K.}} and
\bibinfo{author}{\bibfnamefont{P.~W.}~\bibnamefont{Chodas}}, \bibinfo{year}{1993},
\bibinfo{journal}{IAU Circular} \textbf{\bibinfo{volume}{5807}}.

\bibitem[{\citenamefont{Yungelson and Livio}(2000)\citenamefont{Yungelson and Livio}}] {Yungelson:2000}
\bibinfo{author}{\bibnamefont{Yungelson}, \bibfnamefont{L.~R.}} and
\bibinfo{author}{\bibfnamefont{M.}~\bibnamefont{Livio}}, \bibinfo{year}{2000}, \bibinfo{journal}{Astrophys.~J.} \textbf{\bibinfo{volume}{528}}, \bibinfo{pages}{108}.

\bibitem[{\citenamefont{Zahnle}(1996)\citenamefont{Zahnle}}] {Zahnle:1996}
\bibinfo{author}{\bibnamefont{Zahnle}, \bibfnamefont{K.}}, \bibinfo{year}{1996}, 
  in \emph{\bibinfo{booktitle}{The Collision of Comet Shoemaker-Levy~9 and Jupiter}}, edited by
  \bibinfo{editor}{\bibfnamefont{K.~S.} \bibnamefont{Noll}},
\bibinfo{editor}{\bibfnamefont{H.~A.} \bibnamefont{Weaver}}, and
\bibinfo{editor}{\bibfnamefont{P.~D.} \bibnamefont{Feldman}},
  (\bibinfo{publisher}{Cambridge University, Cambridge, England}), p. \bibinfo{pages}{183}.

\bibitem[{\citenamefont{Zahnle and MacLow}(1994)\citenamefont{Zahnle and MacLow}}] {Zahnle:1994}
\bibinfo{author}{\bibnamefont{Zahnle}, \bibfnamefont{K.}} and
\bibinfo{author}{\bibfnamefont{M.-M.} \bibnamefont{MacLow}}, \bibinfo{year}{1994}, \bibinfo{journal}{Icarus} \textbf{\bibinfo{volume}{108}}, \bibinfo{pages}{1}.

\bibitem[{\citenamefont{Zeldovich and Novikov}(1964)\citenamefont{Zeldovich and Novikov}}] {Zeldovich:1964}
\bibinfo{author}{\bibnamefont{Zeldovich}, \bibfnamefont{Ya.~B.}} and
\bibinfo{author}{\bibfnamefont{I.~D.} \bibnamefont{Novikov}}, \bibinfo{year}{1964}, \bibinfo{journal}{Sov.\ Phys.\ Dokl.} \textbf{\bibinfo{volume}{9}}, \bibinfo{pages}{195}.

\bibitem[{\citenamefont{Zepf}(1997)\citenamefont{Zepf}}] {Zepf:1997}
\bibinfo{author}{\bibnamefont{Zepf}, \bibfnamefont{S.~E.}}, \bibinfo{year}{1997}, \bibinfo{journal}{Nature} \textbf{\bibinfo{volume}{390}}, \bibinfo{pages}{377}.

\bibitem[{\citenamefont{Zlatev, Wang, and Steinhardt}(1998)\citenamefont{Zlatev, Wang, and Steinhardt}}] {Zlatev:1998}
\bibinfo{author}{\bibnamefont{Zlatev}, \bibfnamefont{I.}},
\bibinfo{author}{\bibfnamefont{L.}~\bibnamefont{Wang}}, and
\bibinfo{author}{\bibfnamefont{P.~J.} \bibnamefont{Steinhardt}}, 
\bibinfo{year}{1998}, \bibinfo{journal}{Phys.\ Rev.\ Lett.} \textbf{\bibinfo{volume}{82}}, \bibinfo{pages}{896}.

\end{thebibliography}

\begin{table}[ht]
\caption{The early history of the Hubble constant. Unless otherwise shown, stated error bars were 10\% or less, or not given.}
\bigskip
\begin{tabular}{lclc}
\hline
\hline
When     &\quad &Who                      &Numerical value\\
\hline
1927     &      &Lemaitre                 &600\\
1929     &      &Hubble                   &530--513--465\\
1931     &      &Hubble \& Humason        &558\\
1936     &      &Hubble                   &526\\
1946     &      &Mineur                   &330\\
1951     &      &Behr                     &250\\
1952     &      &Baade, Thackeray         &270\\
1956     &      &Humas, Mayall \& Sandage &180\\
1958     &      &Holmberg                 &134\\
1958     &      &Sandage                  &150--75--38\\
1959     &      &McVittie                 &227--143\\
1960     &      &Sersic                   &125\\
1960     &      &van den Bergh            &125\\
1960     &      &van den Bergh            &125\\
1961     &      &Ambartsumyan             &140--60\\
1961     &      &Sandage                  &113--85\\
1964     &      &de Vaucouleurs           &125\\
1968--69 &      &de Vaucouleurs           &100\\
1969     &      &van den Bergh            &110--83\\
1968--76 &      &Sandage \& Tammann       &\enspace50\\
1972     &      &Sandage                  &\enspace55\\
1979     &      &de Vaucoulers            &100\\
\hline
\hline
\end{tabular}
\end{table}

\begin{table}
\caption{$H_0$ from secondary methods (the Key Project)}
\bigskip
\begin{tabular}{lcc}
\hline
\hline
       &      &Error\\
       &      &(random, systematic)\\
Method &$H_0$ &(\%)\\
\hline
36 Type Ia SN, $4000<cz<30,000$ km s$^{-1}$ &71 &$\pm2\pm6$\\
21 TF clusters, $1000<cz<9000$ km s$^{-1}$  &71 &$\pm3\pm7$\\
11 FP clusters, $1000<cz<11,000$ km s$^{-1}$ &82 &$\pm6\pm9$\\
SBF for 6 clusters, $3800<cz<5800$ km s$^{-1}$ &70 &$\pm5\pm6$\\
4 Type II SN, $1900<cz<14,200$ km s$^{-1}$ &72 &$\pm9\pm7$\\
\hline
\hline
\end{tabular}
\end{table}
\clearpage

\begin{turnpage}
\begin{table}
\caption{Mean absolute $B$, $V$, and $I$~magnitudes of nine SNe~Ia with known Cepheid distances, without and with corrections for decline rate and color}
\bigskip
\begin{tabular}{llccccccccc}
\hline
\hline
\multicolumn{1}{c}{SN} &\multicolumn{1}{c}{Galaxy} &$(m-M)^o$ &$M^0_B$ &$M^o_V$ &$M^o_I$ &$\Delta m_{15}$ &$(B-V)^o$ &$M^\mathrm{corr}_B$ &$M^\mathrm{corr}_V$ &$M^\mathrm{corr}_I$\\ 
\multicolumn{1}{c}{(1)} &\multicolumn{1}{c}{(2)} &(3) &(4) &(5) &(6) &(7) &(8) &(9) &(10) &(11)\\
\hline
1937C	&1C 4182 &28.36(12) &$-$19.56 (15)	&19.54 (17)	&...	&0.87 (10) &$-$0.02 &$-$19.39 (18) &$-$19.37 (17) &...\\
1960F &NGC 4496A &31.03 (10) &$-$19.56 (18) &$-$19.62 (22) &... &1.06 (12) &\enspace0.06 &$-$19.67 (18) &$-$19.65 (22) &...\\
1972E &NGC 5253 &28.00 (07) &$-$19.64 (16) &$-$19.61 (17) &$-$19.27 (20) &0.87 (10) &$-$0.03 &$-$19.44 (16) &$-$19.42 (17) &$-$19.12 (20)\\
1974G &NGC 4414 &31.46 (17) &$-$19.67 (34) &$-$19.69 (27) &... &1.11 (06) &\enspace0.02 &$-$19.70 (34) &$-$19.69 (27) &...\\
1981B &NGC 4536 &31.10 (12) &$-$19.50 (18) &$-$19.50 (16) &... &1.10 (07) &\enspace0.00 &$-$19.48 (18) &$-$19.46 (16) &...\\
1989B &BGC 3627 &30.22 (12) &$-$19.47 (18) &$-$19.42 (16) &$-$19.21 (14) &1.31 (07) &$-$0.05 &$-$19.42 (18) &$-$19.41 (16) &$-$19.20 (14)\\
1990N &NGC 4639 &32.03 (22) &$-$19.39 (26) &$-$19.41 (24) &$-$19.14 (23) &1.05 (05) &\enspace0.02 &$-$19.39 (26) &$-$19.38 (24) &$-$19.02 (23)\\
1998bu &NGC 3368 &30.37 (16) &$-$19.76 (31) &$-$19.69 (26) &$-$19.43 (21) &1.08 (05) &$-$0.07 &$-$19.56 (31) &$-$19.55 (36) & $-$19.31 (21)\\
1998aq &NGC 3982 &31.72 (14) &$-$19.56 (21) &$-$19.48 (20) &... &1.12 (03) &$-$0.08 &$-$19.35 (24) 
&$-$19.34 (23) &...\\
\quad Straight mean & & &$-$19.57 (04) &$-$19.55 (04) &$-$19.26 (06) & & &$-$19.49 (04) &$-$19.47 (04) 
&$-$19.16 (06)\\
\quad Weighted mean & & &$-$19.56 (07) &$-$19.53 (06) &$-$19.25 (09) & & &$-$19.47 (07) &$-$19.46 (06) 
&$-$19.19 (09)\\
\hline
\hline
\end{tabular}
\end{table}
\end{turnpage}
\clearpage

\begin{figure}
\caption{Hubble Space Telescope in orbit photographed by the crew of the Space
Shuttle Columbia, just after being released during mission STS-109 to
service and upgrade Hubble. March~9, 2002. Credit: NASA.
http://spaceflight.nasa.gov/gallery/images/shuttle/sts-109/html/s109e5700.html
}
\end{figure}

\begin{figure}
\caption{Comet P/Shoemaker-Levy 9 (1993e) ``String of Pearls,'' HST/WFPC2, January 1994.
Credit: NASA, H.~A.\ Weaver (JHU) and T.~E.\ Smith (STScI).\hfil\break
http://hubblesite.org/newscenter/archive/1994/13/
}
\end{figure}

\begin{figure}
\caption{Series of images showing the impact of Comet P/Shoemaker-Levy~9 (1993e)
fragment~W on Jupiter, HST/WFPC2, July 22, 1994.
Credit: NASA and the HST Comet Impact Team.
}
\end{figure}

\begin{figure}
\caption{The light curve from the impact of fragment~R at 2.3~$\mu$, adapted from \textcite{Graham:1995} and \textcite{Zahnle:1996}.}
\end{figure}

\begin{figure}
\caption{The viewing geometry of a typical SL-9 impact \cite[see text; adapted from][]{Zahnle:1996}.}
\end{figure}

\begin{figure}
\caption{Flat Projection of Comet P/Shoemaker-Levy 9 (1993e) Fragments~D (left) and~G
Impacts on Jupiter, HST/WFPC2, July~18, 1994. Two rings of propagating waves can be seen.
Credit: NASA and the HST Comet Impact Team.
http://hubblesite.org/newscenter/archive/1994/36/
}
\end{figure}

\begin{figure}
\caption{Globular Cluster 47~Tucanae (NGC~104), HST/WFPC2, July 1999.
Credit: NASA and R.~Gilliland (STScI).
http://hubblesite.org/newscenter/archive/2000/33/
}
\end{figure}

\begin{figure}
\caption{Light curve of the slight dimming of the star HD~209458 due to a
planet passing directly in front of it, HST/STIS, April-May 2000.
Credit: NASA, T.~M.\ Brown, D.~Charbonneau, R.~L.\ Gilliland, R.~W.\ Noyes, \& A.~Burrows.
}
\end{figure}

\begin{figure}
\caption{Montage of four images showing jets from young stars
HST/WFPC2. \hfil\break
\emph{Top Left}: HH~47
Credit: NASA and J.~Morse (University of Colorado).\hfil\break
http://hubblesite.org/newscenter/archive/1995/24/\hfil\break
\emph{Top Right}: DG~Tau~B
Credit: NASA, C.~Burrows (STScI), J.~Krist (STScI), K.~Stapelfeldt (JPL),
and the WFPC2 Science Team. 
http://hubblesite.org/newscenter/archive/1999/05/\hfil\break
\emph{Bottom Left}: HH~34
Credit: NASA, J.~Hester (Arizona State University), and the WFPC2
Investigation Definition Team. 
http://hubblesite.org/newscenter/archive/1995/24/\hfil\break
\emph{Bottom Right}: HH~1-2
Credit: NASA, J.~Hester (Arizona State University), and the WFPC2
Investigation Definition Team. 
http://hubblesite.org/newscenter/archive/1995/24/
}
\end{figure}

\begin{figure}
\caption{HH 30 disk/jet, HST/WFPC2.
Credit: NASA, C.~Burrows (STScI), J.~Krist (STScI), K.~Stapelfeldt (JPL),
and the WFPC2 Science Team.\hfil\break
http://hubblesite.org/newscenter/archive/1999/05/
}
\end{figure}

\begin{figure}
\caption{A bright one-sided microjet emerging from a proplyd located
   below Orion's Bright Bar. Adapted from Bally \emph{et~al.} (2000).
}
\end{figure}

\begin{figure}
\caption{Protoplanetary disks (Proplyds) in the Orion Nebula (M42),
HST/WFPC2. 
Credit: NASA, C.~R.\ O'Dell (Vanderbilt University), and M.~McCaughrean
(Max-Planck-Institute for Astronomy).
http://hubblesite.org/newscenter/archive/1995/45/
}
\end{figure}

\begin{figure}
\caption{Evolutionary tracks of 1~M$_\odot$, 5~M$_\odot$, and 25~M$_\odot$ stars in the 
Luminosity-Effective Temperature (Hertzspring-Russell) diagram. Thick segments mark long, nuclear burning, evolutionary phases \cite[from Iben 1985; adapted from][]{Prialnik:2000}.}
\end{figure}

\begin{figure}
\caption{The Luminous Blue Variable star Eta Carinae, HST/WFPC2, 1994. 
Credit: NASA and J.~Hester (Arizona State University).
http://hubblesite.org/newscenter/archive/1994/09/
}
\end{figure}

\begin{figure}
\caption{Rings around Supernova 1987A, HST/WPFC2, February 1994.
Credit: NASA and C.~Burrows (ESA/STScI).
http://hubblesite.org/newscenter/archive/1994/22/
}
\end{figure}
\clearpage

\begin{figure}
\caption{Planetary Nebula MyCn18, The Hourglass Nebula, HST/WFPC2.
Credit: NASA, R.~Sahai, J.~Trauger (JPL) and the WFPC2 Science Team. \hfil\break
http://hubblesite.org/newscenter/archive/1996/07/
}
\end{figure}

\begin{figure}
\caption{Nebula surrounding the symbiotic star system He2$-$104, the Southern Crab Nebula, 
HST/WFPC2, May 1999.
Credit: NASA and R.~Corradi (Instituto de Astrofisica de Canarias, Tenerife,
Spain), M.~Livio (Space Telescope Science Institute), U.~Munari
(Osservatorio Astronomico di Padova-Asiago, Italy), H.~Schwarz (Nordic
Optical Telescope, Canarias, Spain).
http://hubblesite.org/newscenter/archive/1999/32/
}
\end{figure}

\begin{figure}
\caption{A schematic distribution of the planetary nebulae morphology in the rotation ($\Omega$--magnetic energy density ($\sigma$) plane \cite[see text; adapted from][]{Garcia:1999}.}
\end{figure}

\begin{figure}
\caption{An HST-WFPC2 image of the circumstellar ring around supernova 1987A,
obtained in May 2002 with narrow band filters that include H$\alpha$ 6563~\AA\
and [NII] 6584~\AA\
emission line radiation. We can identify a number of bright spots that
correspond to gaseous protuberances on the inner side of the ring that
are currently being hit by the supernova ejecta, thus becoming hot and
luminous.  Because of the narrow width of the filters employed to gather
this image, the ring and the spots appear bright since they are mostly
radiating in emission lines.  On the other hand, both the supernova
itself (a wide patch at the center of the ellipse) and a star (projected
by chance near spot \#6) are not visible in this image because most of
their energy is radiated in the form of a rather smooth continuum. 
Credits: Nino Panagia (ESA/STScI), on behalf of the SINS Collaboration.
}
\end{figure}

\begin{figure}
\caption{Jet and disk of stars and gas in the active galaxy M87, HST/WFPC2, 1994.
Credit: NASA and H.~Ford (JHU).
http://hubblesite.org/newscenter/archive/1994/23/
}
\end{figure}

\begin{figure}
\caption{Black hole mass versus bulge luminosity (left), and the luminosity-weighted dispersion (right). Adopted from \textcite{Gebhardt:2000}.}
\end{figure}

\begin{figure}
\caption{Schematic gamma-ray burst from internal shocks and afterglow from external shock, arising from a relativistic jet produced by the collapse of a massive star. Internal shocks produce $\gamma$-rays and neutrinos, external shocks produce $\gamma$-rays, X-rays, optical and radio emission. Adapted from \textcite{Meszaros:2002}.}
\end{figure}

\begin{figure}
\caption{Hubble Deep Field North, HST/WFPC2, December 1995.
Credit: NASA, R.~Williams (STScI) and the HDF Team.
http://hubblesite.org/newscenter/archive/1996/01/
}
\end{figure}

\begin{figure}
\caption{Hubble Deep Field South, HST/WFPC2, October 1998.
Credit: NASA, R.~Williams (STScI) and the HDF Team.
http://hubblesite.org/newscenter/archive/1998/41/
}
\end{figure}

\begin{figure}
\caption{Optical and near-infrared images of distant galaxies in the Hubble Deep Field North. For each galaxy, the left panel shows a composite of WFPC2 images, while the right panel shows a new-infrared view with NICMOS. Courtesy of Mark Dickinson.}
\end{figure}

\begin{figure}
\caption{The star formation rate density versus redshift derived from the UV luminosity density. Adapted from \textcite{Ferguson:2000}.}
\end{figure}

\begin{figure}
\caption{The redshift evolution of the co-moving stellar mass density. The vertical extent of the boxes shows the range of systematic uncertainty. The bottom two solid lines (on the right-hand scale) show the result of integrating the star-formation-rate histories traced by the rest-frame UV light, with and without corrections for dust extinction. The top two solid lines and the dashed line show theoretical predictions from semi-analytical galaxy evolution models \cite{Cole:2000,Somerville:2001}. Adapted from \textcite{Dickinson:2003}.}
\end{figure}

\begin{figure}
\caption{Joint confidence intervals for ($\Omega_M$, $\Omega_\Lambda$) from SNe~Ia. Regions representing specific cosmological scenarios are illustrated.}
\end{figure}

\begin{figure}
\caption{Hubble diagram of SNe~Ia minus an empty (i.e., $\Omega_M + \Omega_\Lambda =0$) universe compared to various cosmological and astrophysical models. The points are redshift-binned data from \textcite{Riess:1998} and \textcite{Perlmutter:1999}. The observations of SN~1997ff are inconsistent with monotonic evolutionary or dust effects that could mimic an accelerating universe. Adapted from \textcite{Riess:2001}.}
\end{figure}

\end{document}